# Epigraphene : epitaxial graphene on silicon carbide

*Claire Berger[1,2], Edward H. Conrad[1], Walt A. de Heer[1]*

[1] School of Physics – Georgia Institute of Technology, Atlanta, GA 30332, USA
[2] Institut Néel, CNRS – Université Grenoble Alpes, 38042 Grenoble, France

## 1. Introduction and overview

### 1.1. Epitaxial graphene and transferred graphene

Graphene, a single sheet of carbon atoms, and the basic building block of graphite (see Figure 1) has been studied for more than half a century. Last century, academic graphene research focused primarily on surface science of epitaxial graphene on various metal surfaces, as well as on silicon carbide. Only some of its electronic properties were theoretically considered, but none were experimentally probed. In the past decade, initially fueled by its potential for electronics, graphene research has flourished, following two main distinct paths: graphene grown epitaxially on silicon carbide, or epigraphene[1], and graphene that is produced by a variety of methods and is designed to be transferred onto various substrates, known as transferred graphene. Basically, the two materials only differ due to the substrate that they are on, but this difference is fundamentally important. Much research has been devoted to producing large essentially perfect graphene sheets, which was considered by many to be an essential first step for graphene electronics. However, essentially all applications require graphene structures. Graphene electronics is an extreme case that requires highly reproducible graphene nanostructures in order to be technologically interesting. However, most nanolithographic pattering methods are detrimental to graphene. Consequently nanoelectronic graphene devices are not competitive with conventional nanoelectronics. Epigraphene is an exception, and is currently the only type of graphene that is suitable for graphene nanoelectronics. [1] [2]

Historically, the first examples of patterned epitaxial graphene for graphene-based electronics were published in 2004, in a paper titled: "Ultrathin epitaxial graphite and a route to graphene based electronics" [1]. The present review follows the history and the developments of that seminal work, that, incidentally presents the first (monolayer) graphene transport measurements.

In 2005, Novoselov et al. [3] invented a method now generally known as the "Scotch tape method". In this method, adhesive tape is used to cleave (exfoliate) graphene flakes from bulk graphite and transfer them onto oxidized silicon wafers, to demonstrate their field effect properties. In 2004 the same group reported very similar properties in ultrathin graphite flakes transferred on oxidized silicon (using a different deposition method). The paper, titled "Field effect in atomically thin graphite" [4], carefully demonstrates that the measured properties were those of thin graphite (see also similar work by Zhang et al. in 2005 [5]). Nevertheless this paper is widely cited not only for demonstrating graphene properties, but also for the discovery of graphene. The controversial discovery claim was based on the authors presumption, that freestanding graphene should be expected to be chemically unstable, unaware at the time that in 1962 Boehm et al.[6] had already produced and identified freestanding graphene sheets. Boehm also coined the name "graphene" in 1986 [7, 8].

Transferred graphene was originally characterized as "quasi-freestanding" in Ref [3] to distinguish it from other previously known forms of graphene directly grown on substrates including epigraphene. The name reflected the belief at that time that interactions with the SiO$_2$ substrate were negligible. Later research found that substrate induced disorder is significant. In contrast, epitaxial graphene on silicon carbide is widely used to demonstrate the intrinsic band structure of graphene[9-11] (see Figure 2) that is not observable in transferred graphene (see Figure 3 and Section 4).

---

[1] The term epitaxial graphene originally coined for graphene on SiC [1] is now widely used for graphene grown epitaxially – or not – on various metals. For this reason in this review we refer to epitaxial graphene on SiC as epigraphene.

Epitaxial graphene is a graphene film that is directly grown on various crystalline surfaces. The carbon atoms of the graphene layer are orientationally registered and nearly commensurate with the atomic lattice of the substrate surface. The degree of chemical bonding of the graphene to the substrate varies from extremely weak to relatively strong, and the graphene properties are modified accordingly. In contrast, there is in general no coherent atomic registration for graphene films that are transferred onto surfaces, for example on oxidized degenerately doped silicon wafers (graphene transferred on boron nitride single crystals is a notable exception but the registry is not controlled). Various transfer methods have been developed, either by drying graphene flake solutions [4, 6] or by direct mechanical transfer from graphite onto these surfaces (as first done by Novoselov et al. [12] in 2005). The Scotch tape method, has a great advantage for two dimensional electron gas (2DEG) investigations, [12, 13] because the charge density of the graphene layer can be adjusted by electrostatic "back gating", which cannot be done with epitaxially grown graphene. For back gating a parallel plate capacitor is made of a thin SiO$_2$ dielectric sandwiched between a conducting silicon substrate and the graphene layer(s) that is transferred on it. For epigraphene, by virtue of its growth process directly from the silicon carbide crystal, such an insulating barrier does not exist., However the transfer process has the disadvantage of considerable disorder inherent in the transfer.

Epitaxial graphene on silicon carbide (epigraphene) was first identified in 1962 by Badami [14, 15], followed by van Bommel et al. [16] in 1973, in the investigations of the graphite layers that spontaneously grow on silicon carbide when silicon carbide crystals are heated to extremely high temperatures (>1000 C) in vacuum. The growth proceeds by sublimation of Si from the SiC surfaces, resulting in a carbon-rich surface that reconstructs to produce graphene. It is interesting, that graphitic layers on heated SiC crystals were already noted in investigations by G.E. Acheson, who also first synthesized silicon carbide in 1891.[17] It is worth noting also that in 1907 H.J Round produced the first SiC light emitting diode and SiC diodes were used in the first radio receivers [18]. These are harbingers of SiC based electronics.

The invention of graphene-based electronics [19] (patented in 2003) was based on the earlier graphene research in combination with carbon nanotube electronics research. The choice of SiC as the substrate material was motivated by the strict requirements for high-end electronic materials (i.e. Si, Ge, GaAs, SiN, SiC). This requires a platform that is reliably nanopatternable, which demands that the substrate is a single crystal. Single crystal, electronics grade SiC is commercially available and it is currently extensively used in electronics. In the last decade, SiC based electronics has developed significantly including high temperature complementary metal oxide semiconductor (CMOS) technology [20]

As we will see in this review, epigraphene has been shown to satisfy requirements for electronics grade graphene. Moreover, various schemes to modify the properties of epigraphene using nanopatterning methods and by tuning the interaction with the SiC substrate have added significant flexibility as discussed below.

In contrast, most transferred graphene and thin graphitic flake research currently focuses on chemical properties, para-electronics and physical properties like super-capacitors, transparent and flexible conductors, in optoelectronic and photonic demonstrators and as ultrathin membranes [21, 22]. This shift in emphasis is primarily due to the inherent difficulties of reliably producing high-end electronics grade graphene-based materials by transferring graphene (grown epitaxially on metal surfaces or from graphite) onto various substrates as mentioned above. While there is still considerable research on fundamental graphene properties using small mechanically transferred graphene flakes, expectations that transferred graphene will significantly impact high-end electronics have recently faded. The current confusion of the definition of graphene, is not merely one of semantics, but actually confounds graphene research with nanographite research (a decades old, mature field).

## 1.2. Definition of graphene

Graphene was defined [7] by H-P Boehm in 1986, and the nomenclature was officially adopted[23] by the International Union of Pure and Applied Chemistry (IUPAC) in 1994. It reads as follows:

"Graphene is a single carbon layer of the graphite structure, describing its nature by analogy to a polycyclic aromatic hydrocarbon of quasi infinite size. Previously, descriptions such as graphite layers, carbon layers or carbon sheets, graphite monolayers have been used for the term graphene. Because graphite designates that modification of the chemical element carbon, in which planar sheets of carbon atoms, each atom bound to three neighbors in a honeycomb-like structure, are stacked in a three-dimensional regular order, it is not correct to use for a single layer a term which includes the term graphite, which would imply a three-dimensional structure. The term graphene should be used only when the reactions, structural relations or other properties of individual layers are discussed."

Consistent with this definition, "graphite" should be used when graphene layers are stacked in the graphitic (Bernal) structure (see Figure 1), so that "few layer graphene" is incorrect and should be referred to as "thin graphite".

In practice, graphene (including transferred graphene) is usually supported on a substrate. The term "freestanding graphene" is often used to suggest the absence of substrate-induced perturbations. Some even have suggested [24] to redefine graphene accordingly:

"Graphene is a single atomic plane of graphite, which—and this is essential—is sufficiently isolated from its environment to be considered freestanding".

However in practice graphene on a substrate is never freestanding. Substrates always affect the properties and quite significantly so in graphene transferred on $SiO_2$ using the "Scotch tape" method, making this alternative definition ineffective. It makes more sense to adhere to the IUPAC definition of graphene and to apply "quasi-freestanding" relative to a property. For electronic properties, quasi-freestanding implies that the electronic properties are essentially identical to those of an ideal graphene sheet. A recent publication in the authoritative journal Carbon suggests the usage of more precise definitions [25].

The adhesion of graphene to many substrates involving van der Waals forces and/or electrostatic forces usually minimally affects its chemical and electronic properties (see below). But the adhesion to the substrate can be strong, involving significant chemical bonding of the carbon atoms in the graphene layer to the substrate. In these cases the electronic structure (as well as planarity) will be significantly modified. Finally, graphene can also be functionalized, in which cases atoms or molecules are chemically bound to graphene carbon atoms, whereby the electronic structure is typically significantly modified. Consequently chemical functionalization and chemical bonding to a substrate are closely related. Below, examples of all three forms of graphene are presented in the context of epitaxial graphene on silicon carbide. Properties of "structured graphene" ribbons and islands are also discussed.

### 1.3. Graphite, freely suspended graphene, and graphene isolated on substrates

Until 1987, graphene was known as *monolayer graphite*, emphasizing its nature as a structural unit of graphite. Graphene was already known to be one of the most chemically and mechanically stable materials in nature, with a cohesive energy of ≈7eV. This extreme stability of the individual graphene sheets results from the $sp^2$ bond, as first described theoretically by Pauling [26]. Pauling demonstrated that the $sp^2$ hybridization produces three extremely strong symmetrically arranged coplanar bonds (the sigma bonds), which explains why polycyclic aromatic hydrocarbons (including graphene) are flat and rigid: it requires energy to bend these structures.

In addition to the in-plane sigma bonds, each carbon atom has an atomic $p_z$ orbital that extends above and below the plane. Overlap of these orbitals produces π-bands that are responsible for the electronic properties of graphene.

Graphene is so stable that it requires temperatures exceeding 4000º C (as occur, for example, in electric arcs) to convert it into fullerenes, nanotubes, soot and other curved graphitic structures, as was experimentally demonstrated in the late 1989 in the famous experiments by Krätschmer et al.[27] and Ebbesen et al.[28].

In contrast, the interlayer bonding of the graphene layers in graphite is extremely weak, on the order of 30 meV per carbon atom [29]. This small value (less that 1% of C-C bond energy in graphene) represents the difference in energy of a carbon atom in graphite compared to a carbon atom in a free graphene sheet. The weak adhesion explains why graphene layers are so easily peeled from a graphite crystal in the "Scotch tape method" for example. This anciently known property underlies graphite's name (graphite, from *graphein*, the Greek word for to write). Consequently, for all practical purposes, graphite's exceptional chemical and mechanical stability is entirely due to graphene's stability, and the added stability due to the crystal substrate is negligible. This explains the well-known fact, that almost all chemical and physical properties of graphene and graphite are practically identical.

Nowadays the most common forms are: aqueous graphene suspensions produced from exfoliated graphite[6, 30-32]; chemical vapor deposition on various metal substrates (for a review, see [33]); epitaxial growth on SiC by vacuum sublimation[15, 34]; deposition on various substrates by mechanical exfoliation of graphite [3].

The suspensions of graphene flakes, produced and isolated by H-P Boehm in 1962 were, identified with, for those times, novel electron microscopy methods, revealing that they indeed were graphene monolayers. These tour-de-force measurements were published in the most prominent German Science journal of the time [6]. As a carbon chemist Boehm knew that his work did not amount to the discovery of graphene, nor proof of its stability. This work was performed mainly as an academic exercise to demonstrate that freestanding graphene flakes could be made and measured.

Later, graphene was epitaxially grown on many single crystal metal surfaces, by heating metal samples in the presence of carbon containing gasses (chemical vapor deposition (CVD)). In these experiments performed in ultrahigh vacuum, the atomic and electronic structure can be probed using a variety of surface science probes. In 1997 Gall et al. [33] recognized the quasi-freestanding nature of epitaxial graphene on various metal surfaces, as well as the fact that they were natural two dimensional crystals. In their article titled "Two Dimensional Graphite Films on Metals and Their Intercalation", the first sentence of the abstract reads:

"Two-dimensional graphite films (2DGF) on solids are wonderful objects, real nature-made two-dimensional crystals…. A special attention is paid to intercalation of 2DGF — a process when foreign atoms and even molecules (fullerenes $C_{60}$ molecules) spontaneously penetrate between graphite film and metal substrate."

In this paper, and many other last century papers, the quasi-freestanding nature of the graphene layer is explicitly demonstrated, as it was by Forbeaux et al., in early investigations of epigraphene, who realized that the epitaxial graphene layers appeared to be "floating above the substrate".[35]

Several investigators pursued controlled mechanical exfoliation of graphene with explicit attempts to produce graphene by peeling layers from graphite. In 1999, Ruoff and coworkers published a paper titled: "*Tailoring graphite with the goal of achieving single sheets.*" [36] The process involved producing micron sized graphite islands on a substrate and mechanically peeling layers off with an atomic force microscope. The process produced ultrathin graphite flakes but not monolayers. In 2004 Novoselov and co-workers ultrasonicated islands produced by Ruoff's method resulting in suspension of micron-size flakes.[4] The suspensions were dried on oxidized silicon wafers. The electrical properties of micron sized ultrathin graphite flakes were measured. After Novoselov and coworkers measured transport properties of graphene flakes produced by the "Scotch tape method" in 2005 [3], Novoselov et al. [12] and Zhang et al. [13] simultaneously published quantum Hall effect measurements in graphene.

Graphene was first exfoliated chemically by separation of the layers through strong oxidation (graphite oxide, also called graphite paper), dispersion in aqueous solution followed by reduction [6]. Graphene can also be directly exfoliated in solution by ultrasonication in various solvents [31, 32] or using surfactants, similarly to carbon nanotubes and for stabilizing polymers [37], resulting submicron square (<10 $\mu m^2$ in area) graphene flakes.

## 2. The electronic band structure of graphene

Graphene is a semimetal. This means that there is no energy gap between the valence band and the conduction band. Graphene is therefore often referred to as a gapless semiconductor. It is quite curious that such a material could nevertheless be interesting for electronics applications that typically involve semiconductors with significant energy gaps (i.e. for Si the band gap is 1.1 eV). In fact, even gapless graphene can be used in certain electronic applications, while limited, like RF electronics. However, developing methods to "open a gap" in graphene is considered to be essential for digital electronics.

In order to understand the band structure of graphene, it is important to realize that the hexagonal graphene lattice is actually a triangular lattice with two carbon atoms in each unit cell, one called A and the other called B (see **Figure 1**). While both A and B are chemically equivalent, they are not equivalent from a crystallographic point of view, which leads to an interesting degeneracy resulting in two-component wave functions as explained below.

### 2.1. Tight-binding: graphene

The electronic structure of graphene is remarkably well described in the nearest neighbor tight-binding model involving only the $p_z$ orbitals[38, 39] A general review can be found in Castro-Neto et al. [40] where this model is used extensively. In this model the $sp^2$ bonding is represented by a quantum mechanical overlap integral $\gamma_0 \approx 3$ eV, which represents the overlap of $p_z$ orbitals centered on neighboring carbon atoms (see Figure 1b); when the $p_z$ orbitals of two neighboring atoms are in phase, the bond energy is $+\gamma_0$ when they are out of phase it is $-\gamma_0$, with values in between when the phase is between these extremes. In this way, waves develop with a wave vector **k**. These quantum mechanical "phase waves" that oscillate at frequency $\omega$ have energy $E=\hbar\omega$ and momentum $p=\hbar \mathbf{k}$, are the "quasi particles" that transport energy, momentum, and charge in graphene at a speed $v_F = d\omega/dk$ (much like a water wave transports energy and momentum, while the water itself only moves up and down). Quasiparticles with positive energy are called "electrons" those with negative energy are called "holes".

A plot of E(k) is the band structure as shown in Figure 2a

$$E(k) = \pm \gamma_0 \sqrt{\left(1 + 4\cos(\frac{\sqrt{3}}{2}ak_x)\cos(\frac{1}{2}ak_y) + 4\cos^2(\frac{1}{2}ak_y)\right)}$$

Where a= 0.246 nm is the graphene lattice constant (the C-C distance is 0.142 nm). For charge neutral graphene, i.e. at E=0, the Fermi surface consists of a hexagon of 3x2 points (K and K' points; $|\mathbf{K}|= 2\pi a/3$).

It is instructive to consider the case for which $k_x=0$ (i.e. a cut from along the $\Gamma$ K direction, $\Gamma$ corresponds to the k=0, see Figure 2a) for which

$$E(k_y) = \pm\gamma_0(1+2\cos(a k_y /2))$$

This shows two simple cosines that intercept at E=0 for $k_y=2\pi a/3$. Negative energies corresponds to bonding states (the $\pi$ bands), for which all of the $p_z$ orbitals at **k**=0 (the $\Gamma$ point) are in phase, while the positive energies corresponds to the anti-bonding states (the $\pi^*$ bands) where $p_z$ orbitals at **k**=0 of the A atoms are out of phase with those of the B atoms. At the K point (E=0) the bonding states and anti-bonding states are degenerate. The group velocity of the quasiparticles at the K-point is $v_F=a\gamma_0\hbar^{-1}\sqrt{3}/2 \approx 10^6$ m/s.

Near the K and K' points the band structure is reflection symmetric with E(k) cones that touch at the charge neutrality point where the energy E=0. These so-called Dirac cones are also reproduced in the simplified Hamiltonian and eigen energies, near the charge neutrality point:

$$H_k = v_F\, \sigma \cdot k^*; \qquad E(k) = \pm\, \hbar v_F k^* \qquad (\text{Eq. 1})$$

Where $k^*=|\mathbf{k}-\mathbf{K_i}|$ the Fermi velocity $v_F = \sigma$ are the 2x2 Pauli spin matrices $\sigma_x$ and $\sigma_y$ and **k** is the two component momentum vector, $k_x$ and $k_y$ and $\mathbf{K}_i$ represent the K and K' points. In this approximation near E=0, the quasi-particles move at a constant velocity independent of their energy, as for photons (or more appropriately, neutrinos, see below). This property is of fundamental importance for graphene electronics, since it sets a scale for quantum confinement effects. In particular, since the Fermi wavelength writes

$$\lambda_F = 2\pi/k^* = 2\pi\hbar v_F/E_F$$

for $E_F$=25 meV (corresponding to room temperature) one gets $\lambda_F$ = 160 nm. Hence, generally, room temperature quantum size effects are expected to be relevant for graphene structures up to this size (which is much larger than for typical metal structures).

## 2.2. Tight-binding: graphite

Wallace's paper [38] concerned calculating the band structure of graphite. In the graphite Bernal structure, graphene layers are stacked on top of each other, where the A atoms on one layer are directly above B atoms in the two neighboring layers, and the B atoms of that layer are above the centers of the carbon hexagons of neighboring layers (see Figure 1b-c). Consequently, the $p_z$ orbitals of the A atoms (but not the B atoms) in that layer weakly overlap with the B atoms in the neighboring layers. Due to this alignment, the A and B sublattice degeneracy is lifted, causing the electronic structure of graphite to be significantly modified near E=0, even though the interactions are weak. The Dirac cones near E=0 evolve into four hyperbolic bands, two that touch and two that are separated by an energy on the order of the interlayer overlap integral $\gamma_1 \approx 0.4$eV. Specifically, the eigen energies, near K points and for $k_z$=0, are then

$$E = \pm\gamma_1 \pm \sqrt{\gamma_1^2 + \left(\hbar v_F |k - K_1|\right)^2}$$

In this way the graphite quasi particles attain a small mass in the graphene plane (as well as a large mass perpendicular to the plane). The complete model (see for example Ref [41] involves 13 parameters that have been fitted to experiment (the McClure Slonczewski-Weiss parameters [39, 42]).

## 2.3. Ab initio methods

Both the graphene and graphite π-band structure are well described in the tight-binding approximation (of which the nearest-neighbor version discussed above is the simplest). These models provide analytical solutions that are appealing from a physics point of view, but they involve several empirical parameters. Modern ab initio computer-based simulations are used [43] that do not use free parameters. Density functional methods, of which the local density approximation (LDA) is the most commonly used variety, and the *GW* approximation, which stands for Greens function *G* of the Coulomb interaction *W*. This method is more accurate because it includes the long-range electron correlation effects that are absent in LDA. While numerically more accurate these computations typically do not provide analytical solutions resulting in some loss in transparency of the underlying physics. Ab initio calculations are particularly important in more complex problems, for example, when substrate effects, functionalization, edges and correlations are considered.

## 2.4. Relativistic interpretation

At the other extreme is the relativistic interpretation of the graphene band structure. Divincenzo [44] first recognized that in Wallace's nearest neighbor, tight-binding Hamiltonian, (Eq. 1) very close to the K points is formally identical to the Weyl Hamiltonian (also known as the massless Dirac Hamiltonian). The Weyl Hamiltonian describes neutrinos as massless spin ½ Fermions (see also Ando et al. [45] and Khveshchenko [46]). For neutrinos, σ in Eq. 1 represents the neutrino spin (its intrinsic angular momentum). For graphene σ is a two component vector (called pseudo-spin) that quantifies the π and π* components of the wave functions. Furthermore, in the neutrino case, c = 3×10$^8$ m/s (ultra-relativistic) replaces $v_F$=c/300 (non-relativistic speeds) in graphene. Note that relativistic interpretation of the nearest-neighbor tight-binding approximation at |**k**-**K**$_i$|≈0 introduces no new physics beyond the nearest neighbor tight-binding formulation. Moreover, it fails to describe graphene well away from E≈0 [40]. The analogy with relativistic quantum mechanics is formal, but beneficial because it allows mathematical developments in relativistic quantum mechanics to be applied to graphene. Unfortunately, the "relativistic nature" of quasi particles in graphene has led to considerable confusion, especially in the popular science literature.

## 2.5. Physics near the K point

The electronics properties of graphene near the K points are particularly interesting. There, the Fermi wavelength $\lambda_F = 2\pi v_F/k^*$ diverges as $k^* \to 0$. Moreover, since graphene is neutral at the K point, the electric fields of the quasiparticles are not screened. This implies that the quasiparticles strongly interact with each other resulting in long-range interactions. Therefore they cannot be treated as approximately independent particles (as is usually the case in metals and semiconductors),. Furthermore, due to the strength of the interaction, the usual theoretical many body techniques, based on perturbation theory, fail near the K point. This can be understood as follows: in quantum electrodynamics calculations, the perturbation expansions involve powers of the fine structure constant $\alpha = e^2/\hbar c \approx 0.0073 \ll 1$, typically resulting in converging expansions. However, for graphene $\alpha = e^2/\hbar v_F \approx 2.2 > 1$, so that the expansions tend to diverge. It is generally expected that these interactions result in a broken symmetry ground state. Furthermore, lack of screening near the K points is also generally expected to greatly increase the Fermi velocity. There has been some recent evidence for this effect [47]. In any case, it is clear, that the graphene must be exceptionally pure for effects close to the K point (with diverging Fermi wavelengths) to be observed. Impurities like charge puddles, random strain, and other defects, even at low concentrations n, will distort the pristine graphene properties at wavelengths $\lambda \approx 1/\sqrt{n}$.

The many body effects near the K point have for a long time been one of the most interesting unanswered questions in graphene physics and they are starting to be addressed experimentally.

## 2.6. Influence of the substrate : epitaxial graphene

In graphene grown epitaxially on a substrate, the graphene lattice is (almost) commensurate with the substrate lattice, thereby producing a higher order (super cell) modulation of the lattice (examples of higher order patterns for graphene on the buffer layer – see Section 4.1)– are given in Figure 4a-b). The unit cell of the (super cell) is much larger than that of graphene. In general, within a super cell, neither A or B graphene atoms are singled out. Consequently, the graphene A-B sublattice degeneracy is not lifted. Furthermore, if the interaction strength of the substrate to the graphene layer is weak (i.e. van der Waals like and no significant $sp^3$ bonding), then the electronic structure near the K points is not greatly perturbed. Consequently, in those cases, the graphene may be considered to be quasi-freestanding regarding the electronic properties at the K-point, as discussed by Gall et al. [33] . However, the super cell periodicity will cause energy gaps on the order of the interaction strength at periodic intervals in the Brillouin zone [48]. In other cases the bonding to the substrate is so strong that the $p_z$ orbitals of the graphene layer hybridize with the substrate atoms, producing $sp^3$ bonds with the substrate. [33] These chemical bonds significantly affect the graphene electronic properties.

Graphite, or "Graphene on graphite", discussed above can be seen as a special case of epitaxial graphene. But it is exceptional because in this case A-B symmetry is lifted. In Figure 2b-c, the effect A-B symmetry lifting is demonstrated by band structure measurements of AB and ABC stacked epigraphene on SiC (stacking defined in Figure 1a-d).

## 2.7. Role of stacking: multilayer epitaxial graphene

Graphene multilayers grown on SiC in some cases are not Bernal stacked as they are in graphite, but rather, adjacent layers are rotated by about 30 degrees (see Section 4.2.3). In this case, even though the interlayer interactions are very similar to those in graphite, the A-B sublattice degeneracy is not lifted (as it is in graphite) and consequently, the Dirac cone structure is essentially unperturbed [9] In this way, these multilayers behave like a stack of decoupled graphene layers [49-54]. Multilayer epitaxial graphene (MEG) has been studied in great detail, and has provided a wealth of information on the properties of graphene close to the Dirac point (see Section 4.2).

## 2.8. Influence of substrate: transferred graphene

Mechanically exfoliated graphene is graphene transferred from graphite to a secondary substrate. The method requires that adhesion of graphene to these substrates is greater that the adhesion of graphene to graphite, otherwise graphene would not transfer. Note that when these transfers are performed in ambient conditions (not high vacuum), contamination trapped between the graphene layer and substrate is unavoidable, giving rise to charge inhomogeneity (charge puddles) [55-57]. Furthermore since graphene is flexible, significant stress results from the transfer process on the non-flat substrates. This random stress causes disorder in the electronic structure of the graphene [57-59]. Both sources of disorder affect the electronic properties in particular near E=0 in an energy range that is comparable to the energy of the interaction with the substrate and characteristic energies related to the disorder (typically hundreds of meV) [56, 59]. This disorder may obscure the physics near the Dirac point. [59]

In order to mitigate some of these problems, graphene has more recently been deposited on single crystals of boron nitride [48, 60](a semiconductor with an atomic structure that is similar to graphene), producing a quasi-epitaxial form of graphene that is considerably better ordered than graphene deposited on amorphous substrates.

### 2.9. Graphene nanostructures

While most research has focused on the electronic properties of extended (quasi two-dimensional) graphene sheets, graphene nanostructures are technologically more relevant and theoretically quite interesting in their own right. As mentioned above, quantum confinement effects are manifested in relatively large graphene nanostructures even at room temperature. In fact, graphene electronics was originally motivated [1, 19] by the predicted properties of graphene nanoribbons that have much in common with carbon nanotubes.

However, the advantage of graphene ribbons compared with carbon nanotubes is that ribbons can be seamlessly interconnected using graphene leads [1], so that entire graphene integrated circuits could in principle be cut out from a single graphene sheet. In this way, major hurdles hindering the development of carbon electronics based on carbon nanotubes (i.e. placement and interconnects) could be overcome.

The first tight-binding calculations performed on graphene nanoribbons were performed by Fujita et al. [61] and Nakada et al. [62] in 1996, who showed that graphene ribbons could be semiconducting or metallic, depending on the width W and the structure of the edges, as shown Figure 5. These simple calculations show that relatively large bandgaps due to width confinement, of the order of Eg=1eV /W (the width W is in nm), can be achieved by patterning narrow ribbons, using, for example, nanolithography methods. While realization of these nanostructures is certainly quite difficult, because of the small sizes and high accuracy required, the overall paradigm of a monolithic structured graphene sheet as a foundation for carbon circuitry is a new direction in electronics (see Section 5.5).

Because the Fermi velocity in graphene is nearly constant, the energy levels $E_{n,m}$ of a rectangular graphene structure, of length L and width W are closely related to the normal modes of a optical wave guide. Consequently,

$$E_{n,m} = hv_F \sqrt{(n/L)^2 + (m/W)^2}$$

Eq. 2

This is a reasonable approximation but it ignores the particular properties of the edge states mentioned above (Figure 5). It does show, however that when L and W are relatively small, at low temperatures and in charge neutral graphene, only the lowest modes are occupied. For example, at T=4K (corresponding to liquid He) a charge neutral sample of 10μmx10μm has $E_{1,0}$(= 4 meV) >>$k_B$T, so that only the n=m=0 mode will be occupied. A 100 nm wide ribbon will only have the m=0 modes occupied and therefore it behaves like a one dimensional wire, even at room temperature (see Section 5.5).

## 3. Silicon carbide and epitaxial graphene on silicon carbide

### 3.1. G.E Acheson: Silicon carbide, Graphite and Graphene

Silicon carbide is a synthetic compound, first mass-produced by E.G. Acheson in 1891 (who called it Carborundum) by heating aluminum silicate with carbon. This process was patented in 1893 and Acheson founded the Carborundum Company. Silicon carbide was immediately industrially interesting because of its extreme hardness, so that it was used as an abrasive. Interestingly, Acheson also discovered that when carborundum was heated to a high temperature it produced extremely pure graphite. He patented his graphite-making process in 1896 and founded the Acheson Graphite Company in 1899. The company initially produced graphite for incandescent light filaments and electrolysis anodes and electrodes for arc furnaces. Interestingly Acheson also discovered a way to produce ultrafine graphite colloidal suspensions in water, that he produced in his Acheson Industries Company under the trade-name AquaDAG (DAG: Deflocculated Acheson Graphite). These "graphene-like" conducting inks continue to be used extensively in the electronics industry to produce conductive coatings inside electronic vacuum tubes and cathode ray tubes ever since the early days of radio and television. It is now used for the electronic, automotive, aerospace, appliance, medical, metalworking, die casting and energy storage industries [32].

### 3.2. Electronics grade silicon carbide

Crystalline silicon carbide is a wide band gap semiconductor that occurs in more than 150 crystalline forms (polytypes) (see for instance Ref. [63]). The most relevant polytypes for electronics and epigraphene are hexagonal 6H (band gap: 3.05 eV) and 4H (3.23 eV) and cubic β -3C (band gap: 2.36 eV) (see Figure 6). The structures consist of Si-C bilayers which stacking determines the polytype. The 4H-(aSiC=3.0805Å and cSiC=10.0848 Å[64]) and 6H-hexagonal crystals consist of 4 (6, respectively) Si-C bilayers (see Figure 6). The 4H commercial crystals are generally considered of higher quality. The hexagonal planes are polar, that is the (0001) face is made only of Si atoms (the bulk terminated surface Si atoms have one dangling bond per atom), and the opposite (000-1) has carbon atoms only. Single crystal wafers of these two polytypes with diameters up to 150 mm are produced by several companies. These crystals are produced by the physical vapor transport method (a modified version of the original Lely process), at elevated temperatures above 2000ºC in a closed graphite crucible. SiC powder, placed in the hot zone of the growth cell (e.g. 2200ºC), sublimes and re-crystallizes in the colder zone (e.g. 2150ºC) at a SiC seed. Control of various defects ("micropipes" and screw dislocations) has plagued SiC electronics industry from the outset, however improvements in the quality as well as reduction in cost have been dramatic in the past decade, and this trend is expected to continue.

Electronics grade SiC is commercially available, produced as semi-insulating and degenerately n-doped SiC. For semi insulating SiC, the Fermi level is pinned in the band gap for example by doping with vanadium. SiC is widely used in electronics, especially for discrete components like light emitting diodes and high voltage Schottky diodes. Recently, high voltage (1200 V) MOSFETs have become commercially available, and high temperature CMOS integrated circuits have been demonstrated.[20]

### 3.3. Epitaxial graphene growth by silicon sublimation

Epitaxial growth of graphene on SiC is non-conventional. Epitaxial growth of graphene on metals typically involves the decomposition of a carbon-containing gas on single crystal metal surface, after which the carbon atoms arrange into the graphene structure that registers with the crystalline metal surface. However, epitaxial graphene growth on SiC is different and typically does not involve an external source of carbon. Rather, silicon sublimes from the hot SiC surfaces, resulting in a carbon rich surface that then forms a graphene layer on the SiC surface.

Van Bommel et al. [16] was among the first to study this graphitization of SiC (as used by Acheson) as a surface science problem (see also [14]). The motivation for these early studies was to understand the surface phase transitions of clean SiC as a function of temperature. For this, 6H-SiC crystals were heated in ultra-high vacuum and their surfaces were measured with low energy electron diffraction (LEED) and Auger electron spectroscopy as a function of annealing temperature. It was discovered that as temperature increases a monolayer graphite layer grows and stable graphite film ultimately formed on the SiC surface [16]. This early work also showed that graphene grows differently on the two polar surface of hexagonal (4H/6H) SiC: the (000-1) carbon terminated face (C-face) and the (0001) silicon terminated face (Si-face) (see Figure 6 and schematics Figure 7d). On the Si face the graphitization proceeded with an initial phase at $1000°C$ involving a carbon rich reconstructed SiC surface with a 6 $\sqrt{3}$ R30 crystal structure (buffer layer, see Section 4.1.1). This is followed by the formation of an epitaxial graphene layer at higher temperatures and ultimately thin (Bernal) graphite at 1500ºC. The C-face, on the contrary, showed signs of different graphene-SiC orientations. While in the years from 1975-2000 there were many studies of Si-face graphene, very few were conducted on C-face graphene because of its perceived disorder (for a review of the earlier work see Hass et al. [65]). As discussed in Section 4.2.3, the C-face is in fact ordered but the multiple layers are non Bernal stacked, which has important consequences for its electronic properties.

It should be noted that graphene layers growth by silicon sublimation in ultra-high vacuum (the growth method used in earlier studies) typically results in lower quality films that have a so-called "Swiss cheese" nanostructure as shown in Figure 4(d). For this reason, alternative graphitization methods have been developed [34, 66].

### 3.4. Confinement controlled sublimation (CCS)

In free sublimation of silicon in vacuum, the sublimation rate is controlled only by the temperature. If one considers a silicon carbide crystal in a background of silicon vapor, then equilibrium will be established when the rate of absorption of silicon atoms on the surface equals the sublimation rate. Consequently, this equilibrium temperature depends on the silicon vapor pressure at the SiC surface [67]; in equilibrium the formation of graphene is arrested.

Free silicon sublimation (in UHV) produces defective graphene layers because the dynamics controlling the formation of graphene at a given temperature cannot keep up with the rate at which free carbon is produced at the surface, resulting in poorly developed graphene layers. Consequently, either slowing down the rate of carbon liberation, or increasing the temperature so that the graphene formation process is speeded up should produced better graphene. At the same time, growth over the entire sample surfaces will be uniform if the growth processes occur close to equilibrium.

The confinement controlled sublimation method [34] is designed to provide growth conditions close to equilibrium. In this method, a small silicon carbide crystal is placed in a graphite ampule that is supplied with a small calibrated hole (Figure 7b). The assembly is placed in a vacuum chamber that it is uniformly heated (typically using an radio frequency induction heater, as shown in Figure 7c) to a given temperature (in the range of 1500-1700ºC). The silicon that evaporates from the SiC surface will dwell inside the ampoule for relatively long times, before it can escape out of the hole. Consequently, a silicon background pressure builds up and growth occurs in quasi-equilibrium conditions. The rate of graphene formation is directly related to the rate at which the silicon escapes. By confining silicon, the growth temperatures can be increased. Ultimately, the quasi-equilibrium growth at elevated temperatures assures a uniform graphene layer extending over the entire crystal surface. Optimum growth conditions (temperatures, growth times, and hole size) are empirically determined (see sections 4.1.2 and 4.2.2).

### 3.5. Growth in Ar atmosphere (Edison light bulb method)

The rate at which silicon is depleted from the surface, and therefore the graphene growth rate, can also be controlled by heating the silicon carbide crystals in an inert gas background. While the silicon sublimation is unchanged, these atoms must diffuse through the Ar gas in order to escape. This diffusion process results in a Si density gradient in the gas, where the Si density is greatest at the surface. Hence, close to the surface the number of atoms leaving the surface is approximately equal to those returning to it. As in the confinement controlled sublimation method, the return flux of silicon atoms (that depends on the Ar pressure and temperature) and therefore the graphene growth rate can be adjusted. This method pioneered by Emtsev et al. [66] has been successfully applied to produce relatively uniform epigraphene/SiC at the full wafer scale [68-70].

While this method (which is similar to the way tungsten evaporation in incandescent light bulbs is controlled) appears to be similar to the CCS method (see for instance surface studies in Figure 8a-b), there is a fundamental difference. Whereas in the CCS method growth occurs under essentially uniform conditions over the whole surface, in Ar growth, it does not, because it is determined by diffusion of silicon atoms in the Ar gas. Growth rates depend on the boundary conditions and therefore sensitively on the chamber geometry, location on the sample, and critically on Ar flow and convection. Typically the growth temperature of a monolayer epigraphene on the Si-face ranges from 1100ºC in UHV [35] to 1550ºC in the CCS method (depending on the crucible hole size [34]), up to 1650ºC [66] and 2000ºC in one atmosphere of argon [71].

Variations of the high pressure growth have been implemented, for instance by supplying Si with (di)silane gas [72], or by providing external carbon atoms from gas (chemical vapor deposition, see for instance [73]) or solid carbon sources (molecular beam epitaxy [74]). In a recent development, higher mobility samples on the Si-face for quantum Hall effect metrology (see Section 5.2.3) were obtained by balancing both carbon etchant ($H_2$) and carbon supply gases in a process combining chemical vapor deposition and SiC thermal decomposition. [75, 76]

## 4. Structure and band structure of epitaxial graphene on hexagonal silicon carbide

Since the early studies started in the 60's [14, 16, 35], epigraphene production methods have improved significantly so that higher quality material is now produced [34, 66]. The improved material, coupled with the ability to use a variety of modern surface science techniques have allowed more detailed electronic and structural information to be uncovered. No other form of graphene has the crystallinity, epitaxial registry and large-scale patternability that allow a broad range of surface analytical tools necessary to study and modify the properties of graphene. As an example Figure 3 shows a comparison of the surface roughness and K-point Dirac cone energy and momentum spread of various forms of graphene. AFM and STM images in Figure 3a-b show that the surface roughness is an order of magnitude larger in transferred graphene. Angle Resolved Photoemission Spectroscopy (ARPES) measurements show the dramatic difference between epigraphene and transferred graphene on $SiO_2$ (see Figure 3c-d). For large scale graphene grown by CVD on Cu, graphene studies are hindered by rotational disorder and the band gap induced by substrate interactions [77]. At the time of writing only epigraphene is sufficiently ordered to allow graphene's characteristic band structure to be revealed.

Graphene growth on the two 4H-(6H-)SiC polar surfaces are very different. Under identical conditions, graphene growth on the Si (0001)-face, is both slow and orientationally well defined with SiC. On the carbon face, epigraphene grows fast and a graphene layer readily forms. Multilayers subsequently grow with a variety of orientations that are commensurate with the SiC substrate, as can be seen in the LEED patterns of Figure 9 a-b. The status quo of the electronic and crystalline structure epigraphene is reviewed next.

### 4.1. Si-face

On the Si-face, graphene is rotated $30°$ with respect to the SiC <10-10> direction (see **Figure 9**b). [35, 65] Film thicknesses (up to about 5 layers) can be accurately controlled by adjusting growth conditions. The first graphene layer (the so-called buffer layer) is bound to the SiC and the π-bands are modified; instead of a Dirac point, a band gap is observed at the K-point.[78] The second layer (i.e. the graphene layer on top of the buffer layer) shows the text-book graphene band structure. Subsequent layers show characteristic thin graphite behavior.

The first ARPES graphene band structure measurements of the Si-face were performed by Rollings et al. [79] and by Ohta et al. [10]. They studied UHV grown graphene films (1 to 4 layer not counting the buffer layer). Figure 2b-e show the ARPES spectrum at the graphene K-point as a function of film thickness. The linear π-bands (Dirac cone) are clearly observed for the monolayer film (Figure 2b; note that the monolayer film is n-doped by ∼0.45eV due to the SiC substrate. The AB stacking (Bernal stacking) of the second layer, that causes the lifting of the degeneracy of the two graphene sublattices, is evident from the splitting of the π-bands (Figure 2c). The bands are shifted to lower binding energy (∼-0.5eV). Also note that a small band gap has formed. This gap that lies well below the Fermi level can be varied from 0 to 0.25 eV by changing the doping level. [80] The gap originates from the perpendicular electric field due to the difference in doping of the two layers [81]. ARPES shows that the graphene stacking in thicker films is a mixture of Bernal and rhombohedral stacking (see Figure 2d-e). [10, 82]

Electron- phonon, electron-electron and electron-plasmon interactions, are observed in detailed ARPES studies of epigraphene Si face that has the required structural order. Disorder prevents these effects to be observed in other forms of graphene [83-85](for a review see Basov et al. [86] and refs therein). For instance Figure 10 presents non-linearities close to the Dirac point and a kink in E(k) in epigraphene Si-face and potassium-doped quasi-free standing epigraphene. These have been explained by a band renormalization due to many-body interactions (electron–plasmon coupling, plasmarons - see Section 6.4)[84]. Note that in another set of experiments the kink feature was ascribed to a substrate-induced gap.[87]

### 4.1.1. Buffer layer

The existence of the buffer layer (also called the zeroth layer or nanomesh) was not known in the earlier literature. LDA calculations predicted its existence in 2007 [88, 89] and experimental verification soon followed [90, 91]. The buffer layer is a graphene layer that is bonded to the SiC. Little is however understood about its structure, stability, electronic properties and the nature of the bonding. While, LEED shows a $6\sqrt{3}$ periodicity (Figure 9b), early STM measurements showed only a 6x6 reconstruction (see Figure 11a). [92-96] [97] Later STM measurements, using a lower tunneling voltage of ~0.2eV, revealed the $6\sqrt{3}$ periodicity (Figure 11b) [98]. This higher order reconstruction is poorly ordered (with coherent domain sizes of only a few $6\sqrt{3}$ unit cells i.e.~5-6nm). This explains why surface X-ray diffraction (SXRD) finds no evidence of a $6\sqrt{3}$ structure [99]. STM experiments also show groups of adatoms (possibly trimers) at the interface of UHV grown buffer.[90] These adatoms may be responsible for the surface reconstruction, although other studies rule out any significant presence of Si within the interfacial layer [100].

Riedl et al. [91, 101] demonstrated that the buffer layer is structurally a graphene layer. They showed that by intercalating hydrogen between the SiC and the buffer layer, the buffer layer transforms reversibly to a normal graphene layer (Figure 8c). This newly formed graphene layer is simply vertically displaced by 2.1Å along with the other graphene layers on top, if there are any, as shown by high-resolution X-ray reflectivity [102]. Moreover, when the hydrogen is intercalated in a monolayer graphene film, the monolayer graphene electronic structure transforms into that of an AB-stacked bilayer (Figure 8d).

The decoupling of the buffer from the substrate by hydrogen intercalation is an example of the ease that foreign atoms intercalate at the SiC-buffer layer interface. Intercalation and buffer-layer decoupling have been demonstrated with Si [103], Ge [101, 104], $O_2$ [105], $H_2O$ [106], Au[107], Li [108], Pt [109], F [110] and others.

The crystallographic structure of the buffer layer is essentially that of graphene, but interactions with the substrate transforms it into a semiconductor. Early band structure work[35] [78] on UHV grown samples indicated a large band gap, and two non-dispersing surface states $g_1$ and $g_2$ are observed [78] located 0.5 eV and 1.6 eV below the Fermi level (see Figure 11d): the distance from the valence band maximum to $E_F$ is ~2eV indicating that the band gap is at least 2eV wide. However, STS measurements of the band gap shown in Figure 11c gives a much smaller gap value of ~0.4eV. [90] This discrepancy remains to be resolved.

The large unit cell size of the buffer layer impedes an exact determination of its structure. Based on photoemission spectroscopy measurements of the C1s core level, it was argued [78] that one third of the carbon in the buffer layer is bonded to Si in the SiC below. A more detailed x-ray standing wave enhanced XPS (XSW-XPS) measurement [100] indicates that of the two C1s peaks corresponding to the buffer, one (S1 at 284.75eV binding energy) is attributed to C-C bonds in graphene, and the other (S2 at 285.55 eV binding energy) is consistent with C-Si bonds. The S1 carbon is 2.39Å above the last Si layer in the bulk SiC while the S2 carbon is closer to the bulk at 2.07Å. The distance of the buffer to the SiC surface is in agreement with transmission electron microscopy [111] and earlier x-ray reflectivity experiments that put a non-flat graphitic layer (0.16 Å rms thickness) 2.3 Å above the last SiC layer [112].

The in-plane lattice constant of the buffer layer is found [113] to be tensile strained relative to graphite. What is surprising is that when the buffer layer is covered with graphene, not only is the buffer layer's strain reduced by more than half, but the buffer and monolayer lattice parameters are incommensurate with each other (epigraphene is compressed compared to graphite). Clearly more experimental work will be required to clearly understand the buffer-SiC structure.

Theoretical predictions of the structure of the buffer or monolayer on the Si-face have been similarly limited by the long computation time when dealing with a very large $6\sqrt{3}$ unit cell [88, 89] Calculations of a fully relaxed $6\sqrt{3}$ graphene-SiC cell starting from a bulk terminated (0001) surface [114, 115] have shown that a superhexagonal mesh develops that is at least consistent with the 6x6 periodicity observed in STM. [93, 98, 115] While the calculated bands [114] show some features that are similar to the measured bands of Emtsev et al. [78] in Figure 11d there are a set of distorted π-bands that appear near the K-point. In the calculations, these bands are due to the π-bands of carbon atoms in the buffer layer that are not bounded to Si atoms in the SiC surface. The non-bonded carbon atoms form a superhexagonal network seen in STM (Figure 11b). Whether or not the predicted superhexagonal distorted π-bands can be observed experimentally remains to be determined.

### 4.1.2. Growth mechanism

The growth of epigraphene is unusual because it proceeds by decomposition of a crystal rather than by adding atoms to a surface. Very few theoretical studies have addressed this question [116, 117]. STM studies [90, 92, 118, 119] agree that the top layer is continuous and that subsequent graphene layers growth below it. STM, AFM, LEEM and cross-sectional Transmission Electron Microscopt (TEM) observations show fingerlike structures on the Si-face (Figure 12a). Based on these observations, growth models have discussed graphene nucleation at step edges. The terraces recede to remove enough Si to form graphene. The key assumption is that the decomposition of a SiC step edge depends on the local curvature of the front, so that the growth is understood from a competition between capillary smoothing and a decomposition driven step edge roughening (see Figure 12c-d). This is described by a linear stability analysis of a step equation of motion as a function of growth process variables: temperature, background Si and inert gas pressure. [116] It has been argued that epigraphene growth is diffusion-controlled and therefore naturally leads to instabilities (such as the finger-like features ofFigure 12a) instead of the advancement of straight steps. [120] However, simple step flow growth can occur for a step height of three SiC bilayers, which correspond almost exactly to the number of carbon atoms required to form a graphene layer [120]. As seen in Section 4.5, patterned steps have been introduced to direct graphene growth at step edges.

Models were further developed [117] for graphene growth on non-planar nano-faceted 6H–SiC substrates; the model parameters are the effective energy barriers for the nucleation and propagation of graphene at the SiC steps. The main result, in agreement with kinetic Monte Carlo simulations, is that the original nano-facet is fractured into several nano-facets during graphene growth, which is related to terrace width distribution. [117].

Because the growth of epigraphene is driven by Si evaporation, atomic transport of silicon and carbon on graphene and in-between layers is critical to the understanding of multilayer growth processes. Results from density functional theory calculations show that Si atoms can move almost freely on graphene and between graphene layers, while C atoms have much larger diffusion barriers. The results provide an explanation of the high Si sublimation rates during the growth of epigraphene even after graphene layers are formed on the surface.[121]

## 4.2. C-face

The main difference between graphene on the Si- and C-face is as follows. Graphene grows very fast on the C-face compared to the Si-face. With the CCS method [34] 5-10 layer films are typical and films thicker than 20 layers are easily grown while single layer films are much harder to achieve [34, 122-125]. There is no evidence of a buffer layer on the SiC(000-1) surface. However SiC interface reconstruction may occur, as the 2x2 and 3x3 reconstruction observed in UHV-grown samples [126]. Subsequent layers show a mostly ordered stacking with a distribution of relative rotations alternating around 0º and 30º rotations. These various rotations correspond to graphene/ SiC commensurate structures. This is clearly shown in the C-face graphene LEED pattern in Figure 9c, where diffuse intensity arcs are seen in stead of the sharp hexagonal pattern of diffraction spots seen in Si-face graphene.

### 4.2.1. Microstructure

In well-formed multilayer samples [34] the top graphene layer is continuous and drapes over the steps on the SiC substrates and no grain boundaries are observed in STM [127]. AFM and STM images show that the graphene film has pleats (also called folds, puckers, ridges, creases, rumples or wrinkles in the literature) [49, 128]. The pleats are typically 1–10 nm high, are typically spaced 3–10 μm apart and are thought to result from the differential expansion of the silicon carbide and graphene and the very weak coupling of the graphene to the substrate. The monolayer pleats can be easily displaced by an AFM tip, in a manner of nanoscale ironing, showing the weak graphene – substrate interaction.[124]

Estimates of the coherent domain size from Raman, STM, and transport measurements indicate that graphene domains are large, 400–1000nm [49, 53], with a lower limit of 300 nm as observed in surface x-ray diffraction (SXRD) [129] (resolution set by the SiC steps that destroy the x-ray coherence). The film roughness is also extremely small, with atomically flat terraces [128] (in contrast to the nm roughness typically observed in transferred CVD films) (see Figure 13a, b, e).

### 4.2.2. Growth

There is indication that graphene can also nucleate on C-face terraces [123, 124, 130], in contrast with Si-face graphene growth that is mostly promoted at step edges. Graphene is more readily formed where Si can escape easily from the SiC substrate, that is at defects[130], screw dislocations in SiC [71] and holes in the graphene film acting as "volcanoes" [131]. Because of this, although large flat and uniform areas are possible (Figure 13c), thickness uniformity is more difficult to achieve on the C-face and thicker films tend to show more thickness variations as demonstrated by LEEM IV analysis[132] (Figure 13d).

Various scheme have been developed to realize a more uniform growth on the C-face. For large-area graphene, the key is to monitor the Si out-diffusion to grow graphene in near equilibrium condition, either by confining the Si vapor in an enclosure (CCS growth [34, 49], see Section 3.4), by applying a noble gas back pressure (similar to the Si-face in [66]), limiting the Si escape with a graphite cap [124] or in between 2 SiC chips [133], or maintaining a Si partial pressure with a external flux of Si (similar to Si-face work in [72]). Excess graphene was also etched during growth with $H_2$[125]. Another approach is to functionalize the SiC surface prior to growth, and it was shown that nitrogen –seeded 4H-SiC surface induces more thickness uniformity on top of C-face narrow mesas.[134]

From a device perspective, it is obviously advantageous to grow nanostructures at predefined locations. To this end structured growth methods have be developed, the most promising of which is to etch patterns into the SiC substrate prior to growth, that serve as template for graphene nanostructures [135, 136]. Preferential growth was also induced in pits and mesas [137] (see also improved uniformity on the Si face of mesas [138]). Capping AlN [139] or removable SiN masks [140] have been successfully used to prevent or enhance graphene growth in the masked area.

### 4.2.3. Rotational stacking

Unlike on the Si-Face, graphene on the C-face is not Bernal stacked. It must be emphasized that this C-face graphene stacking is not "turbostratic graphite". In the literature, turbostratic graphite refers to disordered graphitic materials composed of non-Bernal stacked platelets that are typically on the order of 10 nm in size [141]. These platelets are randomly distributed in an otherwise AB-stacked film.[142, 143] In contrast, in CCS grown C-face graphene multilayers are ordered and very large (order of at least 10's of microns). The first graphene layer that forms is rotationally aligned 30º relative to the SiC. A ±7º rotation is occasionally observed [127]. Thicker layers exhibit well-ordered rotational stacking. SXRD experiments shows that this stacking consists of alternating 30º and 0º± ~7º rotated graphene sheets approximately every other layer. Rotational angles around 0º correspond [144, 145] to commensurate structures between graphene and the SiC(000-1) surface. LEED pattern of the rotated layers produces arcs centered at 0º, which were mistakenly attributed to small disordered grains of HOPG graphite [16, 146]. These observations indicate that properly produced and annealed C-face multilayers are ordered, while otherwise produced multilayers exhibit disorder, that can be extreme. For example, it has been reported that C-face graphene grows as small graphite crystallites similar to HOPG graphene.[147] However, the samples used in those studies were grown in argon at much higher temperatures (1800ºC-2000ºC). This turbostratic structure is most likely due to an amorphous SiC interface layer that forms at these high growth temperatures.[148]

In C-face multilayered graphene it was estimated that AB stacked layers represent less than 19% of the layers in the film.[144] STM,[149] [141] ARPES,[144], μ-LEED[150], Raman spectroscopy[53] and electronic transport[49] all confirm these results. In STM, these commensurately stacked layers show up as moiré patterns with large supercell sizes. Figure 14c-e show an STM image from C-face films showing moiré patterns from three graphene layers commensurately stacked above each other. μ-LEED presents these commensurate rotations as diffraction patterns with a large supercell (see Figure 14a-b). The supercell size varies over several micron length scales but the top layer orientation is always 30º relative to SiC. This indicates that while the commensuration angle varies in the layers below, the top layer is continuous over very large areas.

### 4.2.4. Electronic structure

The electronic structure from C-face graphene is very different from Si-face graphene. Whereas ARPES of Si-face few layer graphene exhibits overlapping, interacting Dirac cones (see Figure 2), ARPES of C-face multilayer graphene exhibits multiple undistorted Dirac cones that are mutually displaced, corresponding to individual graphene layers of the rotational stacked film. Figure 15a shows the band structure from an 11-layer C-face film measured in ARPES. [9] Two rotated Dirac cones are clearly visible. Note also that the graphene Dirac point is very near $E_F$. The limited mean free paths of the photoelectrons in the ARPES experiment only allows the top three layers to be probed. The data shows that that these three layers are undoped (since the Dirac points coincide with the Fermi level) and that the band structure of each layer corresponds to that of a free graphene sheet. (The band structure of bilayer graphene is only rarely observed.)

Standard area-averaged ARPES cannot distinguish whether the two cones in Figure 15a are due to a stack of two graphene sheets or from two uncorrelated graphene sheets that are laterally separated by a distance within the 50μm ARPES beam diameter. k-PEEM (a version of micro-ARPES) only measures the band structure within the illuminated area. Figure 15b shows the k-PEEM from a 2-layer C-face film using a 7μm illumination aperture.[151] Besides the two mutually displaced (rotated) primary Dirac cones, a third set of Dirac cones is also visible. The third set of cones results from an umklapp process, that translate the Dirac cone of one layer by the reciprocal lattice vector of the second layer and visa versa (see Figure 15b) as shown by Mathieu et al.[151]. The observation of this interaction proves that the rotated graphene sheets must be indeed stacked on top of each other. The signal specifically cannot be due to separate rotated grains in the beam.

The observation that C-face graphene multilayers are electronically decoupled has motivated a number of theoretical studies. [129] Interesting effects are anticipated because the stacking that gives the moiré pattern lifts the degeneracy between sublattices in a non-trivial way. In particular a strong reduction of the Fermi velocity is predicted for small rotation angles, both with ab initio and tight-binding methods [152-156]. Interlayer couplings between the rotated layers may introduce a singularity in the energy spectrum due to geometric considerations. [155], and observed peaks in the STS local density of states that scale with the rotation angle were interpreted as such (Van Hove singularities) [157, 158] The fact that thick C-face graphene films are electronically similar to a stack of independent, mutually rotated graphene sheets, warrants the accepted nomenclature multi-layer epigraphene, or MEG.

### 4.2.5. Raman spectroscopy- thickness determination

Raman spectroscopy, that is sensitive to electron-phonon coupling, is a widely used technique to characterize graphene. The spectrum typically presents three peaks, labeled D, G and 2D (see Figure 16). Since the D peak is only Raman active when the lattice symmetry is broken, its presence is indicative of structural defects (e.g. point defects and edges). Figure 16 a shows the Raman spectrum for thick MEG. Figure 16 b shows the Raman spectrum of a C-face graphene monolayer before and after subtraction of the SiC Raman background signal (that is invisible in Figure 16 a due to the film thickness) [159]. The absence of the D peak attests to the quality of C-face epigraphene monolayers and MEG.

The 2D peak profile evolves from a single Lorentzian in graphene to a characteristic shouldered wide peak in graphite. Therefore, the shape and width of the 2D peak are often used to determine the number of layers in thin (Bernal stacked) graphite films [160] as well as in Si-face graphene films, where graphene layers are Bernal stacked. However, on the C-face, the layers are electronically decoupled and the 2D lineshape does not evolve with increasing thickness but remains a simple Lorentzian peak [53](see Figure 16a) which profile cannot be used to determine the film thickness.

The relative intensity of the graphene Raman peaks compared with those related to SiC increase with the number of layers, but these ratios are difficult to calibrate accurately [161, 162]. In LEEM, the interference of the incoming electrons with the electrons that are reflected from the SiC–graphene interface produce oscillations in the reflected electron intensity. The number of layers equals the number of oscillations. [163]. While accurate, this method is complex and therefore cannot be used as a routine diagnostic tool. Auger spectroscopy provides an accurate measurement of very thin epigraphene layers by comparing the ratio of the Si peak intensity with that of C. For thicker layers, ellipsometry gives a very good estimate of the average number of layers [164].

For graphene monolayers, the position of the Lorentzian 2D peak depends on the electronic doping density and strain in graphene [165-168]. In quasi-free standing epigraphene on the Si-face, the 2D peak position is constant within about 1cm$^{-1}$ for electron doping n<5x10$^{12}$cm$^{-2}$ and shifts by 4 cm$^{-1}$ for p-doped graphene [168]. Shifts in the 2D peak position for C-face monolayers have been attributed mainly to strain and to a much lesser extend to doping density [137]. Note that an unexpected Raman intensity enhancement is observed by collecting light through the SiC substrate, in a reverse configuration. The effect is explained it in terms of dipole radiation at the dielectric surface. [168].

However, besides sensitive to strain and doping, shifts in Raman peaks in graphene are determined by a subtle interplay between the phonon and electron energy dispersions [169]. The 2D peak in Si-face epigraphene is blueshifted compared to exfoliated graphene [170] which indicates compressive strain in epigraphene. But, actually the lattice parameter a=2.456Å of epigraphene Si-face [113] is larger than the theoretical value for graphene, indicating expansion. It would be instructive to know the lattice parameter of transferred graphene and compared to the free graphene value. Comparison with HOPG is more subtle because the 2D peaks is composed of 4 peaks reflecting the more complex graphite electronic band structure.

### 4.3. Epigraphene on other faces

There are very few epigraphene studies on other than those on the (0001) and (000-1) faces, due to the lack of available substrates. Contrary to the Si- and C- faces, theses surfaces are non-polar, i.e. they have an equal amount of silicon and carbon atoms, which is quite interesting regarding the difference in graphene growth and morphology between the 4H/6H-C- and Si-faces. As expected graphene was found to grow on the two perpendicular surfaces, so called a-(11-20) and m-(1-100) faces. From XPS, LEED, LEEM, and in agreement with density functional theory calculations, it is concluded that there is no buffer on either face. [171] Differences were observed between the two faces. On the SiC(1-100), micro-LEED reveals multiple rotations similar to C-face graphene (although the stacking has no been studied), while on the SiC (11-20) face, graphene grows with a single orientation. ARPES shows the typical conical bands of graphene, with a negative charge density of 4×10$^{12}$ cm$^{-2}$. This is unexpected since the graphene doping density in the absence of buffer (quasi-free standing graphene) was argued [172] to be due to spontaneous polarization in the polar Si and C-faces. The charge density therefore could be attributed to the work-function difference between graphene and silicon carbide. Transport properties confirm that lightly doped single layers can be grown on the (11-20) face [173], however the heavily doped substrate confine measurements to very low temperature.

### 4.4. Epigraphene on 3C-SiC

Graphene layers have also successfully been grown on cubic β-3C-SiC (Figure 6), both on the hexagonal (111) and the cubic (001) face of commercial polycrystalline substrates or overlayers grown on silicon or 6H-SiC substrates. [174-179]. Note that the commercial production of 3C-SiC substrates is limited. In contrast to the 4H and 6H- SiC, large single crystal 3C-SiC is not available.

Studies on 3C-SiC are nevertheless interesting in several respects. In contrast to hexagonal SiC, 3C-SiC does not spontaneously electrically polarize (see Section 5.1.2), so that graphene on 3C-SiC is found to be quasi-neutral (see Figure 17b)[176, 180]. Step free growth and large area monolayer coverage have also been reported for 3C [179] (Figure 17a). Since thin layers of 3C-SiC can be epitaxially grown on silicon, this has been proposed as a strategy for integration of SiC based epigraphene with Si-based electronics (see Figure 17c and Section 5.3.3).

For 3C-SiC epilayers on silicon substrates, the polar Si-terminated 3C-SiC(111)/Si(111) surface grown in UHV shows graphene Bernal stacking with an interfacial buffer layer, similarly to the 4H- or 6H-SiC(0001) surface. Conversely, the C-terminated 3C-SiC(111)/Si(110) shows a non-Bernal stacking, with the absence of an interfacial buffer layer, consistent with a C-face termination. The quality of these graphene films is poor as shown by large Raman D peaks. The disorder results from Si diffusion through SiC grain boundary (due to a large ∼ 20% lattice mismatch between Si and 3C-SiC) and the lower graphene growth temperature (limited by the Si melting point at 1414ºC). Growing an epitaxial AlN layer on Si prior to SiC growth significantly reduces Si out-diffusion and helps grow higher quality epigraphene. [175]

Graphene was also grown on the 3C-SiC(001) surface, which demonstrates that a hexagonal template is not required for the growth of graphene. LEED patterns show no evidence for a buffer layer (growth in Ar at 1800ºC). [181]

### 4.5. Nanostructured graphene

From the outset of epigraphene electronics research it was clear that graphene nanostructures would be required [1, 2]. Moreover, these structures would necessarily need to be as defect free and reproducible as possible, in order to serve as building blocks for electronic devices. Early calculations [61, 62] had already predicted that quantum confinement would play an important role in graphene ribbons. Depending on the edge structure a graphene ribbon would be semiconducting or metallic. Those calculations predicted that all zigzag ribbons would be metallic while armchair ribbons would alternate between being metallic or semiconducting, depending on their widths (see Section 2.9).

Early work on exfoliated and on epitaxially grown graphene focused primarily on graphene nanoribbons produced by electron beam lithography and plasma etching. Epigraphene nanoribbons were found to be metallic exhibiting quantum confinement effects at cryogenic temperatures [49], which implied that edge scattering was reasonably coherent. Disorder in the ribbon was not large enough to cause localization. On the other hand, charge neutral exfoliated e-beam patterned nanoribbons were invariably found to become insulating at cryogenic temperatures, implying a band gap. In fact, the magnitude of the band gap was determined to be consistent with Eg=1eV nm/W as expected for the quantum confinement gap in specific armchair ribbons. However it was soon found that the low temperature insulating behavior was due to a transport gap that opened due to charge puddles in the ribbons combined with edge disorder [182-187].

To improve edge order, "bottom-up growth" methods were developed to grow graphene nanoribbons. [135, 188, 189] Very narrow graphene ribbons have been grown at steps on Au(788) facets by catalyzing molecular precursors into linear polyphenylenes. [188, 189] While the edge order is essentially perfect [189] so that the armchair edge ribbons are consistent with the predicted [62, 190] finite size band gap, scalability as required for electronics is still lacking. Moreover, the growth is limited to metal surfaces and it is expected that the transfer process to semiconducting substrates will face the same problems as other transferred CVD graphene films. Hence, these efforts have largely been abandoned for large-scale electronics.

Epigraphene, however, provides another solution by growing ribbons on lithographically patterned steps etched into the SiC substrate [135, 137, 191], and exploiting the fact that graphene growth proceeds first on the facet walls of natural step edges on the Si-face of SiC [192]. In order to produce nanoribbons, trenches are first etched in SiC that serve as a template for graphene growth. This allows ribbons and other nanostructures to be accurately defined on a substrate as required for nanoscale integrated circuits. The method allows thousands of parallel ribbons as narrow as 20nm and with well annealed edges to be grown at once over mm2 area (see Figure 18), No further (potentially damaging) post growth processing of graphene is required. In this process, the ribbon width is defined by the trench depth, that is very well controlled using standard plasma SiC etching processes. [135]

Epigraphene also has a distinct advantage because Si-face graphene is epitaxial with the SiC(0001) surface, therefore the ribbon orientation is also predetermined. The graphene's zigzag (ZZ) or armchair (AC) edges, naturally align with the SiC step edge simply by etching steps in the SiC in a given SiC crystallographic orientation. When a trench in SiC is oriented perpendicular to the 〈-1100〉 direction, the graphene that grows has its AC edge parallel with the step edge. Trenches perpendicular to the 〈11-20〉 direction produce ZZ ribbons. For convenience, we will refer to these SiC step edges as AC edge steps (or AC facet walls) and ZZ step edges (or ZZ facet walls) (see Figure 19). Structured sidewall growth has been used to produce a wide variety of shapes (see Figure 18(c-f) [191]. Of particular interest are pillars (or circular pits) around which graphene rings are grown for quantum interference pattern experiments [191].

The structured SiC chip is then heated to 1100 C to allow the vertical etched sidewalls to crystallize into the equilibrium facets onto which the graphene ribbons grow around 1500-1600 C. The exact temperature depends on the specifics of system, such as the Si escape leak in the furnace CCS method [34], or the face to face geometry in the current annealing growth [193, 194].

### 4.5.1. Armchair Edge Sidewall Ribbons

The structure of AC sidewall ribbons has been studied extensively with a number of techniques. AFM shows that the trench walls for AC steps facet into planes with an average facet angle of 28-30°. Early TEM studies of graphene growth on natural AC step edges used exceptionally thick Si-face films (>4 layers).[192, 195]. The graphene was found to tend to nucleate at the bottom of the step edges so that graphene grows thicker on the step facet than on the (0001) surface (Figure 20a). Several edge terminations are observed (see Figure 20a-b).

Under certain CCS growth conditions cross sectional transmission electron microscopy measurements show mini facets near the top edge and bottom of the primary facet, where the sidewall merges into the (0001) surface on top and bottom of the sidewall.[111] The composite scanning transmission electron microscopy (STEM) image of Figure 20c shows that the graphene ribbon continuously drapes over the entire sidewall. The ribbon appears to be anchored to the edges (or mini terraces) at the mini facets. At those places, the SiC-graphene distance is slightly smaller than that of the SiC-buffer layer spacing. Note that nm-wide ribbons have been predicted at SiC step edges on 6H-SiC.[117]. Ultimately, the graphene ribbon merges into the buffer layer on top (and often at the bottom) of the sidewall

ARPES from the sidewall ribbons gives additional information on both the electronic and topographical structure of these ribbons. [196] Since the facet wall normal is tilted relative to the (0001) plane, the graphene on the facet is tilted by the same angle (see Figure 21a). The step edge the K-point of the facet graphene can be measured (see Figure 21a) by appropriately rotating the sample. ARPES shows that the graphene that grows on the sidewalls has a different doping than the graphene on the (0001) surface. While the monolayer graphene on the (0001) is n-doped by ~0.45eV, the graphene on the facet wall is only slightly p-doped (<70meV). The facet angle can be determined from the relative rotation angle between the (0001) Dirac cone and the facet Dirac cones. The facet contains both (-1106) and (-1107) planes, similar to those seen in the STEM studies in Figure 20c.

In addition to the (0001) surface and sidewall Dirac cones, there is a range of angles, consistent with the position of the mini ribbons, where the set of cones appear to be gapped (see Figure 21d). [196] LDA calculations of bent well-ordered nm-wide ribbons predict an energy gap in good agreement with the observed ARPES and TEM data. Whether the origin of the gap observed in ARPES is due to quantum confinement in nm-wide ribbons or related to other effects (like buffer graphene near the step edge or the mini-ribbon (0001) terraces that are tightly bound to the SiC substrate), it is nonetheless remarkable that a band gap opens in a structure entirely made of graphene (see Section 5.4). In any case, the measured gap confirms that these nanoribbons have smooth edges, comparable to the atomically precise ribbons fabricated by bottom-up chemistry [189].

### 4.5.2. Zig-Zag Edge Sidewall Ribbons

Several structural studies have been focused on graphene growth on commercial off-axis SiC. After annealing, the miscut combined with step bunching produces arrays of ZZ substrate steps [197, 198] to produce ZZ sidewall ribbons with typically >3 layers. Electron energy loss spectra in STEM indicates that the graphene on the sidewall is decoupled from the SiC (see Figure 20d).[198]

Single graphene layers can be grown on ZZ sidewalls. Transport measurements made on ZZ ribbons produced by current annealing have been found to be metallic and even demonstrate ballistic transport properties [136]. However, ARPES measurements of arrays of ZZ ribbons do not show a Dirac cone, while a Dirac cone is observed for AC ribbons that are produced using the same CCS method. The absence of a Dirac cone in this case was originally attributed to bonding of the graphene layer to the ZZ sidewall facet to produce a buffer layer [199]. It is however also possible that, depending on the growth condition, the ZZ annealed sidewall is not straight because natural SiC ZZ steps have a known instability towards nano-facetting into local AC steps [200] (see Figure 22a-b). A similar instability has been seen in patterned ZZ-edge steps in SiC [199], meaning that instead of a ZZ edge, a series of small AC nanofacets, ±60° relative to the average ZZ step edge direction, would form as shown in Figure 22c. If ARPES requires that the sidewalls are well ordered, which could explain why the graphene signal was not observed using those methods, the short order length scale necessary for the observation in PEEM indicates that the two different growth methods may lead to two different structures.

## 5. Electronic transport properties of epigraphene

As shown below epigraphene presents the characteristics expected for graphene. The large surface area and high structural quality allow for experiments that wouldn't be possible with small flakes or rough substrates and for large-scale integration. In some instances specificities are observed, in particular related to the original rotational stacking on the C-face. Here we only review electronic properties as they pertain to epigraphene, and the reader is referred to review articles for general graphene electronic properties that are beyond the scope of this article.

As explained in the introduction, what made graphene attractive in the first place is its potential for electronics. Extremely high electronic mobility were anticipated from the onset [1], based on the room temperature ballistic conduction of the closely related carbon nanotubes [201]. By choosing the appropriate substrate, mobilites in the range of 100,000 cm$^2$/Vs are achieved in 2D graphene on boron nitride [60], up to more than 10$^6$ cm$^2$/Vs at room temperature in the neutral rotated epigraphene C-face layers [202] or for graphene on graphite at low temperature, and unrivaled room temperature ballistic conduction up to 15 μm in epigraphene nanoribbons. Graphene offers also the possibility to vary the charge carrier and thereby the conduction by electrostatic gating [1, 4, 5] and patterning by standard lithographic techniques is available.

### 5.1. Transport properties

#### 5.1.1. Charge density

The first graphene layer above to the SiC interface is negatively charged, as shown in the ARPES measurements presented in Section 4.1. For the Si-face, the n-doping of about 10$^{13}$ cm$^{-2}$ as measured by photoemission spectroscopy [79, 84] is believed to arise, at least partially, from interface states associated with the SiC interface and the buffer layer.[172] Similar doping of a few is 10$^{12}$ cm$^{-2}$ is found on the C-face [131] by photoemission in UHV. Because of screening on a length scale of about one layer, the charge density decreases from one layer to the next as observed in ARPES [131], in optical spectroscopy [203] and electron energy loss [204] spectroscopy experiments.

More precisely, a charge profile consistent with $n_l = n_{le} e^{-1.1(l-1)/\lambda}$ was determined in C-face multi-layered epigraphene by mid-infrared pump-probe spectroscopy, where l is the graphene layer index from the SiC interface, $n_{le}$=9.6 x 10$^{12}$ cm$^{-2}$ and $\lambda$=1.2 [203]. A full calculation can be found in Ref [205]. As a result, the layers away for the SiC interface in multilayer C-face are quasi neutral and present the properties of ideal graphene close to the Dirac point, as close as 8 meV, that is n~5x10$^9$cm$^{-2}$ (see Section 5.1.3 and 5.2.2).

For graphene exposed to air, the top layer shows sign of contamination by the environment, that results in a counter positive doping. The measured charge density changes sign from negative to positive for epigraphene monolayer C-face left in air typically within an hour, after that time it stabilizes. [123] It has been consistently observed that the effect is reversible by heating above 80-100ºC. After air exposure, ambient-environment Kelvin probe microscopy of Si-face single-layer epigraphene indicates a highly uniform carrier concentration of the order of 10$^{12}$ cm$^{-2}$ with very small carrier fluctuations ~10$^{10}$ cm$^{-2}$ (corresponding to measured rms surface potential variation of 12 meV) [206].

On epigraphene Si-face, when the SiC interface is saturated with hydrogen and the buffer layer becomes the first graphene layer (see Section 4.1.1) a consistent p-doping is observed. The positive sign and the charge carrier doping level are explained by a polarization doping arising from the bulk spontaneous polarization of the pyroelectric SiC substrate [172]. This is because at the interface with graphene, this polarization is equivalent to a negative charge that is balanced by a positive charge in the graphene layer, and to a lesser extend by a positive space charge in the SiC substrate depletion layer. When the epigraphene layer resides on the (non H2-lifted) buffer layer , the presence of donor-like states associated with the SiC/buffer interface is believed to overcompensate the positive polarization doping yielding the measured n-doping of epigraphene Si-face. The polarization doping model was further confirmed by ARPES measurements of the charge densities of quasi-free standing graphene on 6H-SiC(0001), 4H-SiC(0001) and 3C-SiC(111). [180] The SiC spontaneous polarization, and accordingly the epigraphene charge density, decrease from 4H to 6H and to 3C-SiC so that quasi free standing graphene on 3C-SiC is quasi- neutral.

### 5.1.2. Square resistance, mobility and charge density in single layers

In two-dimensional systems, transport is characterized by the square resistance defined by $R_{sq}$ = (V/I) W/L = ($\rho$/ t) where V is the voltage difference between the voltage probes, I the current through the sample, W, L, t are the sample width, length and thickness respectively and $\rho$, the material resistivity. An important property of graphene that sets it apart for ordinary metals is that the charge density can be varied by orders of magnitude, inducing a large change in $R_{sq}$. The square resistance of clean graphene, hereafter simply called resistivity, goes through a sharp maximum at charge neutrality (transport at charge neutrality goes beyond the simple one electron picture). The maximum resistivity value and peak sharpness as a function of charge density depend on the graphene structural quality and source of scattering. In the simplest picture of short range scatterers, a conductivity independent of carrier density was expected because of the linear in energy density of states $\mathcal{N}(E)$[207] [208]. The resistance maximum calls for other scattering mechanisms, including charge impurity scattering, resonant scattering (for a review of transport mechanisms, see for instance Ref. [208]).

For "dirty" graphene samples, at energies away from the charge neutrality point, the conductivity rises linearly with the charge density n [208, 209] $\sigma = \sigma_{res} + Ce(|n/n_{imp}|)$. Here $n_{imp}$ describes the charge impurity concentration, and $\sigma_{res}$ reflects the fact that graphene conducts even at zero carrier density. The sharpness of the conductance minimum can phenomenologically also be captured with $\sigma = e\mu\sqrt{n_{imp}^2 + n^2}$ [210].

Note that with the usual definition $\sigma$= ne$\mu$, the mobility $\mu$ would reach unphysically large values at charge neutrality. Indiscriminate and improper use of mobility equations implies that the mobility is infinite at zero charge density, while in fact, the actually mobility (taking into account the charge disorder) is actually much more modest. In disordered samples zero charge density arises from compensated puddles of electrons and holes and transport proceeds by hoping from puddle to puddle.
When the graphene is not too close to charge neutrality, the mobility $\mu$ can nevertheless be determined from the measurement of $\sigma$ = 1/Rsq and n = 1/(Be$\rho_{xy}$) where $\rho_{xy}$ is the transverse (Hall) resistivity in the linear regime of magnetic field B ($\mu_H$ = $\rho_{xy}$/ B.Rsq).

Note that field effect mobilities are often reported, which are determined by modulating the charge density on either the Si or C faces (see also Section 5.3.1). For this, techniques include counter-doping in air [123], with molecules (such as Tetracyanoquinodimethane [101]), with a charge induced silicon nitride top dielectric coating [211], by metal intercalation at the interface (see Section 4.1.1), lifting the buffer layer by hydrogen intercalation to saturate the dangling bonds on the SiC substrate [91]. In a more controllable way, the charge density can be varied by applying a voltage $V_g$ to a top electrostatic gate (standard planar capacitor) as widely used for epigraphene high frequency transistors (see Section 5.3.1). For this, epigraphene is coated with a thin dielectric ($Al_2O_3$, HfO, SiN, ... ) plated with a metal (typical Au or Al) [212] [137, 213]. The (surface) charge variation $\Delta$n induced on graphene is $\Delta$n = ($\varepsilon 0\varepsilon$/td) Vg/e , where $\varepsilon$ and td are the dielectric constant and thickness of the dielectric respectively. Examples of modulation of the resistance of epigraphene are given in Figure 23. Top gating was also achieved, with a UV light photochemically sensitive polymer [214] and electrolytes [215]. The mobility is related to the slope of the transconductance $dI_{sd}/dV_g$ for a graphene strip where a current $I_{sd}$ flows between source and drain contacts; this is the configuration of a field effect transistor (FET).

Because deposition of a dielectric on graphene tends to reduce the mobility, attempts were made to back-gate epigraphene [216]. High energy nitrogen implantation into SiC produces a thin conducting layer buried into the insulating SiC substrate, to which a voltage can be applied. However interface states limit the gate efficiency and efficient doping was observed only for quasi-free standing epigraphene (H2 passivated). Besides, the gate is inefficient at low temperature (the carriers in SiC are frozen in) and at room temperature it relies on the Schottky-like contact between conducting SiC and graphene ('Schottky capacitor' regime) [216, 217].

For epigraphene on the Si-face mobility values are somewhat limited and on the order of 1500 cm$^2$/Vs at charge density of about 10$^{13}$/cm$^{-2}$, even for Hall bars patterned on single terraces. Reported mobilities increase dramatically by decreasing the charge density n$_s$, as expected, with record low temperature values of µ≈30,000 cm$^2$/Vs for n$_s$≈5x10$^{10}$ cm$^{-2}$ [218, 219]. When the buffer layer is lifted by hydrogen intercalation and graphene becomes p-doped, higher mobilities are observed (roughly by a factor two). It is not clear what causes the increase in scattering when the buffer is present. As for the conductivity, high values can be reached with large charge density obtained with top gating or molecular doping: record low conductivity of 16Ω/sq was reported by intercalating the ionic conductor FeCl$_3$ [220]

Epigraphene on the carbon face shows higher mobilities [2], despite contaminations and monolayers draping over multiple SiC steps [123, 137]. For charge densities of the order of a few 10$^{12}$ cm$^{-2}$, µ=20,000 cm$^2$/Vs was reported early on [123] and up to 39,800 cm$^2$/Vs by reducing the charge density to n= 1.9x1011 cm- (Figure 27a). A top gate induces scattering. Nevertheless top-gated FET mobilities of 7,000-8,000 cm2/Vs are routinely found at room temperature [221] for n~ 1.5x1012 cm-2(see Figure 23b for instance).

### 5.1.3. Charge density and mobility in multilayers

For multilayers on the C-face, the graphene layers retain the electronic structure of single layer graphene, as discussed in Section 4.2.4. Mobility values of the quasi neutral graphene layers in the middle of the stack are the highest measured in any 2D graphene system to date, reaching µ=10$^6$ cm$^2$/Vs [202] for n=10$^{10}$cm$^{-2}$ at room temperature and independent of temperature. [52] The mobility was estimated by infra-red spectroscopy measurements in magnetic field (see Section 5.2.2).

Square resistance of about 200 Ω are commonly measured for about 5 layers MEG on the C-face, with mobility of 20,000 cm$^2$/Vs or more [49]. Because of interlayer screening, a top gate addresses primarily the top layer [212] (see Figure 23c).

For graphene on the Si-face, Bernal stacked layers have a quadratic band structure at the K and K' points (see Section 4.14.1). The electric field perpendicular to the layers that builds up due to of the intrinsic charge density difference causes a small gap to open [80]. This may explain the observation of massive carriers in thin epitaxial graphite in magneto-conductance measurements [222].

### 5.1.4. Scattering at SiC steps : the Si-face

For single epigraphene layer on the Si-face, high charge density is associated with moderate mobility of the order of 1,500 cm$^2$/Vs. Several sources of scattering have been implicated including: the substrate, substrate phonons, SiC steps and single layer-bilayer junctions. Hydrogenated quasi-freestanding graphene shows high mobility, roughly by a factor two.

A significantly higher channel conductance is reported for devices patterned parallel to the intrinsic SiC steps direction compared to across the SiC steps [223] [224]. Similar anisotropic transport has been measured by local four-point probe and scanning probe techniques [225-227]. The anisotropy was attributed to various mechanisms. This may be related to a reduction of the carrier concentration for graphene on the sidewall of steps [136, 223], to a n/p junction at the step edge [196], to the opening of a band gap at the step edge [196] (see Section 4.5) or to the presence of bilayers multilayers on the steps (see Figure 8b). Conversely a careful surface preparation carried before graphitization to reduce the SiC step height notably increases the mobility [193] and very homogeneous device-to-device electronic properties (carrier concentration and mobility) are achieved at the wafer scale when the devices are fabricated entirely on single terrace monolayer domains.[69]

### 5.1.5. High current carrying capability

One of the most remarkable properties of graphitic materials is their high current carrying capability. For instance electromigration breakdown was reported for exfoliated graphene at current densities of 1.6 mA/μm (corresponding to $5\times10^8$ A/ cm$^2$. For epigraphene in vacuum, conductance increases for current densities up to $\sim 1.3\times 10^9$ A/ cm$^2$ due to local cleansing of the graphene channel. At higher current the sample glow reflects local heating above 1200ºC and SiC decomposes. The graphene film breaks down at a critical current density on the order of 1.3-2 x 10$^9$A/cm$^2$ [228], to be compared to 3 x 10$^9$A/cm$^2$ for carbon nanotubes [229].

## 5.2. Transport in magnetic field

### 5.2.1. Low field (anti)- weak localization

As in any metallic diffusive system, quantum interference effects are observed at low temperature. These are corrections to the classical conductivity arising from interference phenomena, when the electronic wave function keeps its phase for multiple scattering events (phase coherence length L$\phi$>>mean free path l). These corrections have characteristic functional forms, which magnetotransport data can be fitted to in order to extract relevant transport parameters. In particular weak localization involves an enhanced probability for an electron to be coherently scattered back to its starting point in a loop trajectory. The enhancement originates from interferences between the clockwise and anti-clockwise trajectories. As a result, the resistance is up to twice as much as with no interference, and any mechanism producing incoherent scattering (temperature) or dephasing (magnetic field, spin scattering, spin-orbit) will destroy the interference and restore the classically lower resistance. Therefore a decrease of resistance is expected in a magnetic field (i.e. negative magnetoresistance that peaks at zero field).

Graphene adds an interesting twist to it. Because the character of the wave function in graphene cannot be changed by long-range scattering (only short-range scattering can locally destroy the equivalence of the A and B sublattices), 180° backscattering of electrons is forbidden and the resistance is low. Dephasing the wave function in a magnetic field can restore backscattering, yielding a positive magnetoresistance (so called weak anti-localization)[230].

Weak anti-localization in graphene was first observed in epigraphene grown on the C-face [50] for a high mobility sample (μ=11,600 cm2/Vs). As seen in Figure 24, the magnetoresistance can be fitted very accurately to weak anti-localization theory with only one temperature dependent parameter, ascribed to electron-electron scattering. Besides providing yet more evidence that multilayer epigraphene behaves like a stack of mono layers, these measurements are consistent with a dominant scattering at low temperature that preserves the pseudo-spin, most probably from long-range potentials arising from charges in the substrate.

On the Si-face, detailed analysis of low-field magnetoresistance and temperature dependence shows evidence for strong electron-electron interaction effects, giving logarithmic dependence. This is difficult however to detangle from the logarithmic dependence arising from the Kondo effect. Even by adding magnetic impurities studies so far are not conclusive. [231]

### 5.2.2. Landau levels spectroscopy on the C-face

As the magnetic field increases the energy levels bundle up in discrete (quantized) Landau levels. A simple picture is that the electron energy is minimized for orbits that are a multiple of the Fermi wavelength $\lambda$. An exact calculation for a standard 2D electron gas (2DEGs) gives $E_N = \hbar\omega_c(N+1/2)$ with the angular frequency $\omega_c$=eB/m* and N is the band index. Because graphene dispersion relation E(k) is not quadratic but linear, the energy of the Landau levels varies as the square root of magnetic field and N: $E_N = \pm c^*\sqrt{2\hbar eB|N|}$, where c* is the effective Fermi velocity. For a review of graphene in magnetic field see Ref. [232]. The square root dependence is characteristic of graphene, and there is no ½ offset. The Landau index level N can take positive and negative values (the electrons and hole levels being identical). The N=0 level is special. It is field independent, pinned at $E_0$=0 and it is equally composed of holes and electrons states. The filling factor reflects the number of filled Landau levels.

The first direct observation of the characteristic $\sqrt{B|N|}$ energy dispersion in graphene was made in multilayered epigraphene by infrared magneto-spectroscopy [51]. Infrared light is absorbed (see Figure 25a) when its frequency matches inter-level separation (according to specific optical absorption rules, namely only transition LL ₋(N+1) →

LN and LL‑N → LN+1 are allowed). Only quasi-neutral graphene can be probed in the infrared light and magnetic field range accessible. In C-face epigraphene, the energy versus field plot in Figure 25b displays text-book graphene behavior and the slope give $c^*=1.1 \times 10^6$ m/s, very close to the expected value for graphene (we refer the reader to [233] and [86] for a discussion of the renormalization of the Fermi velocity due to electron/hole excitation effects and substrate-screening effects). Electron-phonon coupling also manifests itself in the magneto-Raman in epigraphene C-face. The position and width of the Raman G band at 196 meV undergo oscillations (avoided crossings) at magnetic field values for which the G band energy crosses the optically active inter-Landau level transitions.[234]

It is significant that the $E_{\pm 1} \rightarrow E_0$ transition is observed at magnetic field as low as $B_0=40$ mT [52]. To resolve Landau levels requires that the level are minimally broaden by disorder. This simply means that an electron needs to complete a cyclotron orbit before being scattered. Because the cyclotron radius is inversely proportional to B, this translates into the criterion, $\mu B_0 > 1$. The low $B_0$ value gives extremely high mobility for the neutral MEG layers: $\mu > 250{,}000$ cm$^2$/Vs, and from a direct measurement of the level broadening a value of $\mu=10^6$ cm$^2$/Vs, was estimated at room temperature [52, 202]. Moreover, the level broadening varies linearly with energy [202], which provides some insight about carrier scattering; in particular it rules out extrinsic scattering mechanisms, such as resonant scatterers or charge impurities, in the epigraphene layers which are protected from the environment.

A great advantage of graphene compared to conventional 2DEGs is that the electron gas is exposed and not buried at the interface between two semiconductors. It is therefore readily accessible to direct imaging and spectroscopy. Local spectroscopic measurements of the Landau levels were performed using high resolution scanning tunneling spectroscopy (STS) on the top C-face layer at very low temperature (10 mK) and in magnetic field (up to 15 Tesla) [54] (See Figure 26). The Landau levels correspond to peaks in the STS local density of states (dI/dV spectra), and the graphene behavior is confirmed. The splitting of the four quantum states that make up a degenerate graphene Landau level (two levels per valley, two per spin) is studied in great detail as the magnetic field is increased (Figure 26b). Splitting due to broken valley degeneracy occurs first, followed by spin degeneracy as the field increases. Most unexpectedly, states with non-integer Landau level filling factors of 7/2, 9/2 and 11/2, are also observed suggestive of new many-body states in graphene. The graphene layers below the probed top surface may also influence STS results, by serving as a charge reservoir but also by screening interactions between top-layer electrons. They may also have a role in the observed fractional filling factor. [54] (see also [205])

Real space mapping of the two-dimensional spatial distribution of the electronic states of a Landau level was achieved for the first time on a 2DEGs by using scanning tunneling spectroscopy on the top C-face epigraphene layer. The local density of states mapping of the zeroth Landau level shows extraordinary fine details giving much insight into the nature of the Landau levels in graphene. [128] In particular unlike disordered patterns found in conventional quantum Hall systems, an organized pattern of localized states and extended states is observed in epigraphene C-face. STS measurements reveal local splitting of the zeroth Landau level corresponding to the local top bilayers AA, AB or BB stacking arising for the C-face rotational stacking [128]. This observation has important implications for transport phenomena. Spatial modulation of the sublattice symmetry and redistribution of charges between layers [205] may account for observations of fractal-like structure in magnetoresistance measurements of this material [49].

### 5.2.3. High field Shubnikov – de Haas oscillation and quantum Hall effect

When the magnetic field is increased, the Fermi energy $E_F$ crosses the successive Landau levels $E_N$; the density of states maximizes for $E_F=E_N$ resulting in resistance oscillations that are periodic in 1/B. (Schubnikov- de Haas oscillations). The oscillations are very well developed in epigraphene on both the Si-face [1] (Figure 27d) and on the C-face [49](Figure 27a). The charge density ns ($E_F = \hbar v_F \sqrt{\pi n_s}$) can be determined from the period of the oscillations (slope of the index N versus 1/B, see Figure 27a). In the plot of Figure 27a (top inset) the zero intercept with the N axis is a signature of graphene (no ½ off-set) even for MEG. The temperature dependence (Lifshitz-Kosevich equation) gives access to the separation between levels that is also characteristic of graphene in multilayers C-face. The Landau levels smoothly merge into equidistant levels at low field, when the cyclotron orbit size is larger than the ribbon width and confinement effects dominate the energy quantization (particle in a box effect) [49].

At high field or very low charge density, similarly to other 2D electron gases, the resistance oscillations develop into the quantum Hall effect (QHE). The resistance becomes vanishingly small at the oscillation minima, concomitant with plateaus in the transverse conductance that are quantized in units of $e^2/h$: $\sigma_{xy} = \nu e^2/h$ ($e^2/h = 1/25,812.807\ \Omega^{-1}$:). The zero resistance comes from the physical separation of forward and backward moving charges at each edge of the sample. Simply put, since the opposite travel direction cannot be reached, the electrons cannot be scattered and the resistance vanishes. The particularity of graphene is in the unique sequence of plateaus at filling factors $\nu=4(N+\tfrac{1}{2})=2, 6, 10$ for $N=0, 1, 2, \ldots$ instead of $\nu=2N$, $N=1, 2, 3, \ldots$ (without spin splitting) in normal 2DEGs. The factor 4 in graphene reflects spin and valley degeneracy. The second particularity of graphene is the large separation between Landau levels, for instance $\Delta E = E_{N=1} - E_0 = 36\sqrt{B}(meV) \sim 400$ K at 1 Tesla, to be compared to $\Delta E = 1.7B(meV) \sim 20$ K in GaAs/AlGaAs 2DEGs heterostructures. This allows to observe this quantum effect at much lower magnetic field [76, 137] (see Figure 27b) and much higher temperature than the sub-kelvin temperature range of QHE in 2DEGs. In graphene the QHE was observed even at room temperature [235], see also inset of Figure 27b for single layer epigraphene C-face. This particularly sought for in metrology applications for resistance standards based on the quantized e2/h plateaus working at a few Tesla (see Figure 27d) and potentially at liquid nitrogen temperature, instead of the current sub-Kelvin range. Large area high mobility graphene are however required, which epigraphene can provide [76, 236, 237].

The first observation of the QHE in epigraphene was on high mobility graphene on the C-face [123, 137]. Environmental positive counter doping [123] or top gate controlled [137] was used to observe the QHE in low magnetic field ($\nu=2$ plateau below 3 Tesla, see Figure 27b). Progress was rapidly made also on the Si-face with proper charge density compensation (see Section 5.1.1 [236] [237] with a measured quantum Hall resistance quantization accuracy of $3\times10^{-9}$ at 300 mK [76, 236], rivaling with the best 2DEG standards. The relative discrepancy between the quantized Hall resistances in the graphene sample and in a reference GaAs was found as low as $(-2\pm4)\times10^{-10}$, which demonstrates that epigraphene can substitute GaAs QHE resistance standard [76, 238]. This result demonstrates the structural integrity and uniformity of epigraphene over hundreds of micrometers. A particular type of disorder is however necessary to observe the QHE to insure that states in the bulk of graphene are localized, so that the states at both edges are well decoupled. It turns out that the QHE for epigraphene Si-face shows bi-layer inclusions [239], by coupling (or channeling) of states from both edges.

## 5.3. Towards electronic devices

### 5.3.1. High frequency transistors

Digital electronics requires that the gate-modulated current through the transistor channel be switched between on and off values, with a large $I_{on}/I_{off}$ ratio and small $I_{off}$ for minimum loss. Because graphene doesn't have a band gap, the conductance is non zero at all gate voltage and the channel current cannot be turned off velocity (for a review, see for instance [240]). This means that $I_{on}/I_{off}$ is limited to values of a few tens (Figure 23), to be compared to at least 10^4 in CMOS technology, and that poor current saturation is observed at high bias voltage [241]. Nevertheless, graphene can be used for amplification at high frequency, owing to its intrinsic low dimensionality, high carrier mobility and large carrier. Epigraphene field-effect transistors have shown fast progress. For radio frequency transistors, the two most important small-signal figures of merit are the cutoff frequency $f_T$ (the maximum frequency for current amplification, i.e. at which the current gain is unity) and the maximum oscillation frequency $f_{max}$ (that is the maximum frequency for power gain).

The intrinsic properties of epigraphene are quite promising in view of the transconductance $g_m = dI_{sd}/dV_g$ and high carrier velocities[242]. The transconductance describes the efficiency of the gate voltage $V_g$ to modulate the channel current $I_{sd}$ and directly relates to the intrinsic cut off frequency $f_T = g_m/(2\pi C_G)$, excluding parasite resistance or capacitance, with $C_G$ the gate capacitance. Large transconductance up to 2 mS/µm and output current above 5 mA/µm have been reported in epigraphene Si-face [243]. Higher carrier velocities can be achieved with high saturation velocity and high mobility. As a substrate SiC compares favorably with $SiO_2$ because optical phonon, argued to be the main source of hot carrier scattering, have a higher energy [241]. Following this line of argument, THz operation is predicted on SiC substrates.

The highest cutoff frequency (after removal of parasitic effects) in epigraphene made by conventional scalable process is $f_T$=280 GHz on the Si-face ($L_g$=40nm) [243], to be compared with record $f_T$ = 427GHz ($L_g$=67 nm) for a transferred gate on peeled graphene [244], and with world record 0.5-THz $f_T$ in SiGe heterojunction bipolar transistors [245] (Note that the cutoff frequency depends on the gated length $L_g$). However, use in any practical circuit requires high values of the power gain $f_{max}$, which still remains much lower than fT in most graphene devices (see Figure 28c). High fmax values (fmax =70 GHz for Lg=100nm) were only very recently demonstrated (see Figure 28b). This was realized with high mobility epigraphene C-face (room temperature FET mobility µ = 6,000 cm2/Vs; fT=110GHz) and by optimization of the device dual gate design (shown in Figure 28a), with low contact and gate resistances. [221]

Note that multilayer graphene is not as efficiently gated because of electronic interlayer screening that results in a constant conductance in parallel (as shown in Figure 23c). On the Si-face, better performances are obtained when the entire channel lies within a single graphene terrace, avoiding step edge scattering (Sec 5.1.4) [246]. Another challenge for any graphene- based transistor is to develop gate dielectrics that preserve high mobility and charge homogeneity in graphene. With no dangling bonds to adhere to, dielectric coating on chemically passive graphene requires a nucleation layer or a roughened top graphene layer [247]. Atomic layer deposition (ALD) or evaporation of high K dielectrics ($Al_2O_3$, $HfO_2$, $SiO_2$)[248] or $Si_3N_4$ [243] have best RF performance. Progress is also being made with less perturbing coating, like boron nitride (a substrate with no dangling bonds where graphene retains a high mobility), or polymers [214]. The latter however, may induce unwanted hysteretic behavior. Good results (in terms of low leakage, high-capacitance gate-dielectrics and no damage to graphene) have been reported for oxides grown by ALD on graphene seeded by self-assembled molecular layers of perylene-3,4,9,10-tetracarboxylic-3,4,9,10-dianhydride (PTCDA). [213] The PTCDA molecules show long-range order that is not perturbed by defects in the epigraphene or SiC atomic steps. [249] A chemical route using solution-based self-assembling nanodielectrics have been processed on epigraphene Si-face and seem promising.[250]

Contact resistance to graphene is an interesting fundamental problem. The question of how to inject from a 3D to a 2D material while matching energy and momentum is difficult considering that at charge neutrality the graphene Fermi surface is reduced to points in k-space. Similarly to dielectric coatings, metal-to-graphene non-wettability issues are solved by using an adhesion layer (Ti, Cr). The lowest reported contact resistance on epigraphene is less than 100 Ωµm [221] for gold-plated high work function metal Pd or Pt, similarly to carbon nanotubes.

Frequency multiplication and mixing have also been developed for epigraphene. This is based on the nonlinearity of graphene FETs near the Dirac point : the $I(V_g)$ has a voltage square ambipolar behavior at zero bias, unlike in other semiconductor transistors, that greatly suppresses odd-order harmonics in epigraphene-FET devices [251]. Epigraphene -based frequency mixers show excellent linearity [251]. Moreover the observed low phase noise and low 1/f noise in epigraphene-FETs [252] is an important consideration for nonlinear circuits. Ultra-wideband detection (2–110 GHz) was also demonstrated in epigraphene-FETs. Their performance is comparable to or better than state-of-the-art FET-based direct millimeter-wave detection without dc biases applied. [253]

### 5.3.2. Spintronics

Spintronics devices use electronic spin instead of charge to process information. Fundamental to this vision is the possibility of efficient spin propagation over long distances, but despite several decades of intense research, spin transport efficiencies have remained low in metals and semiconductors. Carbon–based materials however are good candidates owing to small spin-orbit coupling and results on high-mobility multilayer C-face epigraphene (µ=17,000 $cm^2$/Vs) indeed presents a breakthrough in the field [254]. Spin transport efficiencies reach 75%, spin signals are in the MΩ range and spin diffusion lengths exceed 300 µm (spin lifetimes >100 ns), which is significantly larger than any other material. For these measurements, the devices were provided with two-terminal high-impedance tunnel contacts (evaporated cobalt on alumina) acting as spin polarizer and analyzer (see Figure 29b). The spin signal is measured by the difference ΔR in the resistance when the two ferromagnetic electrodes have parallel or antiparallel magnetization. The MΩ range signal, as shown in Figure 29c, is to be compared to the few tens of Ohms generally reported in other materials [255]. Similarly, much larger relative spin signals are also measured in C-face epigraphene with ΔR/R≈10%. [254] Micrometer long spin diffusion lengths on C-face epigraphene were also inferred from the resonance width in spin resonance experiments. In that case, the transition between Zeeman-split levels probed by microwave radiation in magnetic field was detected electrically. [256]

On the Si-face, sub-µm spin relaxation length and 1-2 ns spin relaxation time were determined from measurements in a non-local configuration and in Hanle spin precession measurements. In the former experimental configuration, a spin-polarized current is sent between two contacts. The spin accumulation generated at the injection contact diffuses and is detected at two remote spin sensitive voltage contacts. [257] Hanle precession experiments yield an order of magnitude lower diffusion coefficient but higher spin relaxation time on monolayer epigraphene compared to quasi-free standing epigraphene (hydrogenated buffer). This was tentatively explained by localized states in the buffer layer.[258]

### 5.3.3. Large-scale integration – integration with Si

Epigraphene is the only graphene platform where large-scale transistor fabrication was demonstrated. Hundreds of devices [212] were produced on mm size SiC chips as early as 2008 (an example is given in Figure 23c), and more than 10,000 two years later [135] (see Figure 30a). Wafer size patterning of high frequency transistors was demonstrated on a 100 mm wafer (see Figure 30b) in 2011 [68]. An integrated circuit was fabricated on a single SiC chip with a graphene FET that operates as a broadband radio-frequency mixer at frequencies up to 10 gigahertz with thermal stability up to 400 K [259].

As SiC is a large band gap semiconductor widely used in the high power electronics industry, it is interesting to develop schemes to integrate the electronic properties of epigraphene and SiC. One such scheme is to utilize the semiconducting SiC as the transistor channel and graphene as electrodes [215]. A variation of the technique consists in using a quasi-freestanding hydrogenated epigraphene bilayer as the gate, so that the entire device can be carved in one step from epigraphene on SiC (the monolithic wafer-scale electronics scheme is presented in Figure 31). It is expected that this will allow for high currents, high operation speed and high operation temperatures. Current on-off ratios exceeding $10^4$ were indeed observed. [260]

In another promising design [261], the channel between co-planar multilayer graphene source and drain consists of the accumulation layer at the interface of semi-insulating SiC and a surface silicate that forms after high temperature annealing (**Figure 31**). The current flow over the one-dimensional C-face graphene/SiC barrier exhibits on/off ratio over $10^6$, with current densities up to 35 A/m. Significantly, transport is dominated by tunneling at low temperatures, while it is dominated by thermal activation above the Schottky barrier at high temperatures. Tunneling FETs are in fact much sought after to overcome the transconductance thermal limitation.

Another approach is to integrate epigraphene to Si-based electronics. The first Graphene-On-Silicon (GOS) strategy takes advantage of the heteroepitaxy growth of 3C-SiC on Si to produce epigraphene on SiC covered Si substrate (3C-SiC is the only SiC polytype known to grow on Si) [175, 177, 178]. Epigraphene growth on SiC/Si subtrates was discussed in Section 4.4. FETs were produced on GOS with SiC layers on both Si(111) and Si(110) substrates. Besides the usual epigraphene top gates [262] and back gates have also been realized by using of the thin (less than 100nm) heteroepitaxial SiC film on Si as the dielectric [263]. Modulation of the current in the graphene channel is observed. Despite non-negligible leakage currents to the Si substrate through the granular SiC, a room temperature logic inverter was demonstrated.[264]

In another strategy [265], thin monocrystalline silicon layers are transferred onto previously graphitized SiC substrates using well-established industrial Silicon–On-Insulator (SOI) wafer bonding and smart-cut techniques. The transferred crystalline silicon layer bonded on top of SiC/epigraphene is ready for silicon-CMOS devices implementation. It is connected to the sealed epigraphene layer beneath it by metallic vias managed in the sub-micron thin Si layer. The key advantage of the technique is that epigraphene is grown and structured prior to bonding. The process is therefore compatible with epigraphene high temperature growth and preserves epigraphene high structural quality integrity with no degradation. The process produces monolithic integration of graphene on SiC/silicon 3D stacked layers and is fully compatible with Very-Large Scale Integration Technology (See Figure 32a-b).

### 5.4. Band gap

The two main schemes to create a gap in gapless graphene include narrowing the channel width W to open a confinement gap and locally functionalize graphene (sp$^3$ –like covalent bonding). The order of magnitude of the confinement gap Eg=1eV nm/W (see Section 2.9) means patterning at the tens nm scale or below for any practical application. This is usually done by lithography and oxygen plasma etching of 2D graphene, which tends to create disordered edges that severely reduce mobility (see Section 5.5).

Band gaps are reported in all-graphene structures. Epigraphene grown on minifacets a few nm wide that border the sidewall ribbons, as described in Section 4.5, [111] have been associated with the ∼ 0.5 eV wide band gap observed by ARPES [196](see Figure 21). Buckled epigraphene also presents a band gap larger than 0.7 eV from ARPES measurements[266]. Buckling arises in this case from a submonolayer concentration of nitrogen seeded on SiC that pins epigraphene to the SiC interface as graphene grows by thermal decomposition. The buffer layer is also a well-known all-graphene semiconductor, with a band gap of at least 1 eV according to early band structure measurements [35] [78]. Recent ARPES measurements put the distance between the Fermi level in the gap and the top of the conduction band closer to 0.5 eV [267] (see Section 4.1.1). For semiconducting graphene, seamless connection to graphene provides optimum carrier injection, that can be provided in this case by the ballistic ribbon grown on the sidewall facets, as described in Sections 4.5 and 5.5. Transport measurements confirm the semiconducting nature of the buffer layer. [256]

Other seamless planar junctions have been proposed between semiconducting functionalized graphene and graphene structures. These include for instance locally converting an epigraphene channel through oxidation between graphene contacts [268], fluorination [269] or covalent chemistry with nitrophenyl groups [270], where a band gap was demonstrated [271]. The formation of covalent carbon-oxygen/fluorine or molecule bonds locally changes the electronic structure and the transport properties of the epigraphene from metallic to semiconducting.[270] Large on-off ratios up to $10^5$ and current saturation at high bias voltage are measured [269].

Remarkably ferromagnetic behavior was observed at room temperature by functionalization epigraphene with nitrophenyl groups. The interpretation is that magnetic products may be produced when two groups are grafted on the A sublattice, leaving 2 unpaired spins in the B sublattice. [272] Although quite promising, progress may be hindered by inhomogeneous graphene functionalization that consequently reduces the mobility.

Graphite oxide, which is graphene functionalized with epoxy and hydroxyl groups, was fabricated by direct oxidation of epigraphene, either by the Hummers' method [268] or through milder oxidation [273]. Nanopatterning under a biased conducting AFM tip also leads to local graphene oxidation. [274] Interestingly epigraphene (both Si and C-face) resists the very harsh chemical treatment and doesn't exfoliate so that multilayer graphene oxide is produced directly on chips. In that case the layers are very well ordered, exhibit excellent inter-layer registry and little amount (<10%) of intercalated water. [275] This good epigraphene oxide quality makes it possible to reduce locally graphene oxide into graphene with a heated AFM tip. Conducting nanoribbons are drawn down to 12nm wide in an insulating graphene oxide matrix in a single step that is clean, rapid, and reliable [276].

### 5.5. Sidewall Ribbons

As presented in Section 2.9 early theoretical work on graphene ribbons [62, 190, 277] [278] predicted different electronic properties based on their edge orientation: zig-zag ribbons have quasi-non dispersive states close to charge neutrality while two out of three armchair ribbons are gapped, see Figure 5. Following these papers and early proposal for graphene nanoelectronics [1, 2], ribbons a few tens to hundreds of nm wide have been patterned by oxygen etching 2D graphene. It was found that transport in these ribbons is quenched at charge neutrality and at low temperature, resulting in gate-induced current on/off ratio up to $10^6$ with a large temperature dependence [182, 184, 186, 187]. All were attributed to electron localization effects due to edge disorder. Following Eq.2 $E_{n,m} = hv_F \sqrt{(n/L)^2 + (m/W)^2}$, described in Section 2.9, it is clear that only the m=0 modes are occupied in narrow ribbons at low temperature. It is therefore expected that these 1D edge states wouldn't be conducting if too much disorder is present.

On the contrary, as shown in Figure 33, 40nm charge-neutral epigraphene sidewall ribbons (Section 4.5 are found to be metallic, with conductance of the order of $h/e^2$. This demonstrates the presence of conducting states in the confinement gap (as discussed in Section 2.9, Figure 5) and indicates single-channel ballistic transport. [136]

Direct proof of ballistic transport is provided by the measurements of a sidewall zigzag ribbon presented in Figure 33a in a four point configuration, where the current is passed through the outer tips and voltage is measured at the two inner probes (see inset of Figure 33a). The distance d between the nanoscopically sharp tungsten voltage probes in ohmic contact with a single ribbon can be controlled in situ in the built-in electron microscope. It is remarkable that (i) The resistance $R_{4p}$ extrapolates to $R_{4p}(d=0)=h/e^2$ at zero length indicating single channel transport (ii) The resistance per unit length $R'=\Delta R_{4p}/\Delta d$ decreases after in situ heating (a process known to clean graphene), and becomes length independent (bottom trace in Figure 33a), indicative of extremely long mean free path (the transmission of a 1D single conducting channel can be related to a mean free path $\lambda$ according to $R=(h/e^2)(1+d/\lambda)$, as indicated in Figure 33a. (iii) The probes are invasive and interrupt the current flow. In a diffusive conductor a probe merely affects the resistance that is dominated by scattering on multiple impurities and defects. Here it is shown that introducing a passive tip in between two voltage probes doubles the resistance; two passive probes triple it. Similarly four point and two point (where the same tips are used as current leads and voltage probes) measurements give essentially the same result. This is because right-moving charges that enter the passive probe from the left, say, will leave the probe going either left or right with equal probability; hence the transmission probability is half. [136]

It is not understood why only one conducting channel is measured. For short ribbons at room temperature a second channel contributes and vanishes exponentially for length longer than 160nm; similarly the single channel conductance of long ribbons drops to zero as the distance between probes increases beyond 16μm (see Figure 33b). At low temperature, for a fixed contact geometry applied to a curved ribbon (see Figure 33c - inset), the conductance increases with increasing the gate voltage. This can be understood by the Fermi level moving to the upper bands so that many more channels contribute to the conduction. The conduction in the upper bands is found to be small, with mean free path of the order of 50 nm, which is comparable to the ribbon width. From the temperature dependence of Figure 33c it can be seen that each of the curves can be displaced vertically to overlap the others (Figure 33d), indicating that only the resistance at Vg=0 is temperature dependent. This was explained in terms of a two-mode transport model.[136] These exceptional results underscore the importance of well-prepared ribbons, with well-defined and annealed edges.

## 6. Optical and Plasmonic Properties

Recent years have seen a surge in publications dealing with photonic, optoelectronic and plasmonic properties of graphene and graphene–based devices. Concepts and figures of merit of prototype devices have been described in recent reviews [279-281]. Although the challenge remains to scale up from single device to large scale production and integration with existing photonic and electronic platforms [280], potential applications cover a large range from passive components such as transparent electrodes in displays or touch screens, to active elements as proposed in photovoltaic module, photodetectors, optical modulators, fast saturable absorbers for high power lasers, infra-red filters, frequency convertors, for THz emission or detection, plasmonic devices, etc. These exploit various properties of graphene, notably high mobility transport, optical transparency, wave-length independent absorption, and non-linear optical response properties. In fact graphene interacts strongly with light (graphite is black) with a remarkably high absorption coefficient in the near-infrared and visible range: γ=πα= 2.3% per layer (α = $(1/4\pi\varepsilon_0)$ $e^2/\hbar c \approx 1/137$ is the fine structure constant). This is one to three orders of magnitude larger than that of technologically relevant photonic materials such as $In_{0.53}Ga_{0.47}As$, GaAs, or germanium at a wave length of 1.55 μm [281], the wavelength used in most fiber optic telecommunication systems. However since graphene is just one layer, it is usually considered as optically transparent. One optical property that is specific for epigraphene on the C-face, is that the optical conductivity is expected to depend on the layer stacking and the relative orientation of bilayers.[282]

Quite uniquely, graphene's gapless band structure allows charge carrier generation by absorption of light in a broad wavelength range. In neutral graphene an electron-hole pair can be excited by any radiation from THz to ultraviolet, since there is no minimum energy difference between conduction and valence band, as shown Figure 34b (see for instance Ref. [280] and refs therein). Since a photon can only be absorbed when its energy matches the energy difference between an occupied and an empty state, for energies below 2EF in graphene (see Figure 34b) the absorption is therefore blocked (so-called Pauli blocking) and the graphene layer is transparent. This provides in principle a controllable way to tune transparency with doping. Carrier multiplication, promoted by strong electron-electron interaction and weak electron-phonon coupling, and high carrier mobility are favorable for ultrafast conversion of light to electrical signals [280]. Integration of graphene with metal contacts, gates, in cavities, wave guides, imbedded in polymers, in composite stacks, coated with molecules or colloidal quantum dots have been devised to increase optical absorption, and photoresponse [280]. As for epigraphene, studies of the optical properties and realization of photonic devices are however quite limited and were mostly directed towards ultrafast spectroscopy and plasmonic effects that we briefly discuss below.

### 6.1. Ultrafast optical spectroscopy

Ultrafast time-resolved optical pump-probe spectroscopy has been used to study the dynamic response and the hot-carrier relaxation and cooling dynamics in epigraphene with femtosecond resolution. In this technique, an ultrafast optical pump pulse excites electrons and holes to high energies, while a probe pulse, whose frequency can be tuned from the visible [283] through the infrared [203] to the THz frequencies [284, 285], probes the dynamics response (see Figure 34). The electron dynamic response is an important property for all photodetectors and photovoltaic devices that rely on the conversion of light into free electron–hole pairs. Best performances can be are achieved if the photoexcited carriers transfer their excess energy into the production of additional electron–hole pairs (carrier multiplication) through carrier–carrier scattering processes rather than production of heat (emission of phonons). Similarly the ultimate performance of high-speed electronic devices will depend on the energy exchange between large electric fields induced hot carriers and the lattice.

Epigraphene is particularly well suited for ultrafast optical experiments due to its large area and homogeneity. Up to cm-scale surface areas provide sample sizes that are large enough even for the typically large probe spot sizes at frequencies below 300 GHz. The signal is enhanced by the multiple layers in C-face epigraphene without compromising the graphene characteristics since each layer behaves electronically like a monolayer graphene. Finally, transmission experiments benefit from the transparency of the SiC substrate over broad frequency window from the visible to the THz range.

In ultrafast time-resolved THz spectroscopy experiments an ultrafast optical pump excitation generates electron-hole pairs in a non-equilibrium state, while a broadband single-cycle THz probe pulse probes the dynamic THz response of the hot-carrier relaxation and cooling dynamics. The high-energy non-equilibrium electrons and holes thermalize with the background of cold carriers due to very efficient carrier-carrier scattering which forms a single hot-carrier Fermi-Dirac distribution within ~100-200 fs, as observed in epigraphene [203, 283-286]. The hot-carrier distribution is characterized by an elevated transient carrier temperature and a transient Fermi level. As the hot carriers relax and cool, both the transient carrier temperature and Fermi level shift return to equilibrium. The THz carrier dynamics are the result of an interplay between efficient carrier-carrier scattering, which maintains a thermalized hot-carrier distribution, and carrier-optical-phonon scattering, which removes the energy from the hot carriers to the lattice [287, 288]. This dynamics is shown to depend critically on the doping density, ranging from a few picoseconds at doping density n >$10^{13}$cm$^{-2}$ to hundreds of picoseconds at doping densities n <$10^{10}$cm$^{-2}$, due to the vanishing density of states at the Dirac point. In MEG, the low-energy dynamics is governed by a unique cooling pathway enabled by interlayer energy transfer via screened Coulomb interactions between Dirac electrons, with the highly doped layers near the SiC substrate acting as a heat sink for the quasi-neutral top layers [284].

In a special semiconducting form of epitaxial graphene, in which an energy gap of up to 0.7 eV was engineered by buckling [266], the relaxation of the dynamics is enhanced by up to two orders of magnitude, which is attributed to direct electron-hole recombination via optical phonon emission over a broad range of carrier temperatures[285].

These experiments, consistent with results on other forms of graphene [279-281] [289, 290], show that carrier–carrier scattering is highly efficient in a wide range of photon wavelengths and produces secondary hot electrons that can drive currents. This multiple hot-carrier generation is therefore interesting for highly efficient broadband conversion of light into electronic energy for optoelectronic applications.

Applying an external magnetic field leads to a significantly longer relaxation [291], which is attributed to a reduction of electron-electron scattering. This is because in a magnetic field the energy levels bunch into unequally spaced Landau levels (see Section 5.2.2), therefore inter-level carrier–carrier scattering processes (Auger processes) are not allowed because they cannot conserve energy. This result is in sharp contrast with normal 2D electron gases, where the formation of equidistant Landau levels in magnetic field increases scattering . Graphene's zeroth Landau level is special because the three $LL_0$, $LL_1$ and $LL_{-1}$ Landau levels are equally spaced so that Auger processes are allowed [292]278]. Neutral C-face layers provide an ideal system to study carrier-carrier scattering processes involving the $LL_0$ level, and strong Auger processes have been observed by addressing the levels individually with circularly polarized light [292].

Ultrafast optical spectroscopy also gives insight into the electronic structure. Consistent with ARPES measurements [9], the upper limit for a potential band gap is found to be below 1 meV in C-face graphene [203]. The doping density in the successive layers in MEG was also evaluated, corresponding to the energies of Pauli blocking ($E_F \sim$ 350meV, 210meV, 135meV, 90meV) [203], see Figure 34b. Also the polarization of the THz emitted light by coherently controlled photocurrents (see below) indicates that there is some coupling between the graphene layers in MEG although the electronic structure of each layer is that of graphene.[293]

Coherently controlled photocurrents have been produced in MEG. [294] The method uses two phase-locked beams at frequency ω and 2ω. Quantum interference between single-photon and two-photon absorption breaks the lattice symmetry, so that the photoinjected carriers have an anisotropic distribution in k resulting in a net current flow. The current flow direction is controlled by the polarization of the pump beam. The transient current is detected by the electromagnetic pulse at the THz frequencies that it generates. Note that this all-optical injection of current could provide a noncontact way of injecting directional current in graphene. Quantitative studies of the magnitude of the effect show that current decay time in graphene are longer compared to common semiconductor like GaAs.

### 6.2. THz generation

Graphene exhibits a large nonlinear optical response due to its linear energy dispersion that leads to harmonic generation at THz frequencies. However, second-order nonlinear effects important for applications like frequency difference or rectification processes, are forbidden by the centro-symmetric nature of graphene. Generation of coherent THz radiation (0.1 to 4 THz and projected up to 60 THz) via a second-order nonlinear effect was nonetheless achieved in multilayer epigraphene on the C-face where it is induced by the anisotropy of the in-plane photon momentum [295]. This dynamical photon drag effect relies on the transfer of photon momentum to the carriers by the ponderomotive electric and magnetic forces. Interestingly, the effect is related to the usually neglected next-nearest-neighbor coupling of C atoms in graphene and to the asymmetry between the electrons and holes dynamics. These properties are particularly interesting for the generation of ultra-broadband terahertz pulses in compact room temperature THz sources.

### 6.3. Photo-current

The electrical response of epigraphene under illumination was studied for both Si and C-faces. Similarly to the classical Hall effect, the crossed AC electric and magnetic fields of circularly polarized light can create a voltage, what is called the circular AC Hall effect. Graphene centrosymmetric structure prevents other helicity-driven photocurrent effects, and the circular AC Hall effect was observed for the first time in epigraphene Si-face, owing to the large area available for the terahertz (THz) laser radiation. [296]

In one report [297], a significant increased photocurrent on the Si-face compared to the C-face (and CVD graphene) is attributed to the presence of the buffer layer, although the mechanism is unclear. In an attempt to increase photocurrents, asymmetric contact configurations were applied on top of C-face and Si-face epigraphene (Au/epigraphene/ Al [298] and Ti/epigraphene/Pd [299]). The enhanced photoresponse is attributed to the difference in built-in electric fields at the two epigraphene /metal interfaces. Thicker layers show stronger response [298, 299]. In another implementation, an epigraphene strip was locally modified by laser illumination (LEG) creating  epigraphene/LEG/epigraphene Schottky junctions. In the LEG strip, epigraphene is converted into a poor conductor (possibly graphene oxide) and the device show nanosecond photoresponse, promising for photodetectors [300]. In the same line enhanced photosensitivity of epigraphene was also reported after electrochemical oxidation in nitric acid [301]. This is attributed to the formation of deep traps at the electro-oxidized epigraphene interface, which release charge carriers under illumination and prolong the lifetime of the photocarriers. The SiC substrate can also induce photocurrent resonance. This was observed in the spectral region where SiC has a negative dielectric constant.

As SiC is a wide-band gap semiconductor in its own right, epigraphene was also used as the (transparent) contact to probe the photo-generated charge carriers in the semi-insulating 4H-SiC substrate [302], with best response in the ultra-violet.

### 6.4. Plasmonics

An important motivation for plasmonics is to merge the fields of photonics and electronics using nanoscale confined electromagnetic fields. Graphene presents the advantage that its plasmonic properties (frequency, propagation direction) are highly tunable by changing its carrier density [303, 304], structural characteristics, such as the graphene nanostructure dimension, packing density, or number of stacked layers or by applying a magnetic field. Also owing to the low ohmic loss of the material, long lifetimes and high degree of optical field confinement are observed.

Like for other metal surface, collective charge density oscillations of the electron gas (plasmons) can propagate at the surface of graphene. These electromagnetic waves have their own, strictly two-dimensional, band-dispersion relation that is different in graphene than in ordinary metals [86, 305]. Plasmons can couple to light (creating surface plasmon-polaritons, that can convert light into electronic signals), to elementary charge (to create plasmarons), to interband electron–hole (e-h) pair excitation (to create plexciton) and to phonons. All the plasmon coupling mechanisms mentioned above have been observed in epigraphene.

The first observation of plasmons in graphene was realized in epigraphene Si-face by high-resolution electron-energy-loss spectroscopy (EELS). The predicted dispersion relation $\omega_p \propto \sqrt{q}$ at small wave vector $q$ was confirmed as well as a kink in the dispersion in the vicinity of the Fermi wavevector [204, 305]. This dip in the plasmon dispersion was shown to be an intrinsic property of pristine epigraphene and does not depend on defect concentration, on the number of graphene layers or temperature. It is attributed to the decay of plasmons due to resonant coupling between plasmons and e–h pair excitations. [306] At higher $q$ an unusual damping of the plasmon mode is observed, because of the combined effect of phase space limited backscattering for electrons in graphene and enhanced electron-electron elastic scattering.[305] In an unexpected application, epigraphene was used to demonstrate that coating materials with graphene can enhance radiative heat transfer between materials in the near field [293. This exploits the tunability of graphene plasmons that can match the frequency of phonon-polariton or plasmon-polariton resonances in a given material. Best heat transfer is indeed expected when resonances in the opposing bodies match each other. This effect may be interesting for energy conversion devices.[307]

Plasmons in highly charged epigraphene on the Si-face strongly interact with the surface phonons of the SiC polar substrate [204]. In this system, the plasmon and phonon energy bands show avoided crossings. Despite electronic screening from one layer to the next, the plasmon-phonon coupling extends to multilayers.

Bound states of plasmons and electrons/holes (plasmarons) can play a strong role in renormalizing the bands around the Dirac energy [83, 84]. In particular for potassium doped quasi-freestanding epigraphene (hydrogen-lifted buffer layer), ARPES measurements show that the energy band-dispersion close to the Dirac point is reconstructed in a nontrivial way with four distinct bands instead of the usual pair [83] (see Figure 10b). This was interpreted as evidence for many-body effects (electron–electron, electron–plasmon and electron–phonon couplings) in the dynamics of quasiparticles.

Surface plasmons manifest themselves as resonance absorptions in the optical transmission spectra. Excitation of plasmons was shown to dramatically modify the magneto-optical response [308], in particular the giant Faraday rotation observed for the first time in epigraphene [309].

Despite the inefficiency of plasmon excitation and detection by light due to the large wavevector mismatch of light with graphene plasmons, light-matter interactions was demonstrated in graphene nanostructures via confined plasmons. In this case, electromagnetic radiation can excite plasmon resonances in graphene, even in nanostructures having dimensions much smaller than the incident light wavelength. These structures can be either intentionally patterned or arise naturally, such as the step-terrace microstructure of epigraphene on the Si face or the tapered monolayers structures grown on the C-face [290]. In magnetic fields, plasmons can couple with the cyclotron resonance (inter Landau level transition) to form a hybrid mode. Infrared magneto-spectroscopy measurements have taken advantage of large arrays of sub 100 nm patterned quasineutral ribbons ($E_F$<17meV) of high mobility (μ>50,000cm$^2$/Vs) on epigraphene C-face to study this hybrid mode in detail. The observed energy shift exhibits a peculiar energy – magnetic field scaling that distinguishes it from conventional 2DEGs and in highly doped graphene. [310]

Complementary to the spectroscopic results above, coupling of plasmons and light have enabled plasmon fields imaging in real space on epigraphene monolayers [304] (see Figure 35) or multilayer [311] C-face nanoribbons [304]. Strong optical field confinement (plasmon to incident light wavelength reduction by a factor more than a hundred)[310] and long propagation distances are observed [304], and the time evolution of the electric field distribution over the surface [312] could be detected in the THz regime.

Important to the field of plasmonics is the possibility of plasmon tuning, also demonstrated in epigraphene. The plasmon wavelength can be varied in epigraphene over a wide spectral range by slightly changing the incident light wavelength [304], owing to the strong dependence the dielectric constant of SiC on the wavelength and the sensitivity of plasmon to the dielectric environment. The dependence of the plasmon dispersion on the carrier density (proportional to $n^{1/4}$) was measured on a top gated epigraphene Si-face, [313] where plasmon transport was studied by time-resolved electrical measurements of a charge pulse travelling in a plasmon mode. Change in the plasmon velocity was also realized by varying the magnetic field, charge density and top gate screening effect in epigraphene on the Si-face. [313]

## 7. Conclusion

In this review we have presented the status quo of the science and technology of epigraphene. Historically epigraphene is the first graphene platform to be scientifically and technologically developed and it is currently the frontrunner for graphene electronics. Epigraphene surface science research predated transferred graphene research by several decades, the concept of graphene electronics was first proposed and patented based on epigraphene electronics research, and electronic transport properties of monolayer graphene were first demonstrated on epigraphene. Because of its remarkable high quality, it is the only material to reveal the characteristic graphene band structure in ARPES, and epigraphene nanostructures reveal one dimensional unique electronic transport properties that have not been seen elsewhere. In contrast to all other graphene flavors, epigraphene survived on the route to become a viable electronics platform, because only it satisfies the many necessary requirements. After more than a decade of research into epigraphene nanoelectronics, the field is still strong, in contrast to nanoelectronics that is based on transferred graphene that has rapidly declined due to difficulties in fabricating the necessary low-defect nanostructures. We can therefore comfortably predict that it is only a matter of time before graphene electronics is realized. However considerably more work needs to be done before this dream is realized.

## Acknowledgements
Zhigang Jiang, Ted Norris and Momchil Mihnev are thanked for a critical reading of the manuscript. C.B. and W.A.d.H. thank the AFOSR and NSF under grants No FA9550-13-1-0217 and 1506006, respectively. Additional support is provided by the Partner University Fund from the French Embassy. CB acknowledges partial funding from the EU Graphene Flagship program. E.H.C acknowledges support from NSF under grants No. DMR-1401193 and No. DMR-1005880.

## 8. Figure Captions

Figure 1: (a) Graphene is a one atom thick 2 dimensional honeycomb structure of carbon atoms. (b) Two Bernal stacked graphene layers are the constitutive building block of hexagonal graphite (c) Three Bernal stacked graphene layers. (d) ABC stacking of graphene layers makes the unit cell of rhombohedral graphite. A and B atoms correspond to the two sublattices forming the honey comb structure. The tight-binding $\gamma_i$ parameters, corresponding to first, second, etc neighbors, for electronic structure calculations are indicated. (From Ref. [314] ).

Figure 2:(a) Band structure E(k) of graphene, and zoom in close to the K point, showing the two inverted cones. (b)–(e) The $\pi$-bands near $E_F$ for 1– 4 graphene layers, respectively, measured for epigraphene (Si-face) by Angle Resolved Photo-Emission Spectroscopy. The dashed lines are from a calculated tight-binding band structure, with band parameters adjusted to reproduce the measured bands. Red and orange lines are for ABAB and ABAC stacking, while blue lines are for rhombohedral ABC stacking. (From Ref.[10]).

Figure 3: Comparison of epigraphene with different exfoliated graphene materials. (a) AFM images of exfoliated graphene [315]. (b) An STM image of epigraphene grown on SiC-(0001).(courtesy Philip First, Georgia Tech and Joseph Stroscio, NIST-CNST). Note the ten- fold increase in flatness of the epigraphene film. (c) A comparison of the ARPES measured Dirac cone for 2D epigraphene grown on SiC-(000-1)[9] (left), the band structure of a 2D exfoliated film (center), [316] and a 24nm wide epigraphene sidewall ribbon.[196] Note that the energy and momentum scales are the same in all images in (c) and (d). Top panels in (c) are measured E(k) and bottom panels are constant energy cuts through E(k) showing the momentum distribution curves (MDC) of the bands. The $\Delta k$ broadening in exfoliated graphene gives a coherence length, Lc, = $2\pi/\Delta k$ of 1-3nm. The momentum distribution curve width for epigraphene, including the 24 nm wide ribbons, is entirely due to instrument resolution giving $L_c$ >10 times longer than exfoliated graphene. (d) Energy and momentum distribution curves and MDCs of the highest quality exfoliated graphene.[317]

Figure 4: Epigraphene Si-face grown in UHV (a)-(c). Atomically resolved Scanning Tunneling Microscopy image of epigraphene (sample bias = 0.2eV) (From Ref .[126]). (a) Monolayer on SiC showing the moiré superstructure. Atomic resolution STM image of (b) monolayer and (c) bilayer (4x4 nm$^2$ images ; Sample bias: -40 mV). (c) Bilayer epigraphene (from Ref.[118]). (d) Large scale AFM image showing a Swiss cheese morphology (scale bar: 1µm)[66].

Figure 5: Tight –binding band structure calculation for narrow graphene ribbons with (top) armchair and (bottom) zigzag orientation for various ribbon widths as labeled by the number N of row of carbon atoms. Zigzag ribbons have flat band at E=0, and armchair present a gap for specific widths. (e) and (j) N=30.(From [62])

Figure 6: SiC-hexagonal (H) and cubic (C) polytypes. Red: carbon atoms; blue: silicon atoms. The unit cell comprises respectively of 2, 3, 4 and 6 Si-C bilayers. A, B, C refers to the stacking sequence (From Ref. [318]).

Figure 7: Schematics of the confinement controlled sublimation growth method [34]. Contrary to Si sublimation in vacuum (a), in the CCS method (b), the Si vapor is confined in a graphite enclosure provided with a calibrated leak. (c) Photograph of the inductive furnace used for the CCS. (d) Schematics of graphene growth on the two polar faces of 4H and 6H-SiC.

Figure 8: (a-b) Si-face epigraphene grown under 1 atm Ar pressure. (From Ref. [66]) (a) AFM image with a nominal thickness of 1.2 monolayer covering the SiC steps (b) LEEM image revealing macro-terraces covered with graphene. Area covered with 1, 2 and 3 graphene layers have been identified by the presence of 1, 2 or 3 reflectivity minima, respectively. Two and three layers are located at the step edges. (c-e) Hydrogen intercalation after epigraphene growth on the Si-face (c) (left to right) E(k) dispersion measured in ARPES for: pristine buffer layer, after hydrogen intercalation, after annealing at 900C. (d) Same as (c) for a monographene layer. (e) Schematics of the hydrogen buffer lifting from the SiC substrate.

Figure 9: (a) The unit cell structure of 4H-SiC. Filled circles are carbon atoms and open circles are silicon atoms. (b) LEED pattern of monolayer graphene grown on the Si-face. The subscript "G" refers to coordinates in the graphene lattice constant. (c) LEED pattern of 10-layer graphene film grown on the C-face. The diffuse graphene rings are marked.

Figure 10: Renormalization of the band structure due to electron-plasmon interactions (a)-(b) Epigraphene Si-face (n- doped $1.1 \times 10^{13}$ cm$^{-2}$). ARPES measured E(k) along a line through the K point and along the vertical double arrow in the inset of (b). The kink shape is outlined by the dashed line (b) same band but along the horizontal double arrow (in that direction the other band cannot be observed in ARPES). (c) Schematic picture of the renormalized band. (d-e) Same experiment but for potassium doped (n=$1.7 \times 10^{13}$cm$^{-2}$) quasifree standing epigraphene Si-face along the (c) vertical and (e) horizontal directions, showing the strongly modified band due to plasmaron (electron-plasmon coupling) (f) Schematic picture of the renormalized band and (g) comparison with non-renormalized graphene.

Figure 11: (a) Atomically resolved STM images of the 6x6 structure at $U_{tip}$=1.7 V (green dashed line marks the 6x6 cell)(from Ref. [35]). (b) STM at $U_{tip}$=0.2 V showing the 6√3 cell [from Ref. [35]]. (c) Differential conductance measurements obtained on the buffer and monolayer showing a band gap feature of ~0.8eV on the buffer [from Ref. [90]]. (d) ARPES intensity map of the SiC(0001) (6√3x 6√3) R30 surface (hν =50 eV). Vertical dashed line marks the K-point of graphene. Two states, $g_1$ and $g_2$ above the SiC valence band are also marked. (From Ref. [78])

Figure 12: Epigraphene growth model (Si-face). (a) 10μm x 10μm atomic force microscopy image of Si-face epigraphene grown by the CCS method, showing step edge growth. (b) Graphene growth kinetic processes on a vicinal surface by step decomposition (c)-(d) Onset of step instability: once the thermal decomposition of SiC has started a positive feedback mechanism promotes further decomposition and graphitization where it has already begun. (From Refs [116, 117])

Figure 13: C-face epigraphene grown by the CCS method. (a) 2.6nmx2.6nm atomic resolution STM image corresponding to the blue square in (b). (b) 400nm x 400nm STM image. Bottom full height scale: 0.1 nm. The RMS height variation is less than 0.02nm of this large area. (c) Atomic Force Microscopy image showing extended flat regions bordered by continuous graphene pleats. Scale bar : 10 μm. (a) A 50 μm field of view LEEM image of a C-face graphene film with an average thickness of 3-layers. Contrast is due to local layer thickness marked in the figure. (e) 400nm long STM image across two regions of different orientations, showing that the top graphene layer is continuous and flat (rms along the profile line in red : 50pm) (STM images: courtesy Philip First, Georgia Tech and Joseph Stroscio, NIST-CNST).

Figure 14: (a) μ-LEED image of a (√57x√57)$_G$R6.59° superlattice from a commensurately rotated two layer C-face graphene film (E= 96 eV). The image is from a 1μm area [From Ref. [150]]. The principle diffraction rods from two rotated graphene sheets are marked. The inset shows a blow-up of the superstructure unit cell in graphene units. (b) A calculated LEED pattern from the same √57 structure commensurate (log scale). (c) STM topography of moiré patterns on multilayer epigraphene (sample bias $V_S$=0.5 V, tunnel current I=100 pA). (c) Two similarly sized moiré patterns due to three graphene layers result in a large superlattice unit cell. Layers 1 (surface layer) and layers 2 and 3 have comparable rotation angles but in opposite directions. (d) High resolution image of a region in (c). (e) STM image of a different moiré superlattice. (f) High resolution image from within the same area. [141]

Figure 15: (a) ARPES measured band structure of an 11-layer C-face graphene film grown on the 6H SiC(000-1). The ARPES resolution was set at 7 meV at hν=30eV. The scan in $k_y$ is perpendicular to the SiC <10-10>$_{SiC}$ direction (i.e. graphene rotated 0° relative to SiC). Two linear Dirac cones are visible. (b) k-PEEM constant energy surface (BE=1.3 eV) from a 2ML C-face graphene film using a 7μm field aperture. Two sets of rotated graphene Dirac cones are visible (relative rotation of 21.9°). The reciprocal lattice constants of the two BZ are $g_1$ and $g_2$. A third set of cones, on a different radius, are obtained by coherent diffraction of either the $g_2$ set of cones by the principal lattice

vector g1 (dotted line), or the g$_1$ set of cones diffracted by the secondary lattice vector g$_2$ (dotted line) (from Ref. [319]).

Figure 16 Raman spectroscopy of C-face epigraphene. (laser excitation: 532 nm)(a) A 26nm thick multilayer epigraphene (average thickness determined by ellipsometry), showing the two main graphene Raman peaks (G and 2D) and a very low D peak, indicative of very low disorder and extended graphene sheet. Inset: 2D peak fitted by a single Lorenzian peak (FWHM=25cm$^{-1}$, position=2702 cm$^{-1}$). (b) for thinner films the raw spectrum (blue) contains both the graphene and SiC contribution. The latter (black trace) is subtracted by a Non-negative Matrix Factorization method [159] to reveal a pure graphene Raman spectrum (red).

Figure 17: Epigraphene on 3C-SiC. (a) LEEM image of epigraphene on 3C-SiC (111), showing 98% monolayer coverage. The 3C-SiC substrates were grown by sublimation epitaxy on 6H-SiC, and epigraphene was grown at 2000ºC under 1atm Ar (From Ref. [179]). (b) ARPES of epigraphene on n-type 3C-SiC(111) showing quasi-neutrality after hydrogen intercalation (From Ref. [180]). (c) Epitaxial graphene formation on 3C-SiC/Si thin films, LEED pattern shows rotational order on 3C-SiC(100) compatible with C-termination, and Bernal stacking on 3C-SiC(111) compatible with Si-termination. (From Ref. [175])

Figure 18: (a) 6µm field of view (FOV) of a sidewall graphene trench array. (b) An expanded 25µm FOV of (a) shows a parallel array used for area averaged ARPES measurements. (c-f) Examples of epigraphene sidewall structures produced by the structured growth method [135], demonstrating its versatility: (c) an array of parallel ribbons grown on the step edges of a vicinal surface, (d) graphene grown in many orientations on the sidewalls of complex convoluted trenches, (e) a Hall bar with eight transverse voltage probes, and (d) rings grown on the sidewalls of pillars. The top images in (a-d) and the left image in (f) represent the AFM topography; the corresponding bottom and right images are electrostatic force microscopy images showing bright areas where graphene grows at the edges of the etched patterns (From. Ref. [191]).

Figure 19: (a) Schematic definition of AC and ZZ graphene ribbons. (b) shows the orientation of the graphene Brillouin zone with respect to the ZZ and AC directions. (c) and (e) show a schematic of an AC edge step on SiC and the corresponding AFM image of an AC step edge array. (d) and (f) show a schematic of an ZZ edge step on SiC and the corresponding AFM image of an ZZ step edge array.

Figure 20: High Resolution cross-sectional TEM images of graphene layers grown on SiC steps. (a) - (b) Views of different edge terminations of natural sidewall graphene. Graphene layers are observed as dark contrast lines [From Ref. [192]]. (c) False-color image of a sub-facet region near the top of a 30nm deep etched facet sidewall (armchair). The image is an overlay of a high-angle annular dark-field image (HAADF) (red) that enhances the SiC and a low- angle annular dark-field image (green) that enhances the graphene. (d) HAADF atomic resolution scanning TEM image of a ZZ-edge step facet on 4H-SiC(0001) with 5 graphene layers, cross sectioned along the [1-120] direction. Note the difference of the distance between the first graphene layer and the SiC (From Ref. [198]).

Figure 21: (a) Schematic of how sidewall graphene is measured in ARPES. The Dirac cone from the (0001) surface is at an angel φ relative to the (0001) surface normal n$_{(0001)}$. The Dirac cone from the sidewall graphene is at an angle θ relative to the Dirac cone of the (0001) surface. (b) A constant energy cut of reciprocal space as a function of the detector angle, θ, from an AC step array. Facet cones from facets on both sides of the trenches are marked.

Figure 22: Schematic of SiC step flow during H$_2$ etching: (a) AC step edges and (b) ZZ-step edges. (From Ref. [200])

Figure 23: Top-gated graphene on Si and C-faces. (a) Top gated Si-face graphene (40nm SiOHx,(HSQ e-beam resist)), gate effect on the resistivity ρ$_{xx}$ (300K) and Hall effect ρ$_{xy}$ (4K – 5T). Inset: image of the patterned Hall bar, with all graphene contacts, before and after top-gate deposition. On-off ratio=31 (From [320]). (b) Top gated C-face graphene (monolayer with 40nm alumina gate dielectric); 2 point conductivities. The two branches give FET

mobility  μ$_{FET}$= 8700 cm$^2$/(V.s) on the n-side, to be compared to Hall mobility  μ$_{Hall}$ = 7500cm$^2$/(V.s) at n = 1.6 x 10$^{12}$/cm$^2$ (Vg = 0 V), and μ$_{FET}$ ~ 5000 cm$^2$/(V.s) on the p-side. Inset: AFM image of a C-face monolayer graphene (scale bar 5 μm) over SiC steps. The white lines are graphene pleats characteristic of C-face epigraphene. (From ref. [221]). (c) Top gated multilayer C-face graphene (40nm HfO$_2$) of mobility up to 5000 cm$^2$/V.s. Inset: optical image of the multiple devices patterned on a single SiC chip (From Ref. [212]).

Figure 24: Weak anti-localization in multilayer epigraphene (C-face) (a) Low-field magnetoresistance for a sample of 100 μmx1000 μm at various temperatures. Inset: magnetoresistance peak near B=0 at 1.4 K. The electron mobility is 11600 cm$^2$/V.s, and the transport time  τ=0:26 ps. (b) Low-field resistance (open circles) at 4.2 K after subtracting the high temperature background resistance. Solid line is a fit to the model of weak anti-localization, showing excellent agreement with the data. The only temperature dependent parameter is the phase coherence time τ$_\phi$ potted in the inset, that is consistent with electron-electron scattering time. (From ref. [50])

Figure 25: Landau level magneto-spectroscopy. (a) Temperature dependence of the transmission spectra taken at B=0.8T, showing the L$_{-1}$(0) to L$_0$(1) transition up to room temperature. Successive spectra are shifted vertically by 0.15 for clarity. The Lorentzian fits to the data (blue curve) indicate no peak shift or widening up to room temperature (from  Ref. [52]). (b) Position of the absorption lines as a function of the square root of the magnetic field. Dashed lines are calculated energy of the transitions between Landau levels. The transitions L$_{-m(-n)}$ to L$_{n(m)}$ are labeled corresponding to the diagram (color coded) in the inset.

Figure 26: Landau level tunneling spectroscopy. (a) Scanning Tunneling Spectroscopy at low temperature in 5T of multilayer epigraphene (C-face) showing graphene Landau level peaks in the local density of states. Insert 5x5nm STM image. The scale is in pm. (From Ref. [321]). (b) High-resolution Landau level spectroscopy of the fourfold states that make up the N=1 Landau level above 11 Tesla. The large energy separation between peaks corresponds to valley splitting (ΔE$_V$); The spin splitting is indicated by up (down) arrows in each valley. (From Ref. [54])

Figure 27: Magneto-transport in epigraphene C-face (a-b) and Si-face (c-d). (a) Multilayer C-face Hall bar (1μm x 5μm, mobility= 12 500 cm$^2$ /V.s, charge density= 4.5.x 10$^{12}$ cm$^{-2}$). Shubnikov de Haas oscillations are observed in the magnetoresistance (main panel), as outlined once a background is subtracted (bottom panel). The level index is plotted as  a function of 1/B, where B is the position of the magnetoresistance maxima (From Ref. [322]). (b) Monolayer C-face at low charge density (n=1.9x10$^{11}$ cm-$^2$) and high mobility (39,800 cm$^2$/V.s). The ν=2 ρ$_{xy}$ plateau and vanishingly small ρ$_{xx}$ are  observed as low as 3 T at 4K. Inset quantum Hall effect observed at 200K, well above liquid nitrogen temperature (From Ref. [137]). (c) Shubnikov-de Haas oscillation in ρ$_{xx}$ and corresponding ρ$_{xy}$ for monolayer Si-face at high charge density. (From Ref. [1]). (d) QHE for metrology in Si-face epigraphene. At lower charge density, quantization is observed above 3T (main panel). The relative deviation ΔRH/RH of the Hall resistance RH to h/(2e$^2$) and of R$_{xx}$ are plotted in the top and bottom panel respectively. It shows a standard uncertainty around 1x10$^{-9}$ (From Ref. [76]).

Figure 28: High frequency epigraphene transistors. (a) Scanning electron microscopy image of a symmetric dual gate transistor, featuring a graphene channel between source (S) and drain (D) modulated by a gate (G) fabricated on C-face graphene. (b) High frequency characteristics of a 100nm gated channel. The cutoff frequency is f$_T$= 110GHz for the current gain (obtained by extrapolation of theoretical slope of 20 dB/decade to |H21| = 1, where |H21|  = ∂Ids /∂Igs, with Igs the gate– source current) and f$_{max}$= 70GHz for the power gain U=1/2. (From Ref. [221]). (c) State-of-the-art graphene high-frequency electronics (stars) compared with high electron mobility transistors (From Ref. [242]). Red stars: C-face epigraphene FET [221]. Vertical bar indicates record f$_T$ in transferred graphene [244].

Figure 29: Spintronics. (a) Resistance versus magnetic field at 1.4 K presenting a large resistance change ΔR=1.7 MΩ when the magnetization of the two Co electrodes are parallel or anti parallel. High tunnel resistances are important for efficient spin injection (from Ref. [254]). The arrows indicate the direction of the magnetic sweep; the peak shift

is due to magnetic hysteresis of the Co contacts. (b) Scanning electron micrograph of the two-terminal lateral spin valve. The two $Al_2O_3$/Co electrodes (red) are deposited on a 10μm wide epigraphene channel grown on the C-face.

Figure 30: Wafer scale graphene transistor integration. (a) Integration of 10,000 sidewall epigraphene transistors on a 4x6 mm SiC chip, with $Al_2O_3$ gate dielectric. (From Ref. [135]). (b) 100 mm wafers scale FETs utilizing hydrogenated Si-face graphene and $HfO_2$ gate dielectric (From Ref. [68]).

Figure 31: Combined graphene - SiC field effect transistors. (left panels) Semiconductor SiC forms the transistor channel that is provided with graphene ohmic source and drain contacts on the Si-face (made of monolayer graphene + buffer, see (b)). Hydrogenated graphene (bilayer graphene, see (c)) produces the FET Schottky-like gate. The device geometry is shown in the electron micrograph of (d) (From Ref. [260]). (right panel) The SiC/SiOx interface forms the transistor channel on the C-face, as sketched in the inset. Contacts are provided by multilayer epigraphene. Inset (bottom) AFM image of the device, the channel is in between the bright graphene pads. Current on-off ratios of more than $10^6$ are demonstrated at room temperature.(From Ref. [261])

Figure 32: (a) concept of Si-SiC wafer bonding. The top Si wafer supports the CMOS platform and the SiC wafer the epigraphene electronic devices. The devices on top and bottom wafers are connected by metal vias through the thin Si wafer. (b) Realization of the bonding after SiC graphitization with an intermediate $Al_2O_3$ bonding layer. (From Ref. [265])

Figure 33: Ballistic transport in sidewall ribbons (From Ref. [136]) (a) Resistance R versus probe spacing d. The tungsten probes are positioned on a single ribbon in situ in an electron microscope. The different traces correspond to ribbons after various cleansing by in-situ annealing. Linear fits extrapolate to R(d=0)= $h/e^2$=25.8 kΩ,. Slope ΔR/ΔL gives mean free path λ=4.2, 28, 16, 58 μm (top to bottom). Cyan line is consistent with zero slope. (b) Conductance versus probe spacing with the same set up as in (a) for an extended probe spacing range, showing the decay of one conducting channel after d=250nm, and the second at d=17μm. (c) Temperature dependence of the conductance G of a top-gated ribbon with fixed contacts (39nm x 1.6μm). Note the conductance G=0.9 $h/e^2$ at charge neutrality, and the weak temperature and gate voltage $V_g$ dependences. Inset: AFM and EFM images of the ribbon grown on a curved SiC natural step. S, D and G refer to the Source, Drain and Gate area respectively. (d) All the G vs $V_g$ curves are parallel and collapse on a single curve indicating that only the resistance at $V_g$=0 is temperature dependent.

Figure 34: Pump Probe Spectroscopy. (a) Band structure and carrier distribution after optical excitation. The pump photons (red) excite carriers that rapidly thermalize establishing a new hot thermal distribution that slowly cools down (From Ref. [203]. (b) After optical excitation an optical probe probes the carrier distribution. (left) For neutral (gapless) graphene, all transitions are allowed. (right) For doped graphene, only excitations larger than twice the Fermi level can create an electron-hole pair (Fermi Blockade).

Figure 35: Plasmons launched and imaged with SNOM. (a) Schematics of the experimental configuration). An infrared laser light illuminates a metallized AFM tip (yellow). (b) The near-field amplitude image (colour scale) is acquired for a tapered epigraphene ribbon (12 μm long). (c) calculated local density of optical states (LDOS) at a distance of 60nm from the graphene surface.[304]

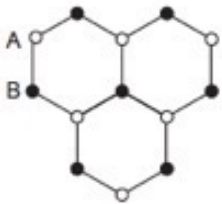
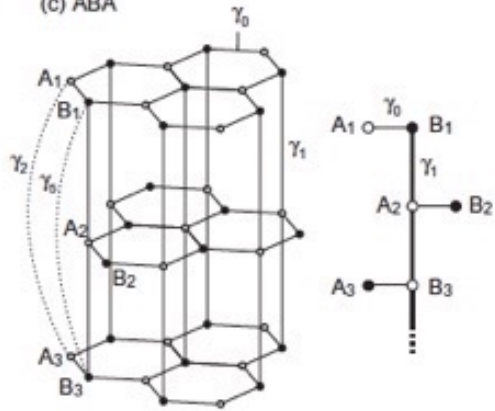
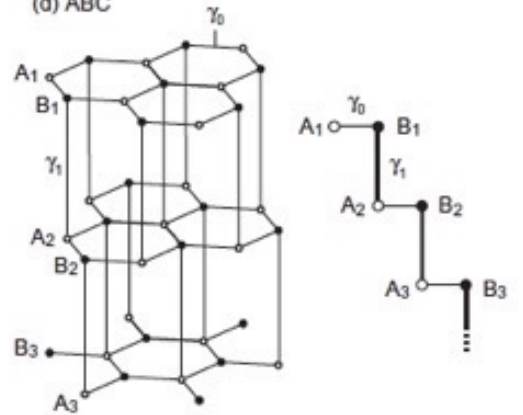
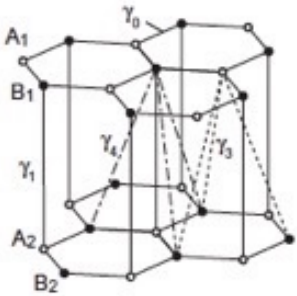
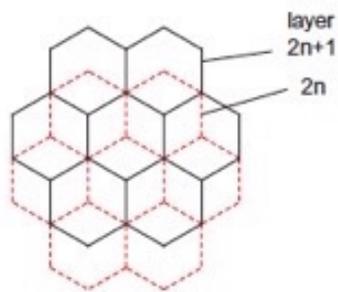
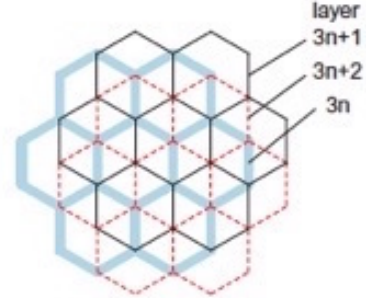

**FIGURE 1**

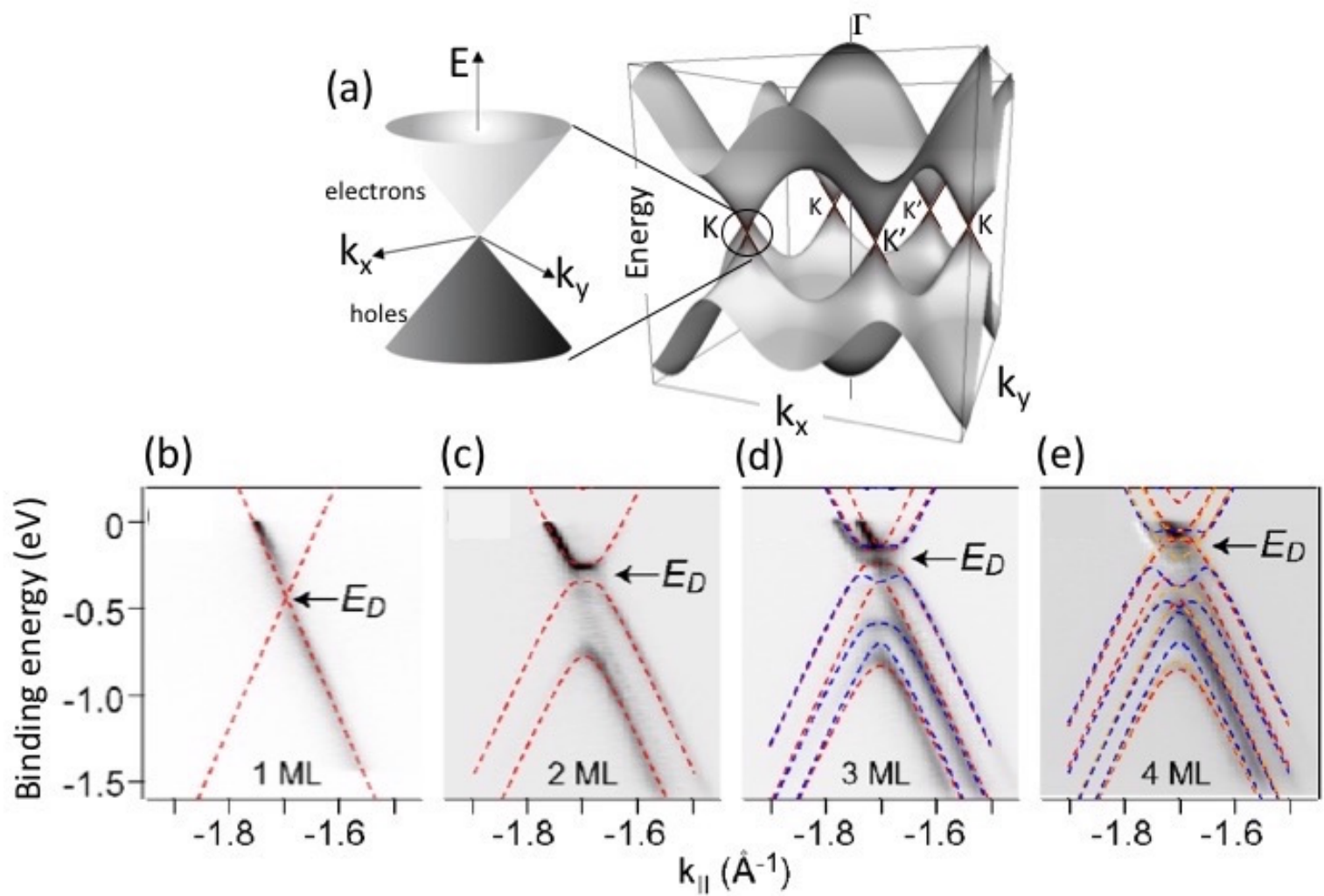

**FIGURE 2**

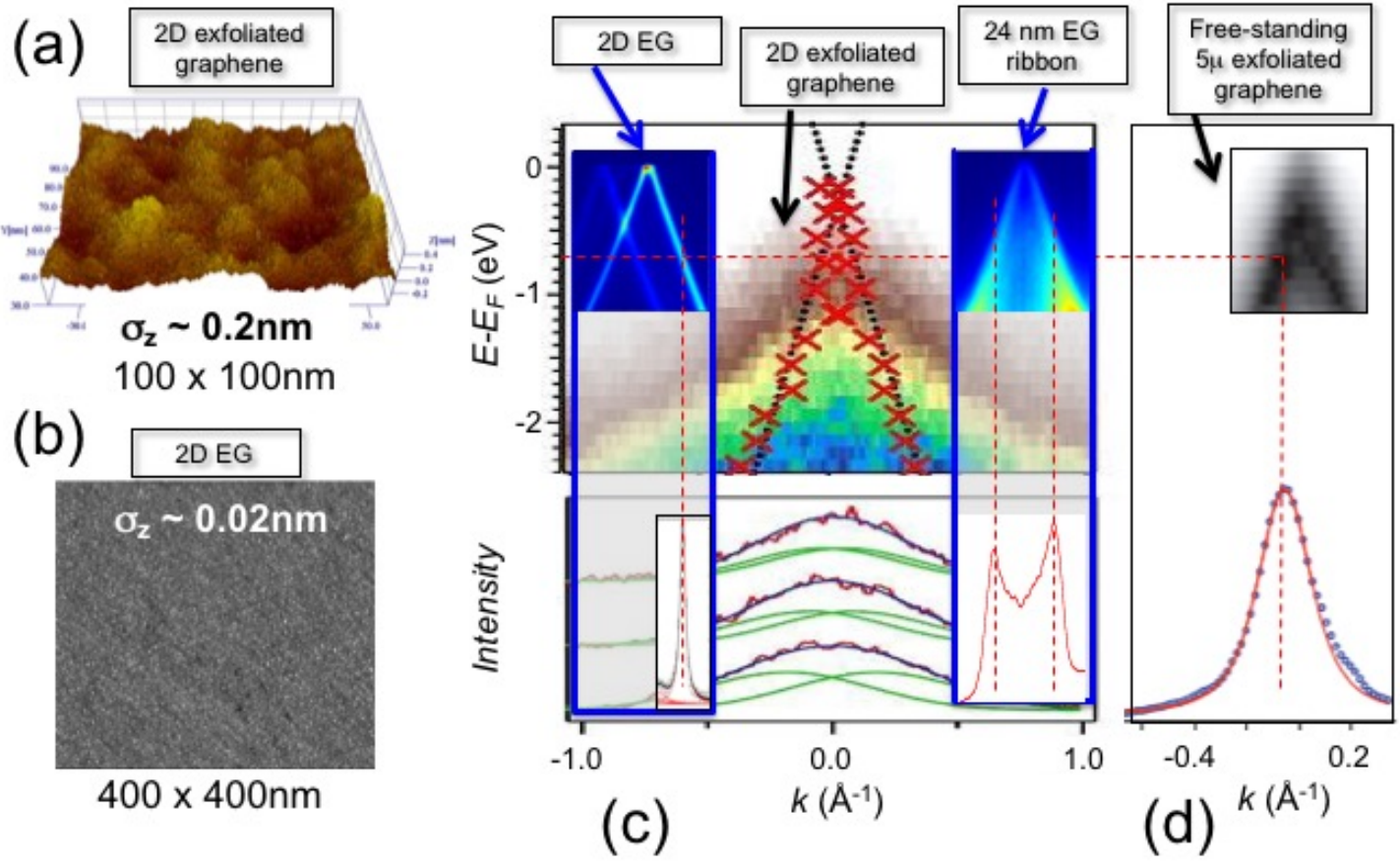

**FIGURE 3**

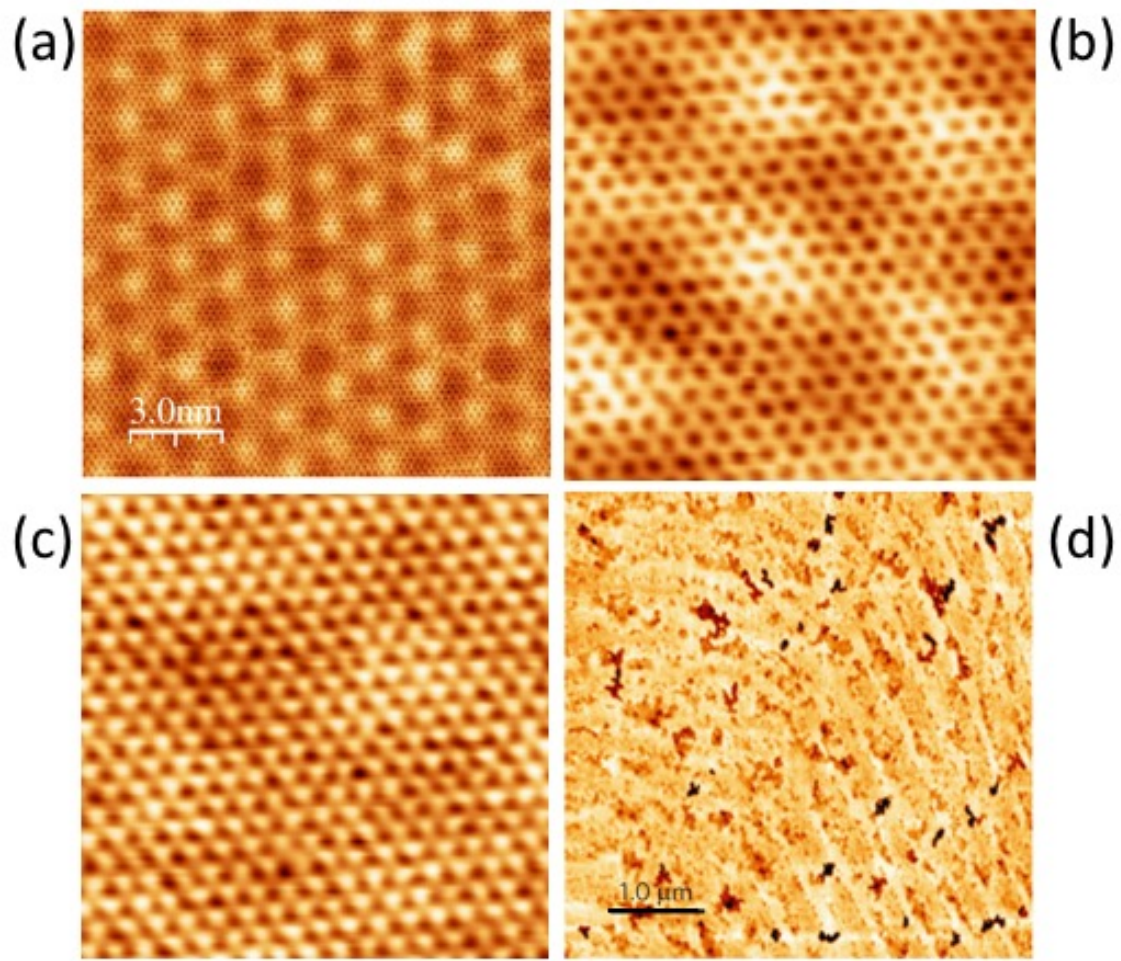

**FIGURE 4**

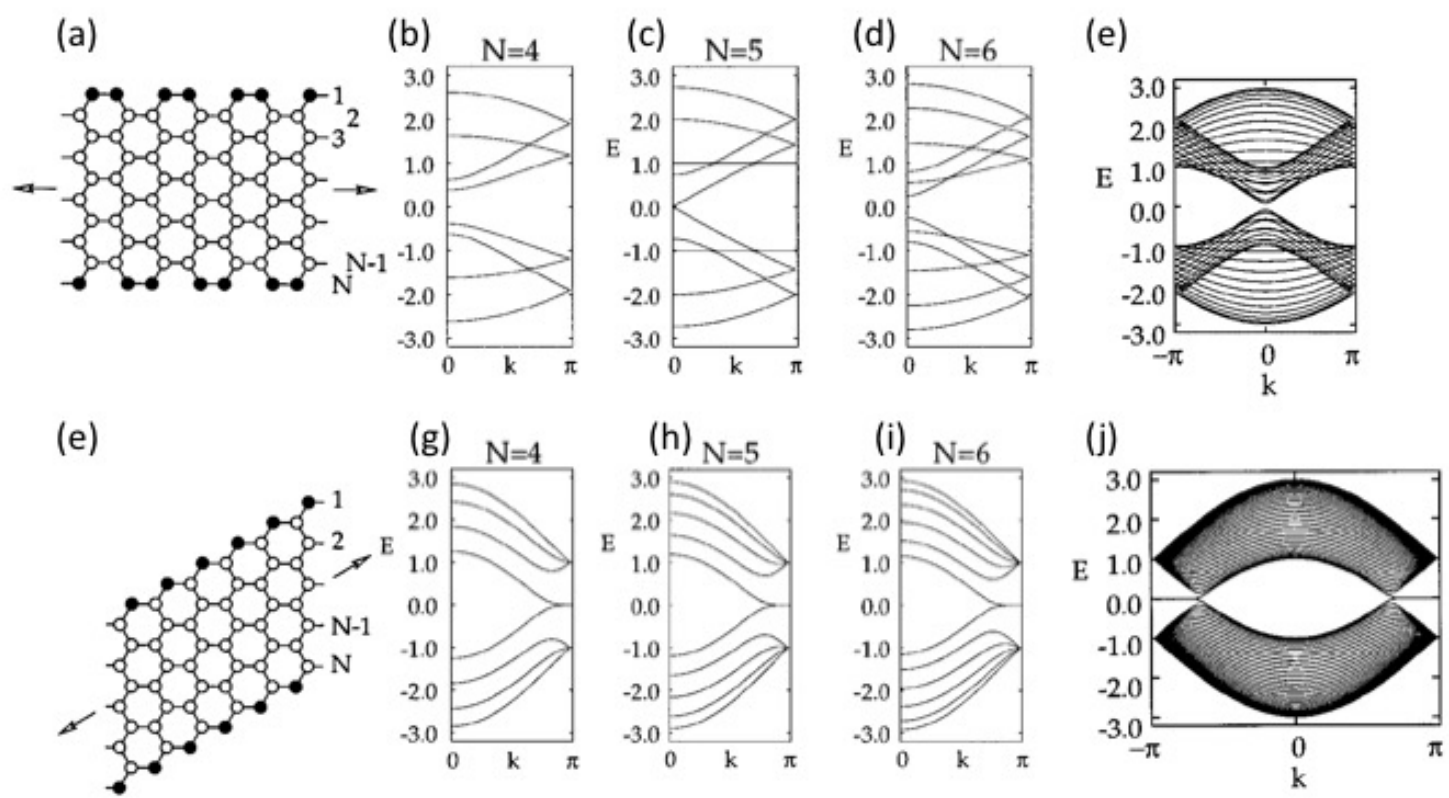

**FIGURE 5**

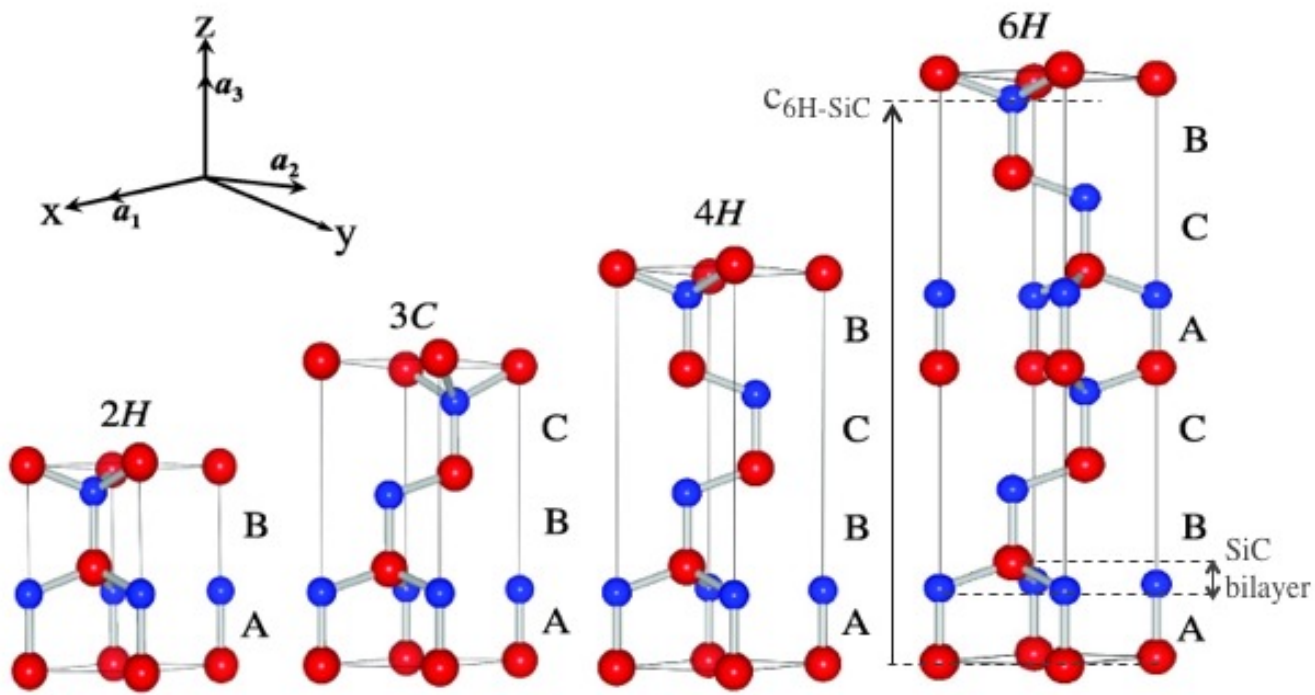

**FIGURE 6**

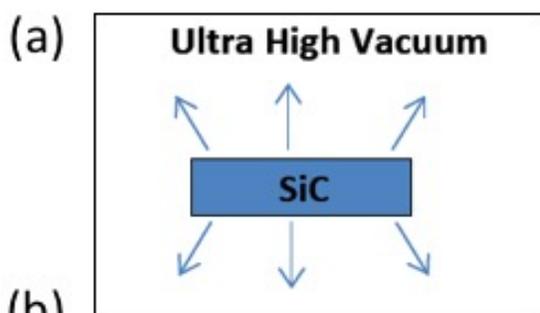
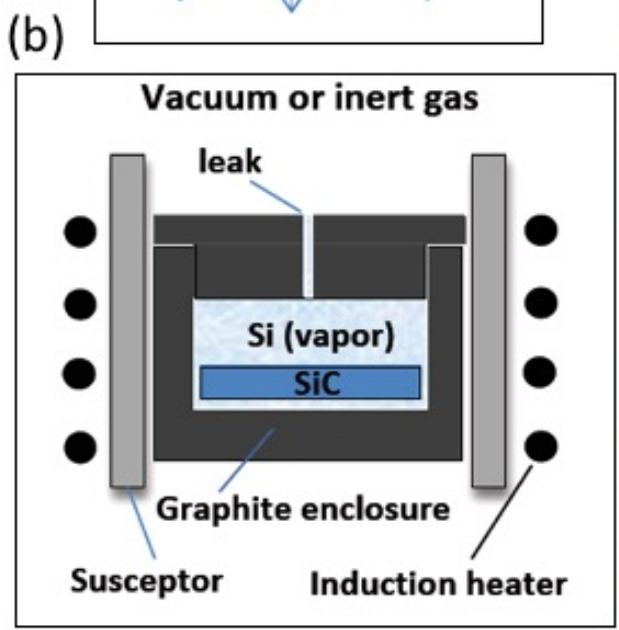
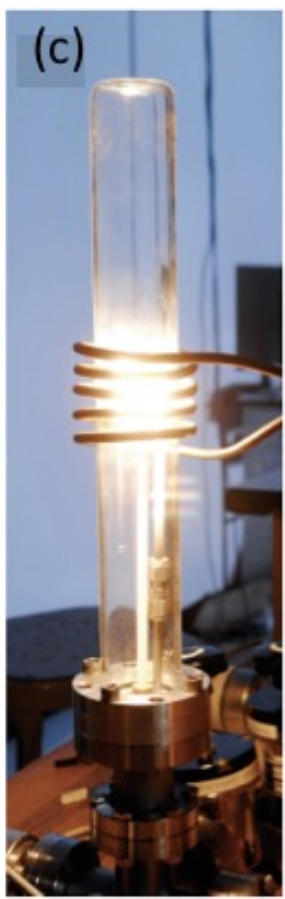
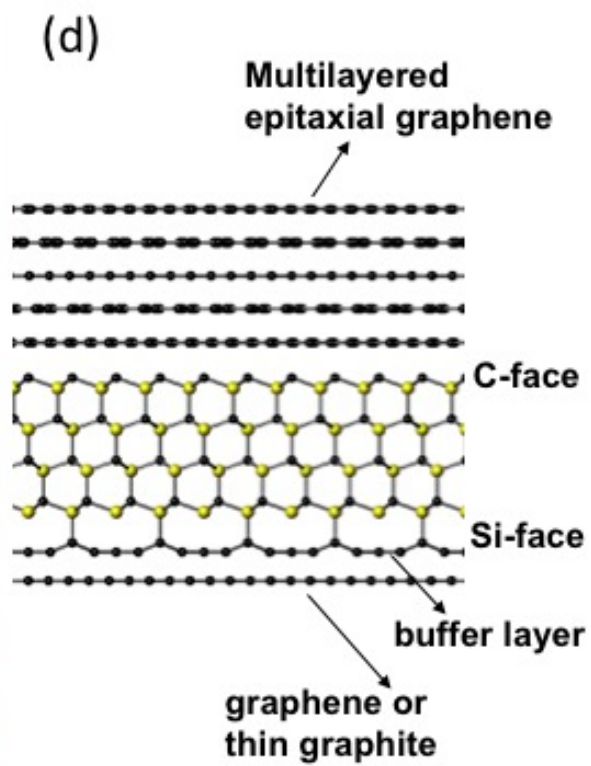

FIGURE 7

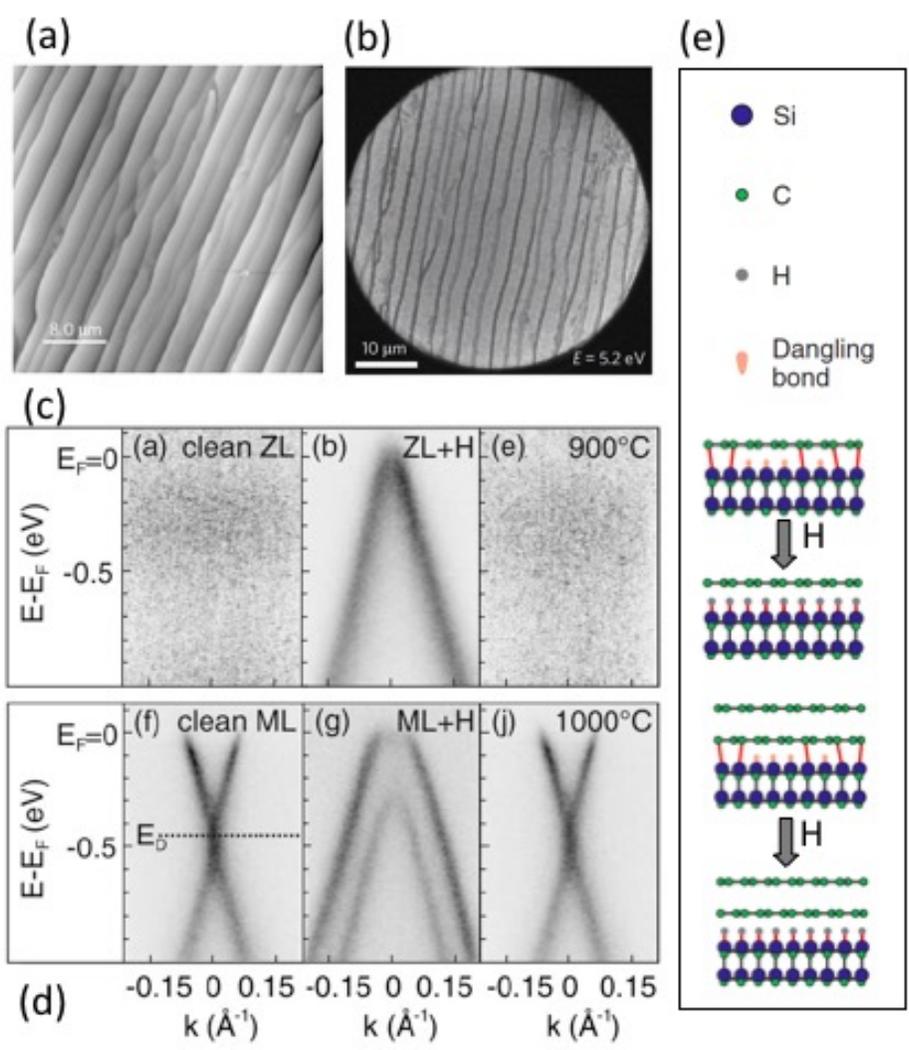

**FIGURE 8**

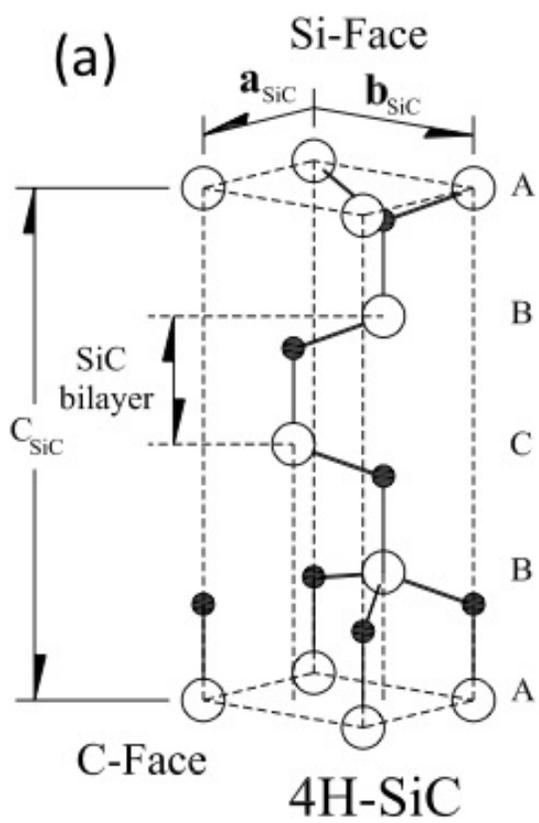
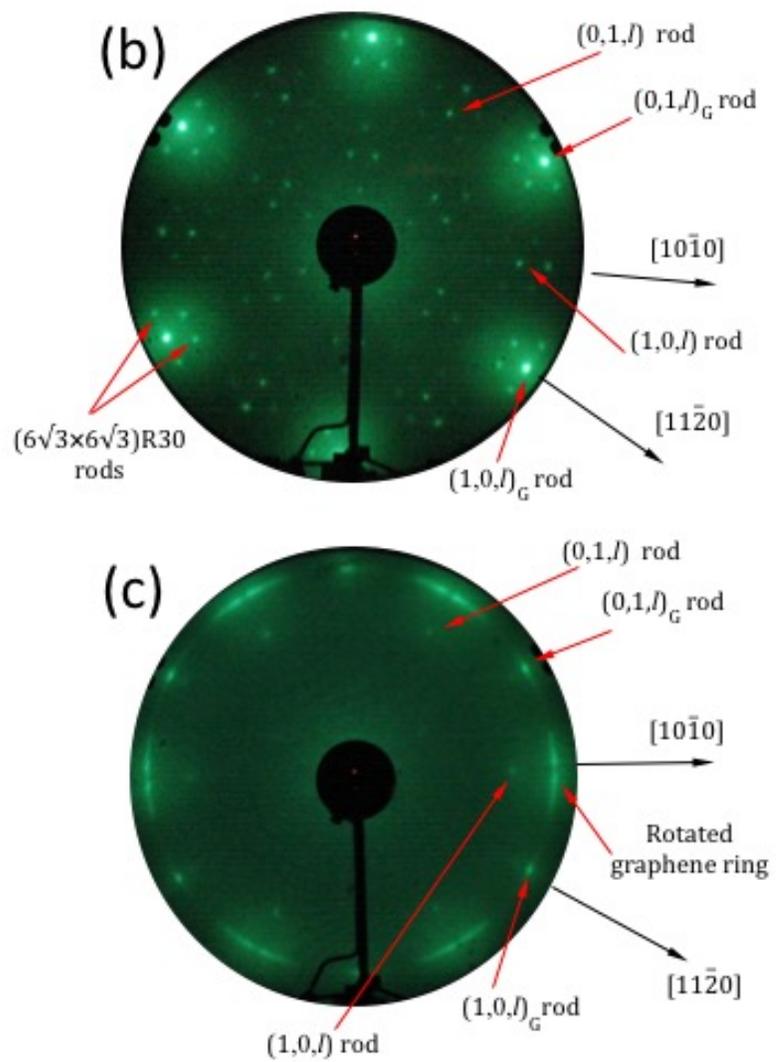

**FIGURE 9**

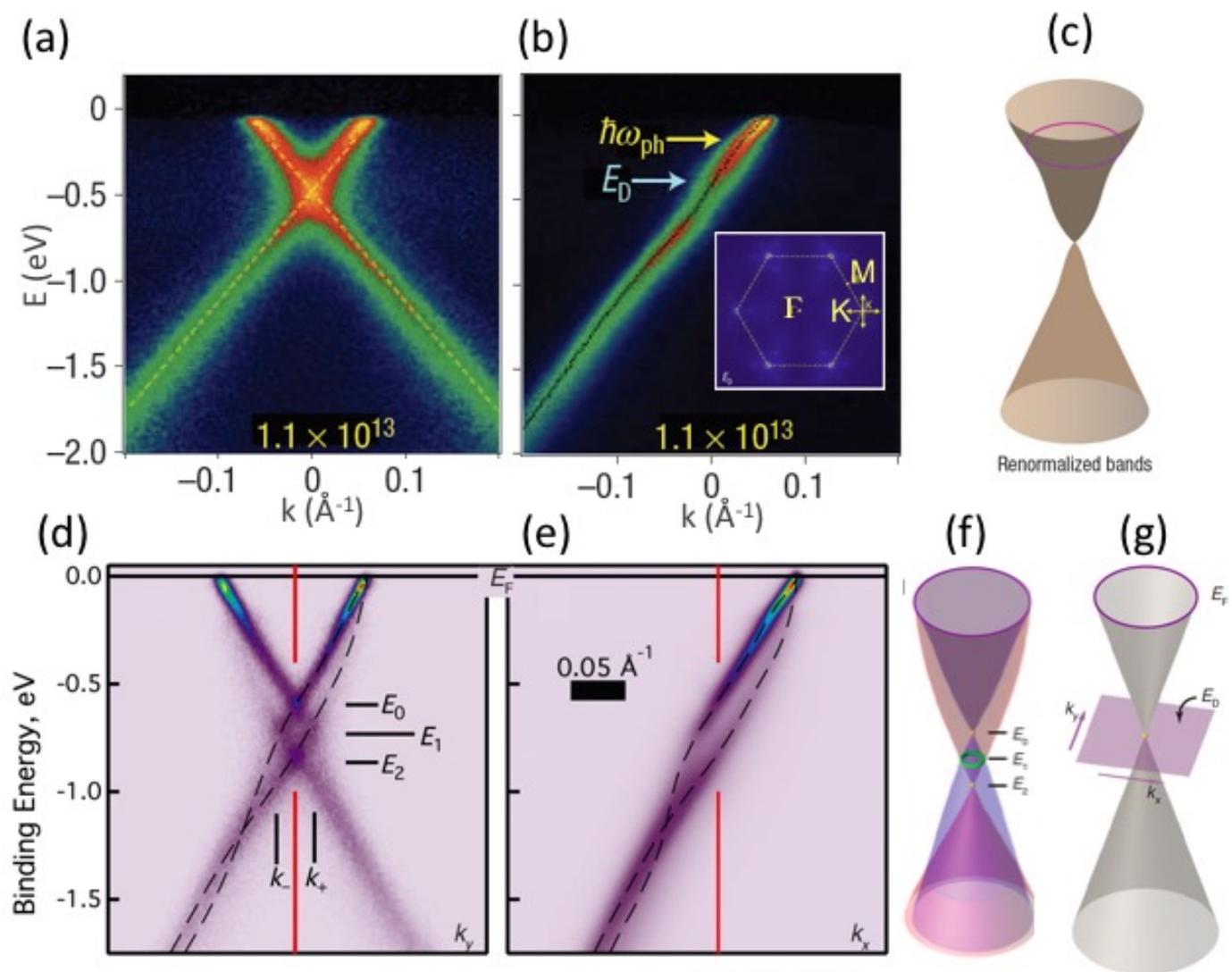

**FIGURE 10**

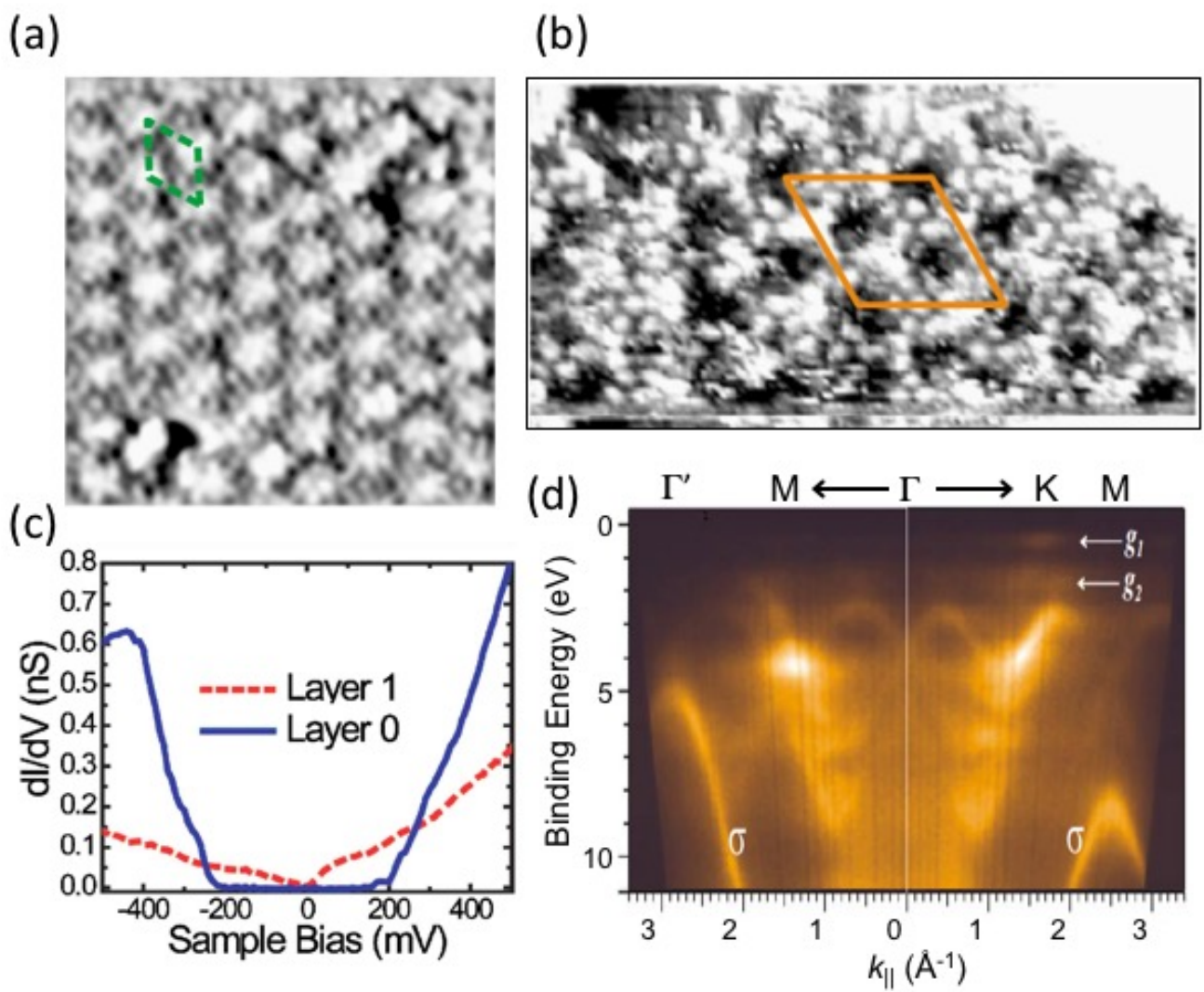

**FIGURE 11**

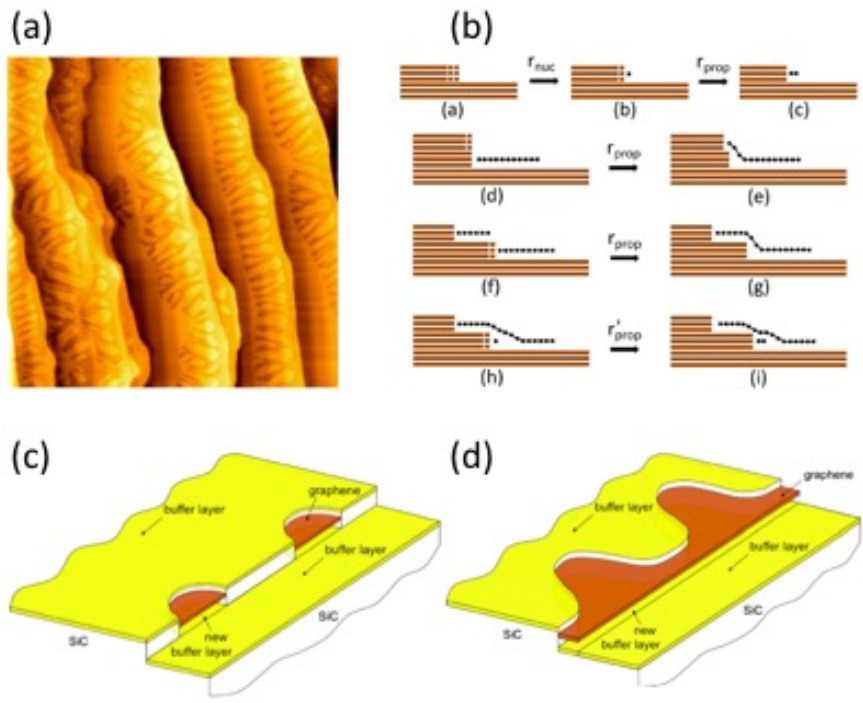

**FIGURE 12**

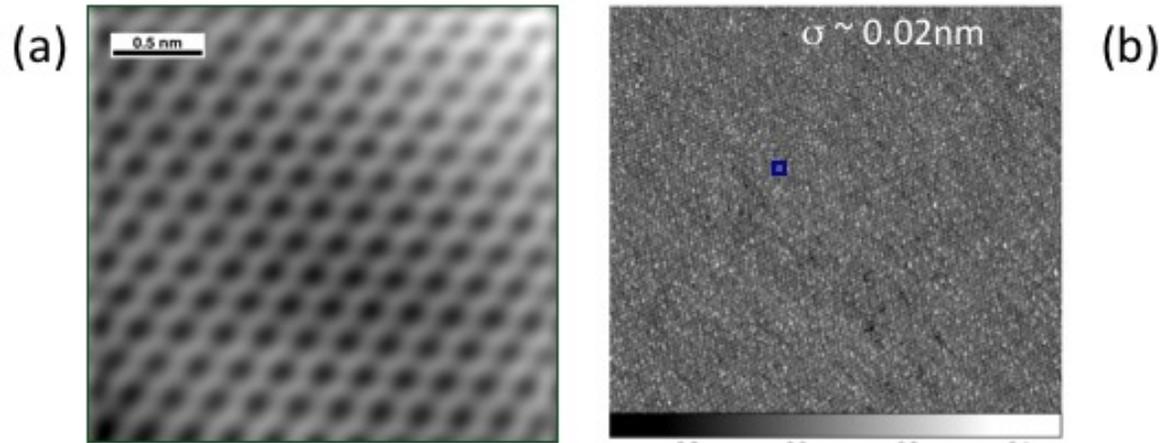
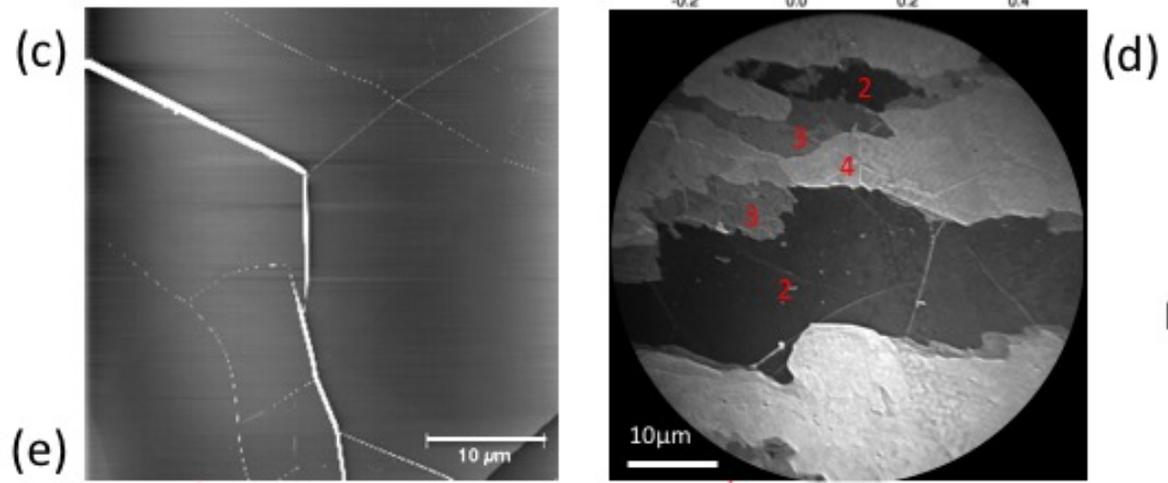
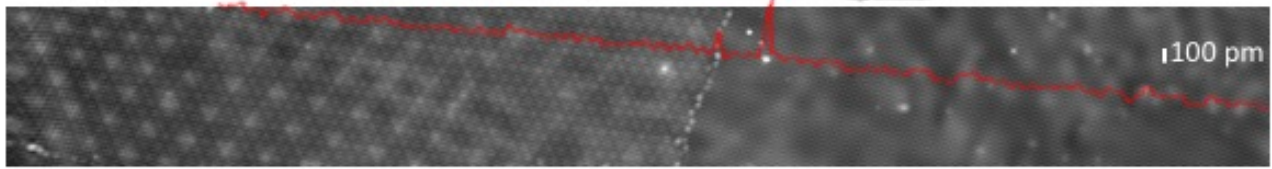

FIGURE 13

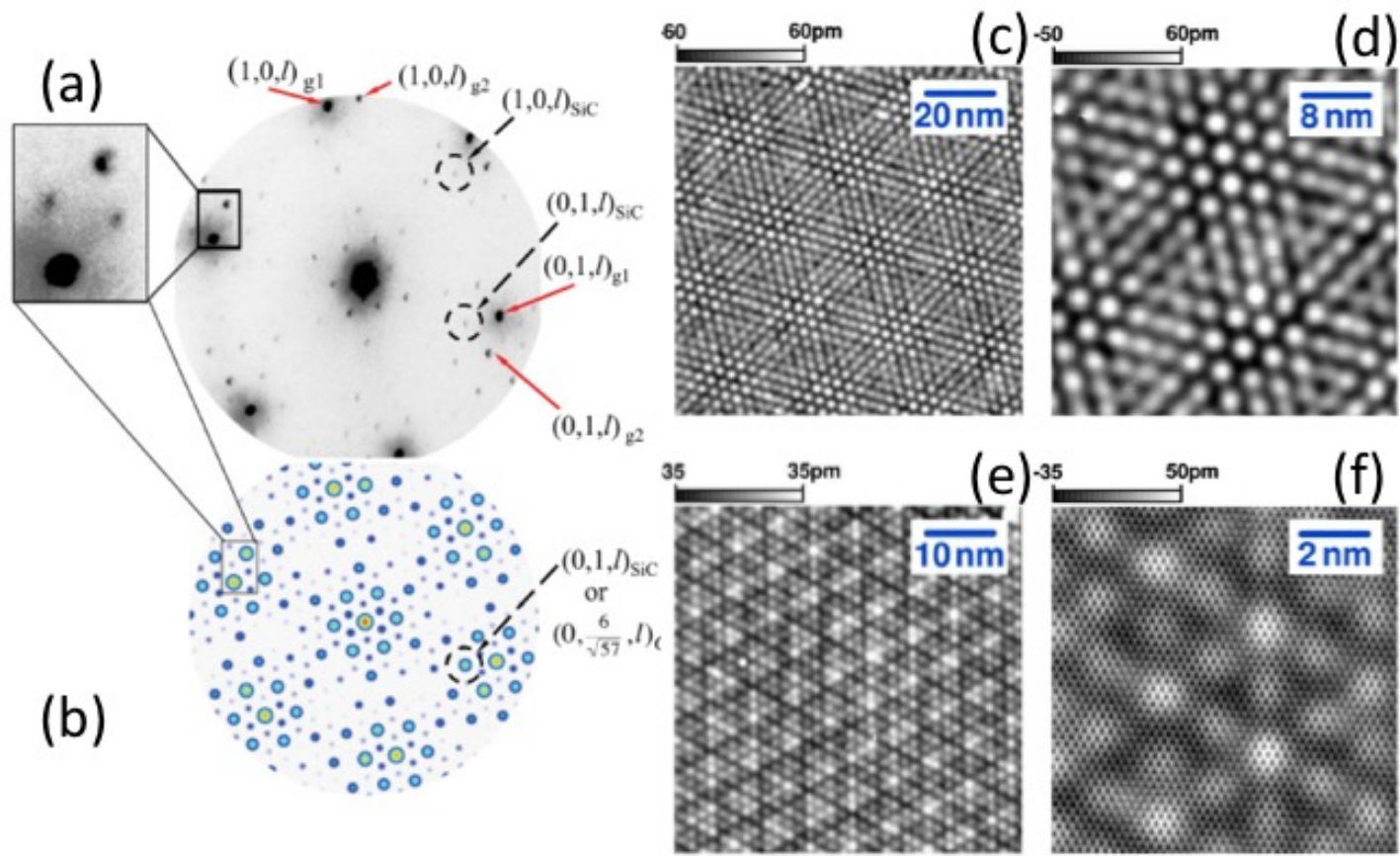

**FIGURE 14**

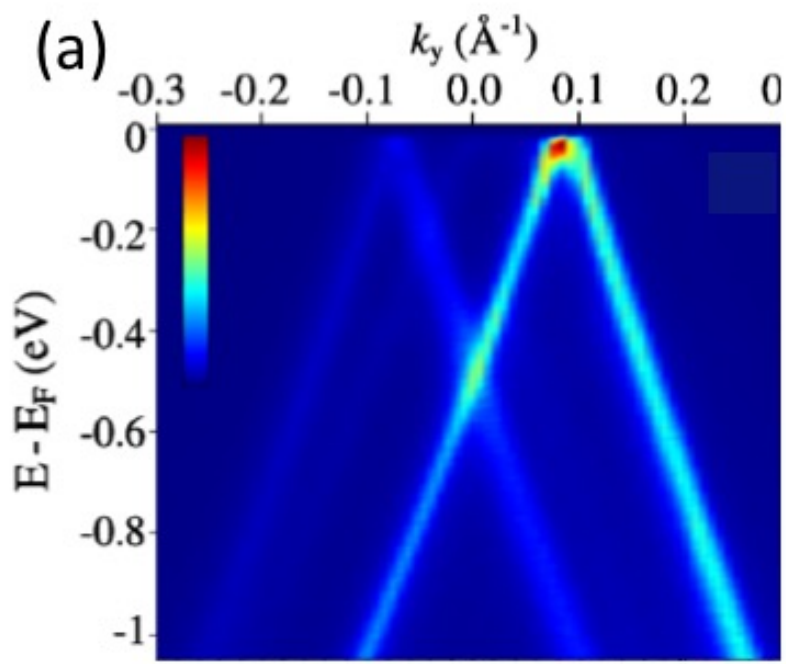 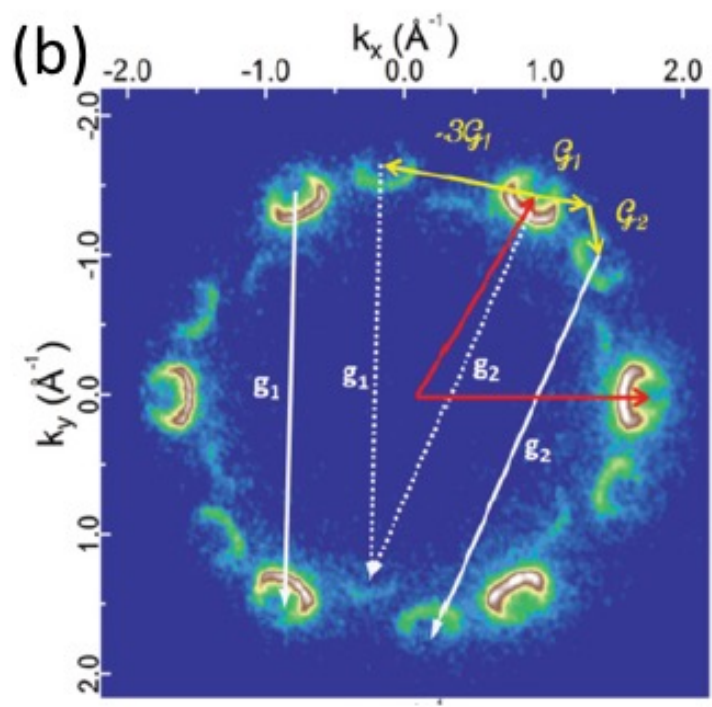

**FIGURE 15**

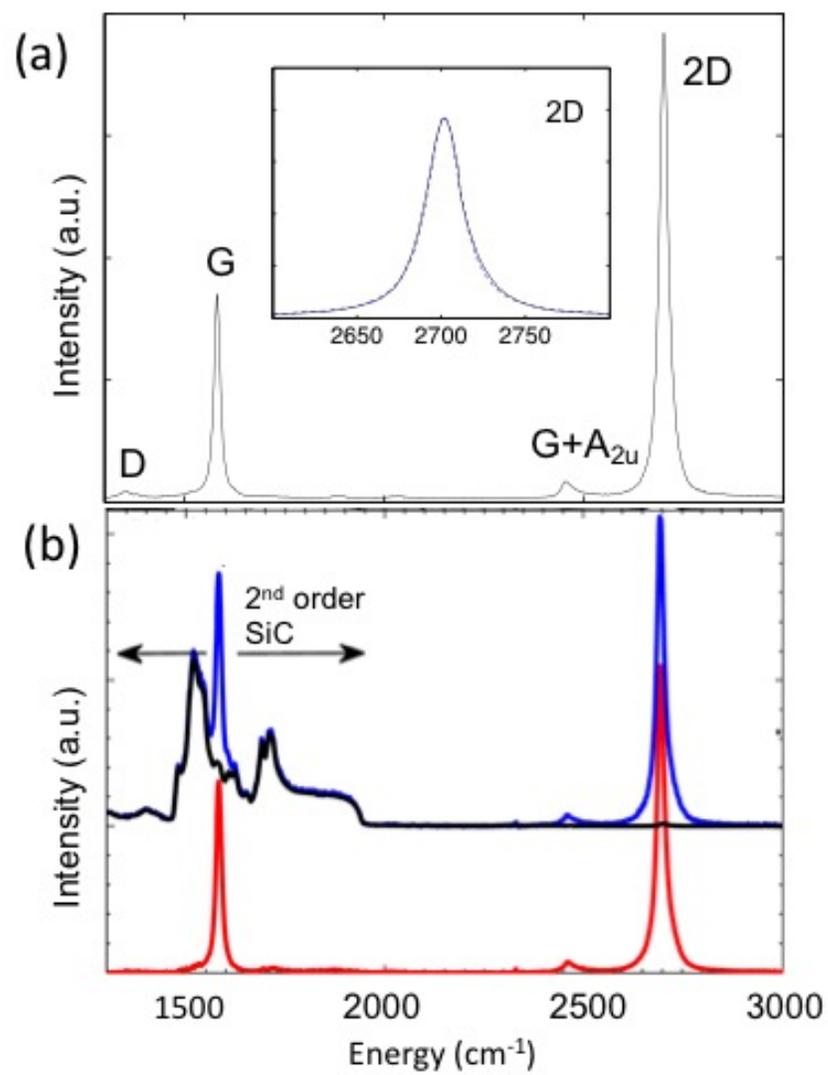

**FIGURE 16**

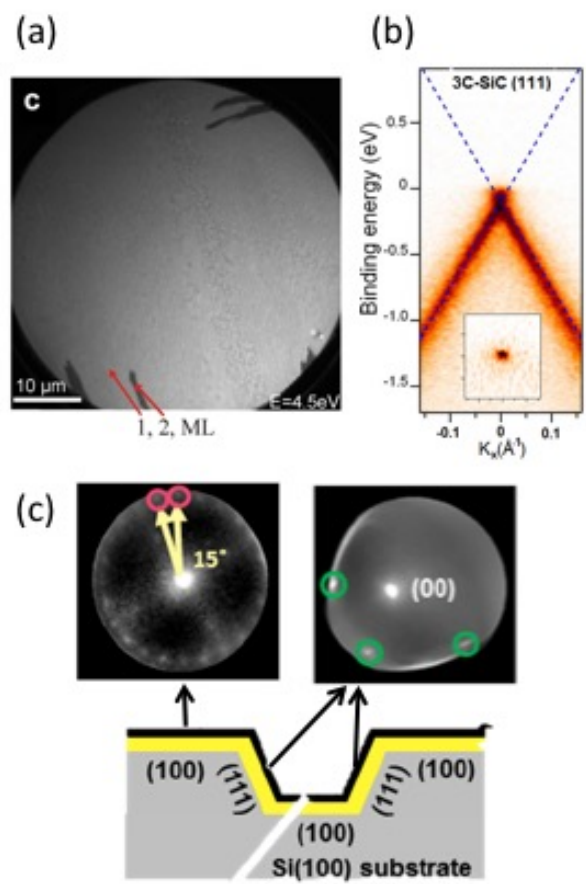

**FIGURE 17**

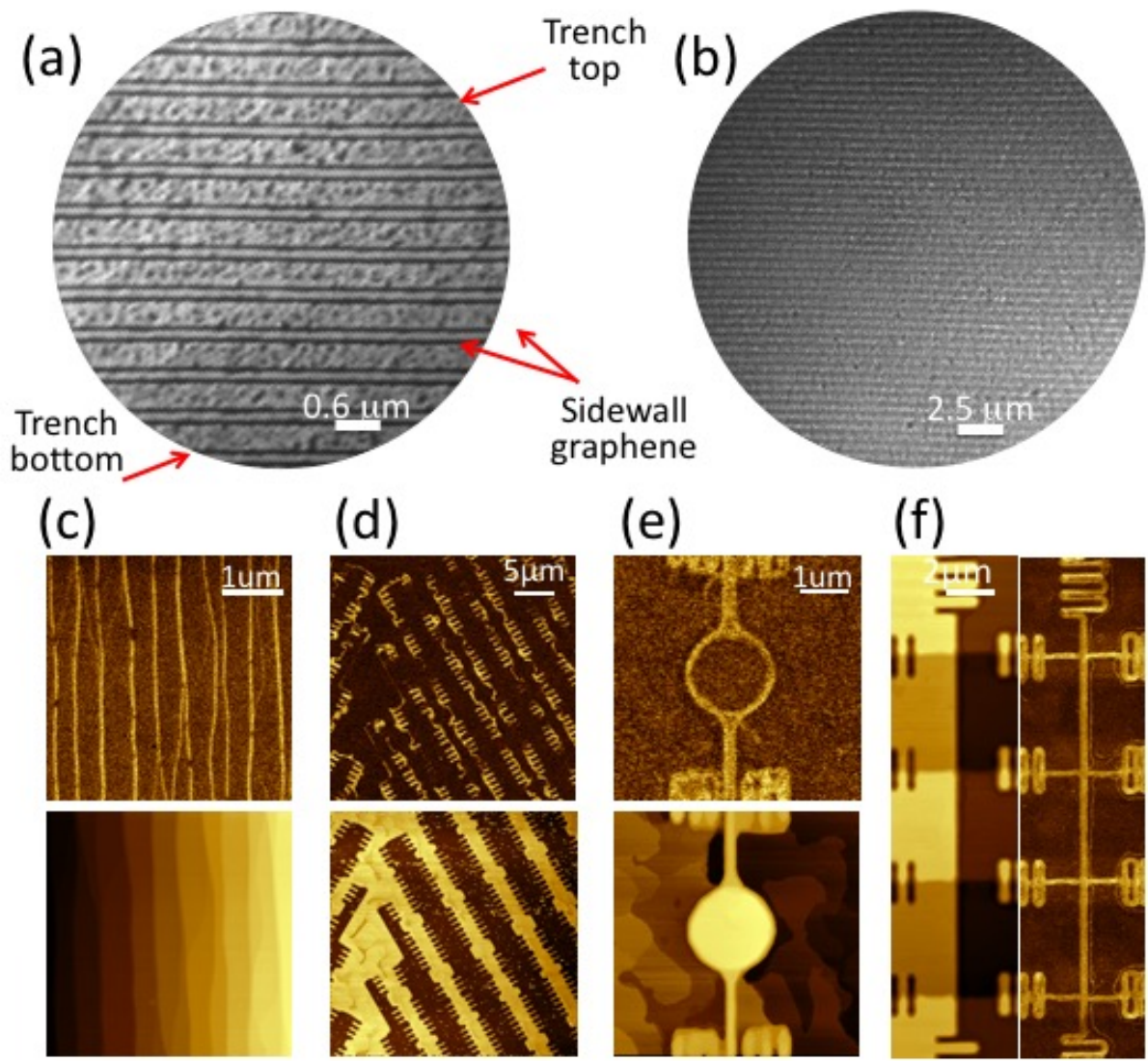

**FIGURE 18**

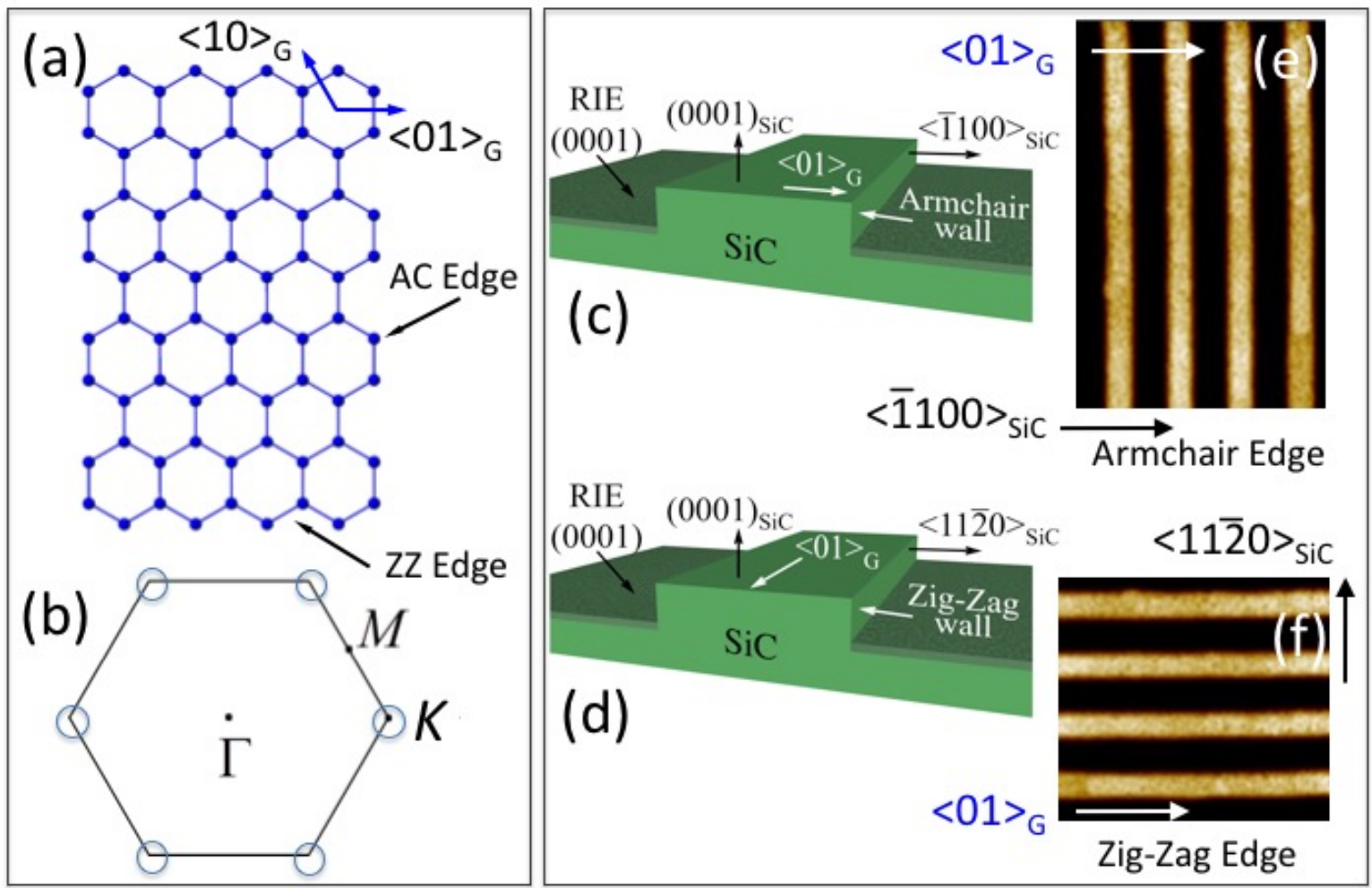

**FIGURE 19**

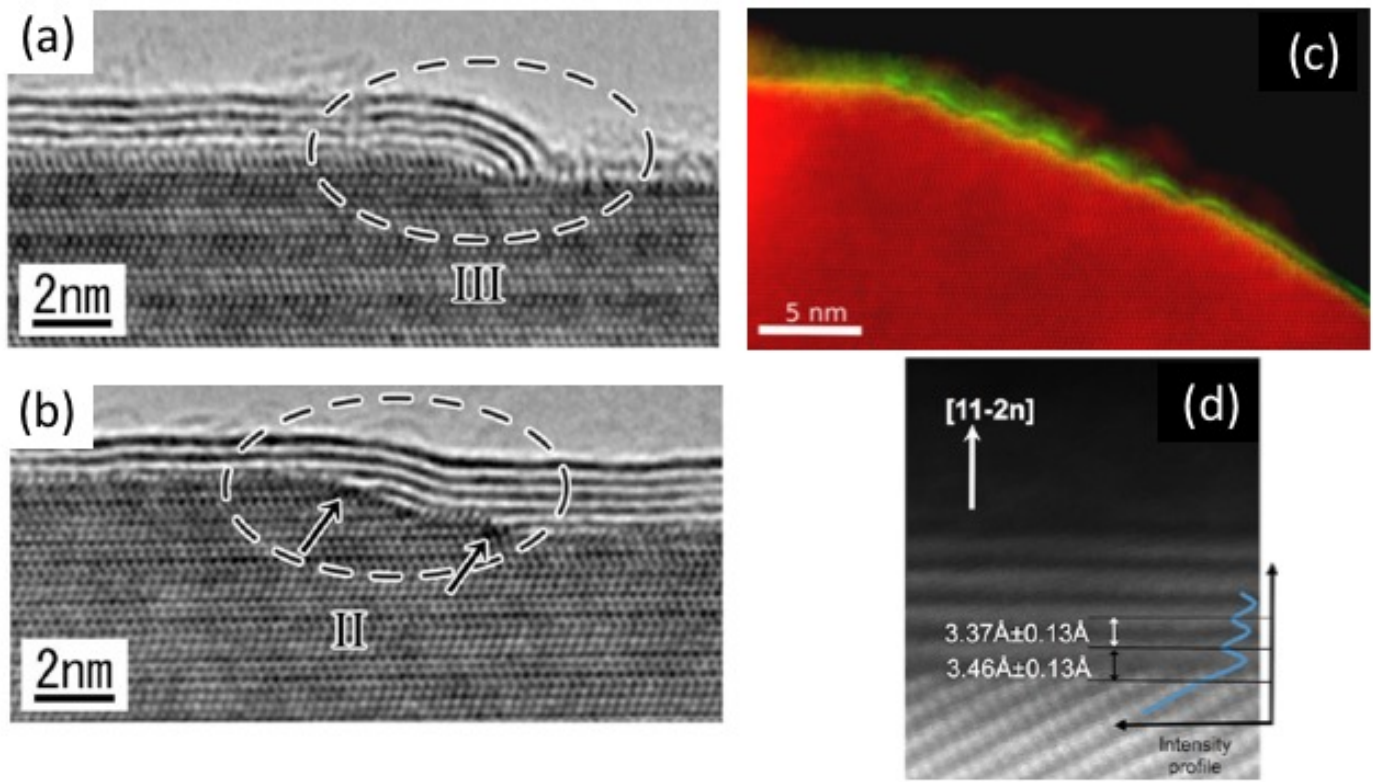

**FIGURE 20**

**FIGURE 21**

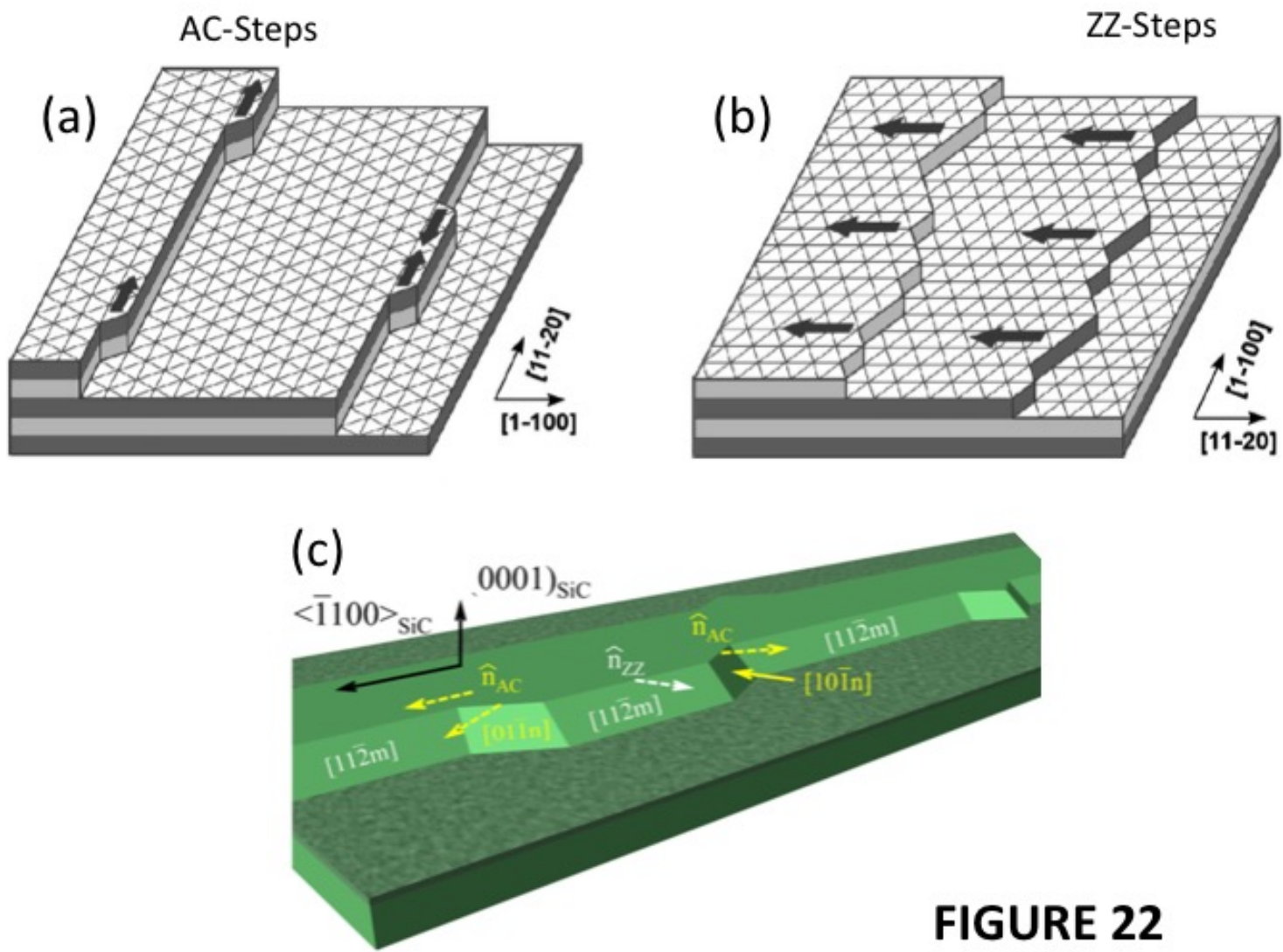

FIGURE 22

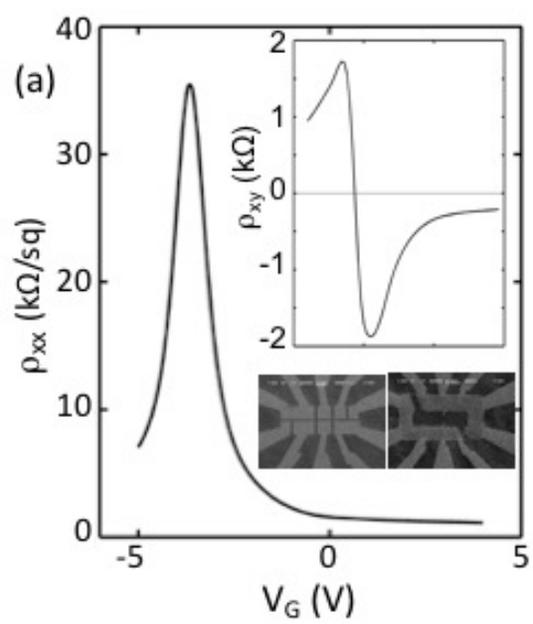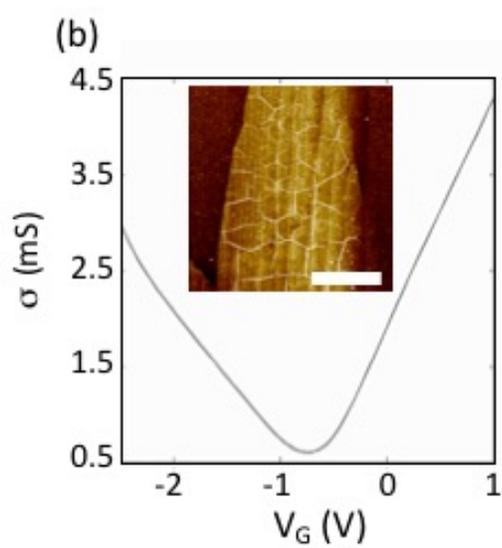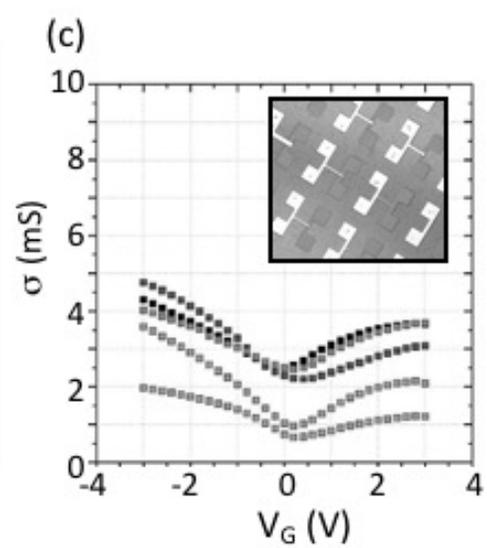

FIGURE 23

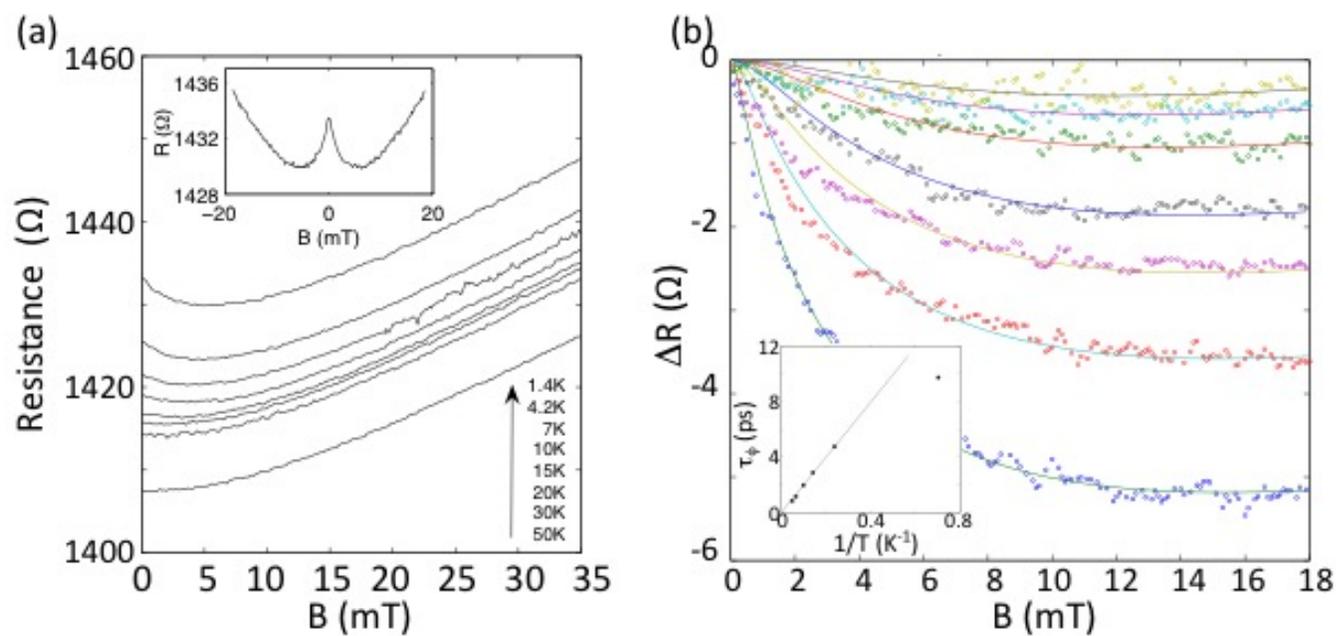

FIGURE 24

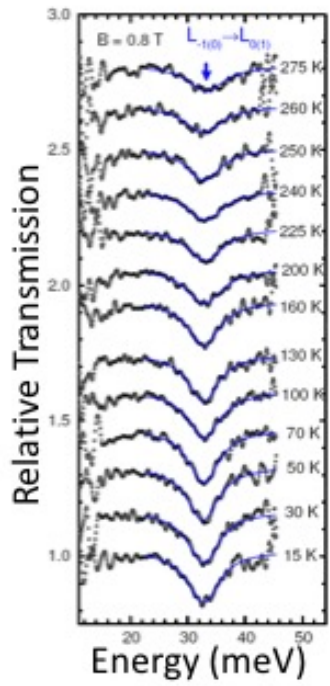 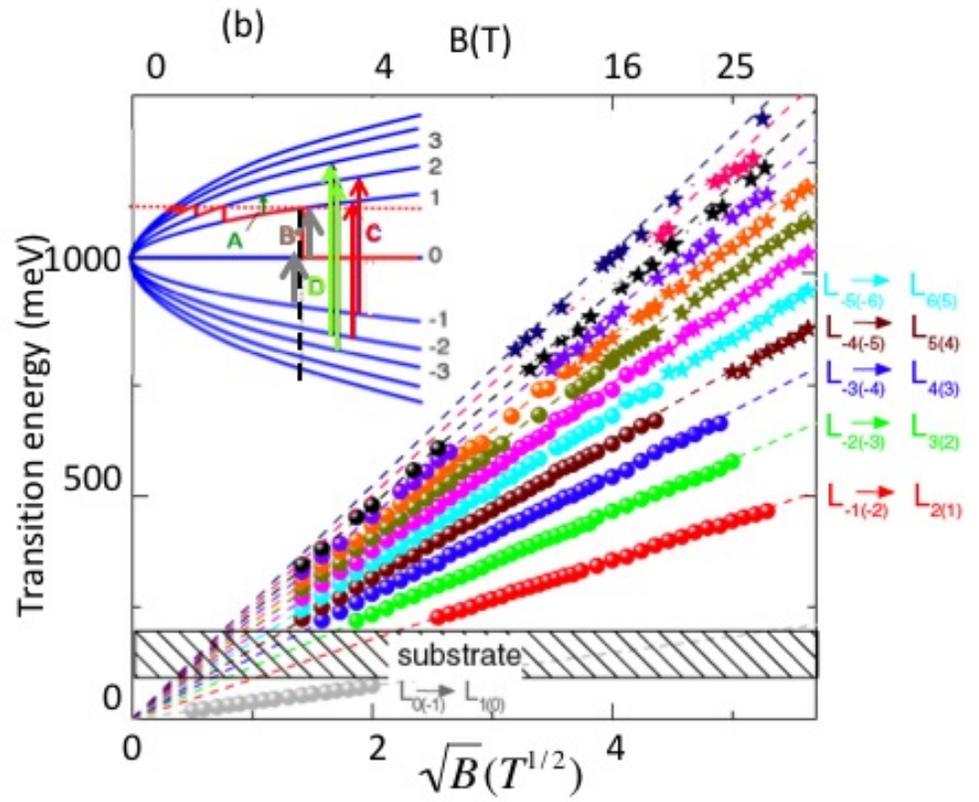

FIGURE 25

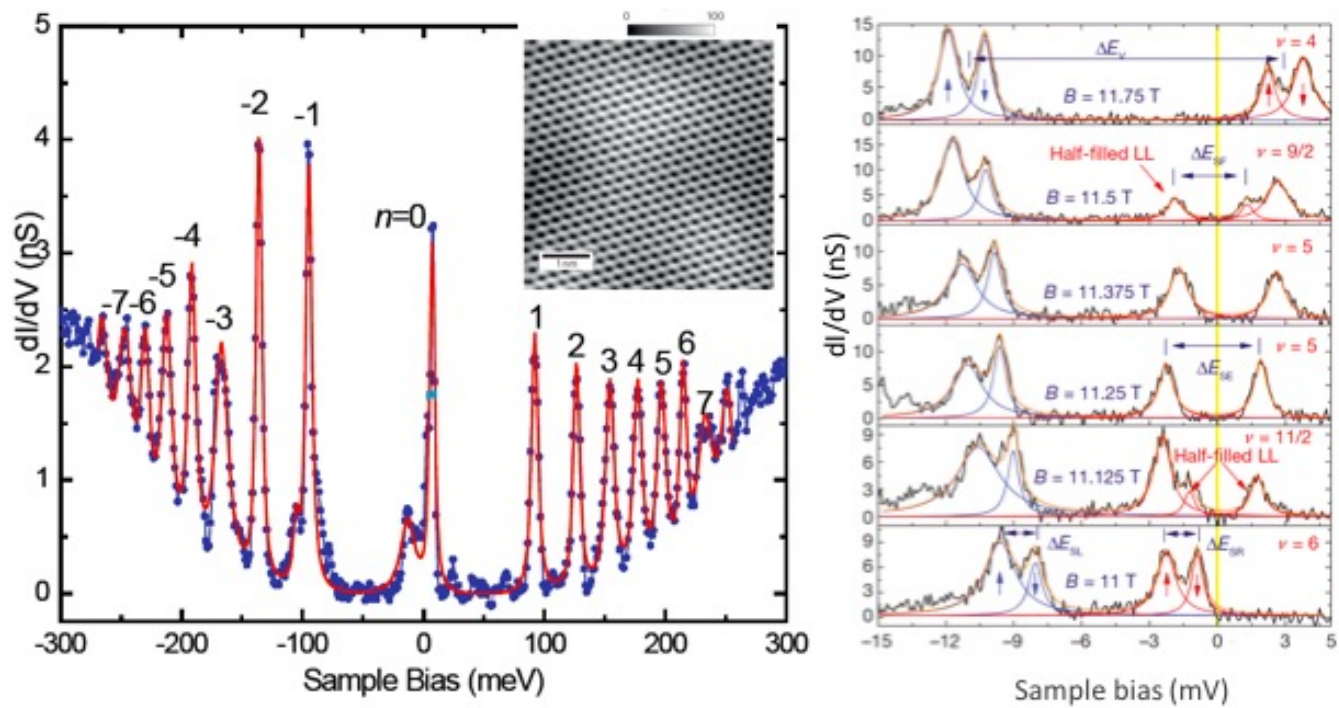

FIGURE 26

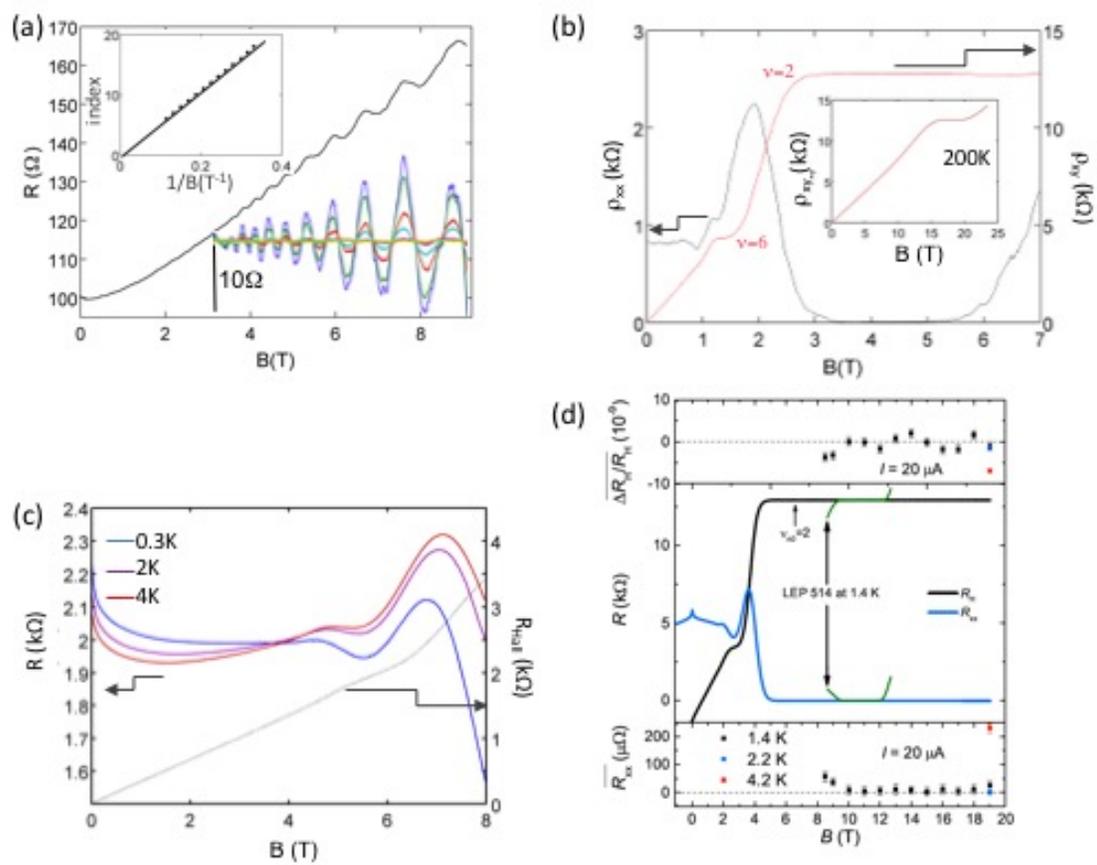

FIGURE 27

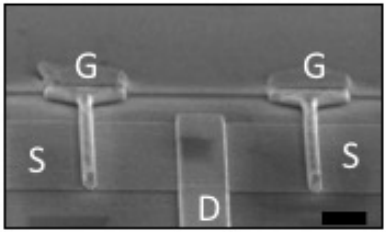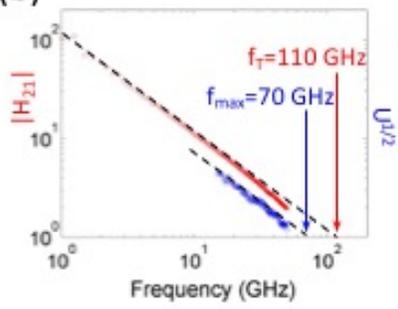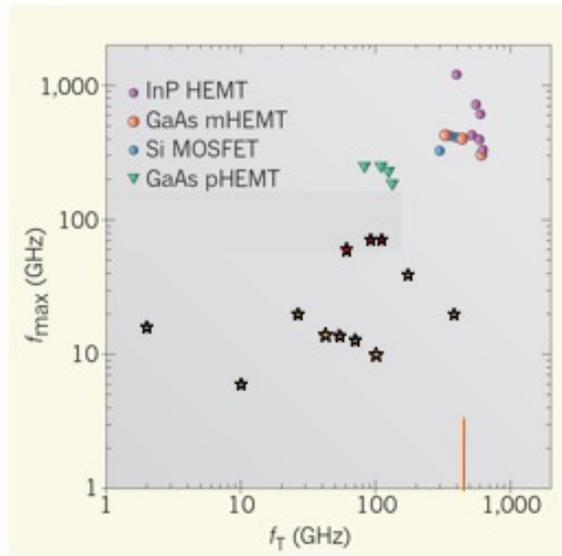

FIGURE 28

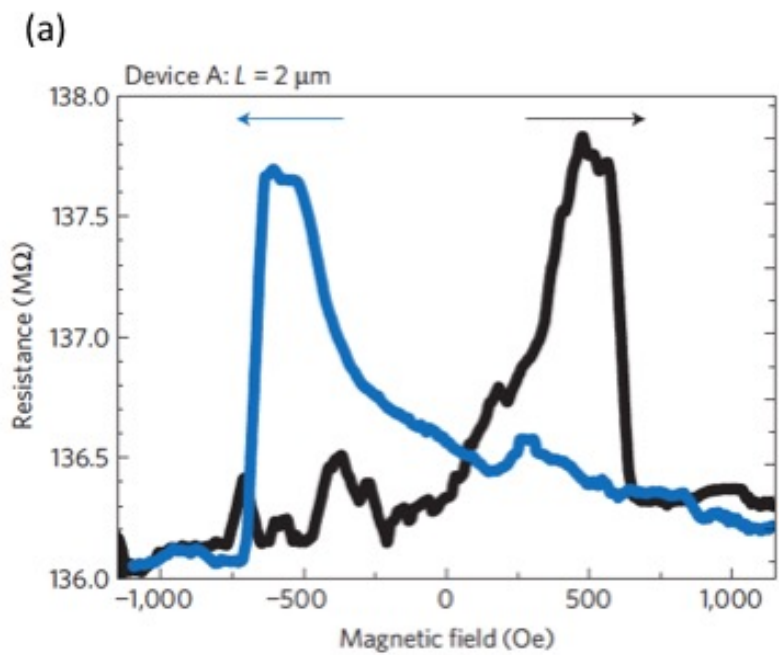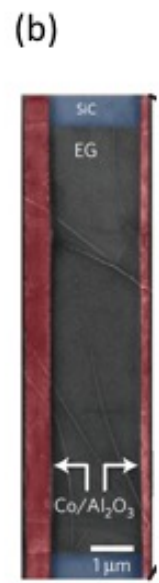

FIGURE 29

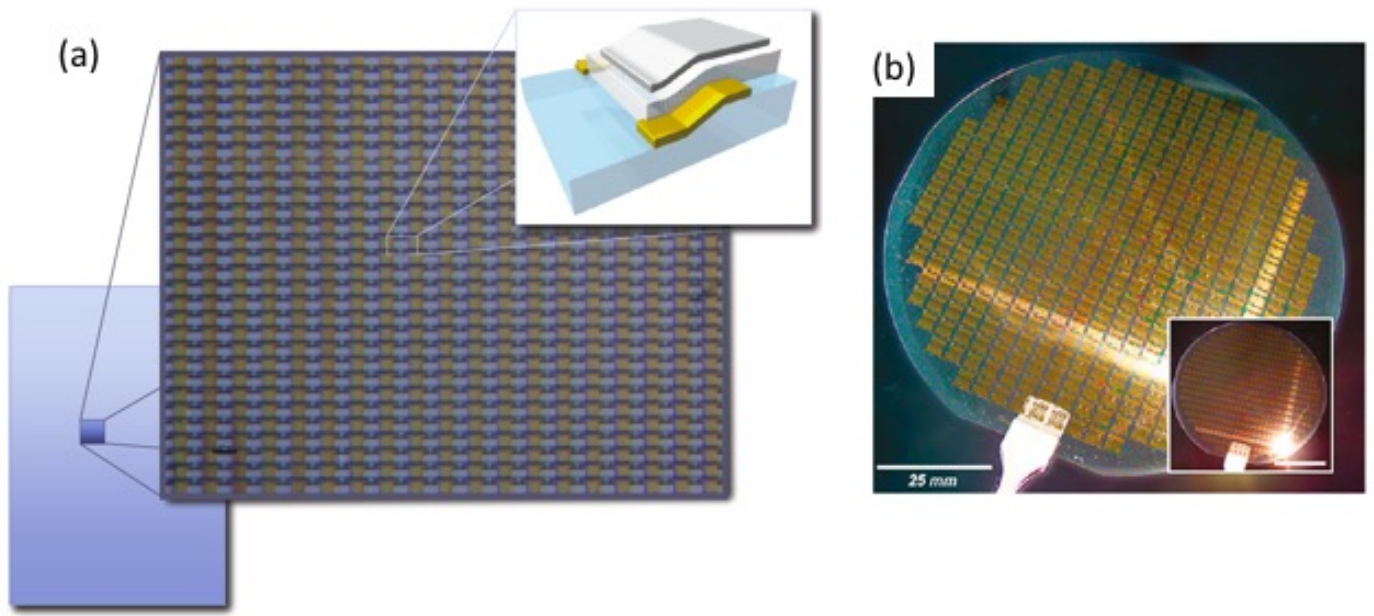

FIGURE 30

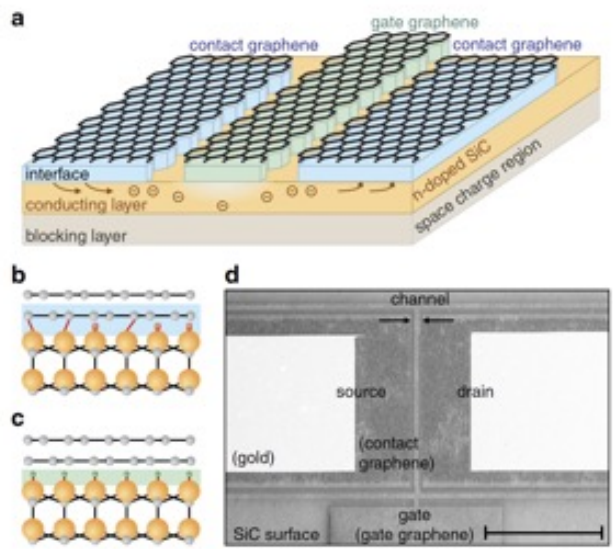
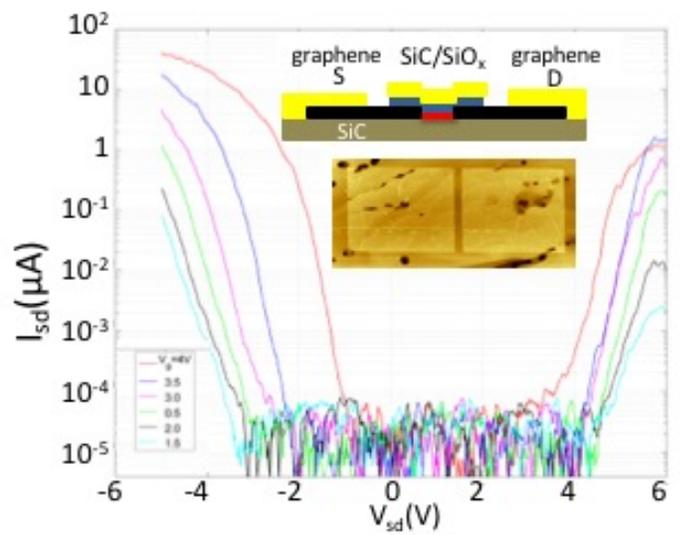

FIGURE 31

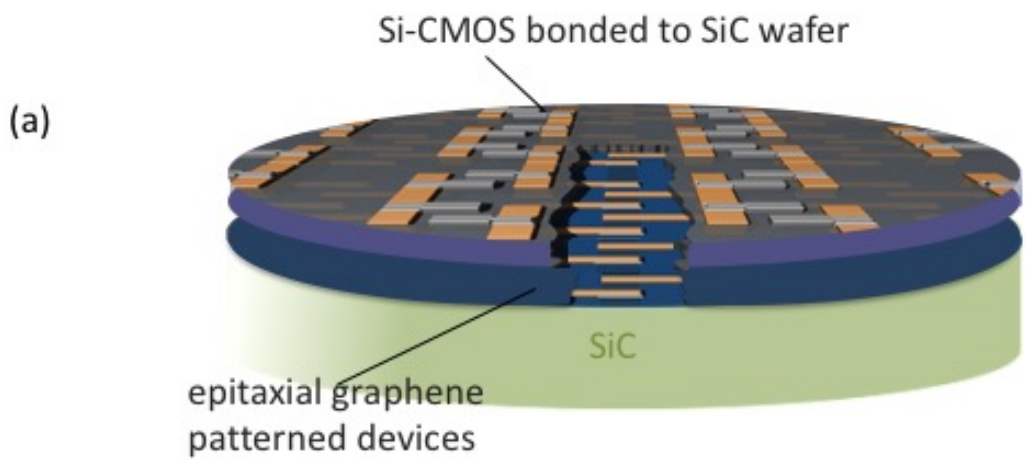

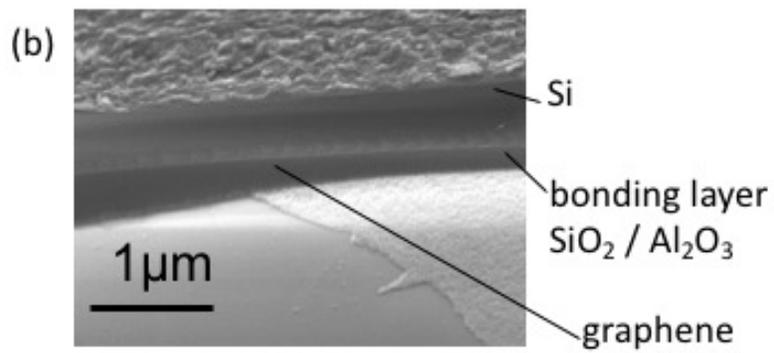

FIGURE 32

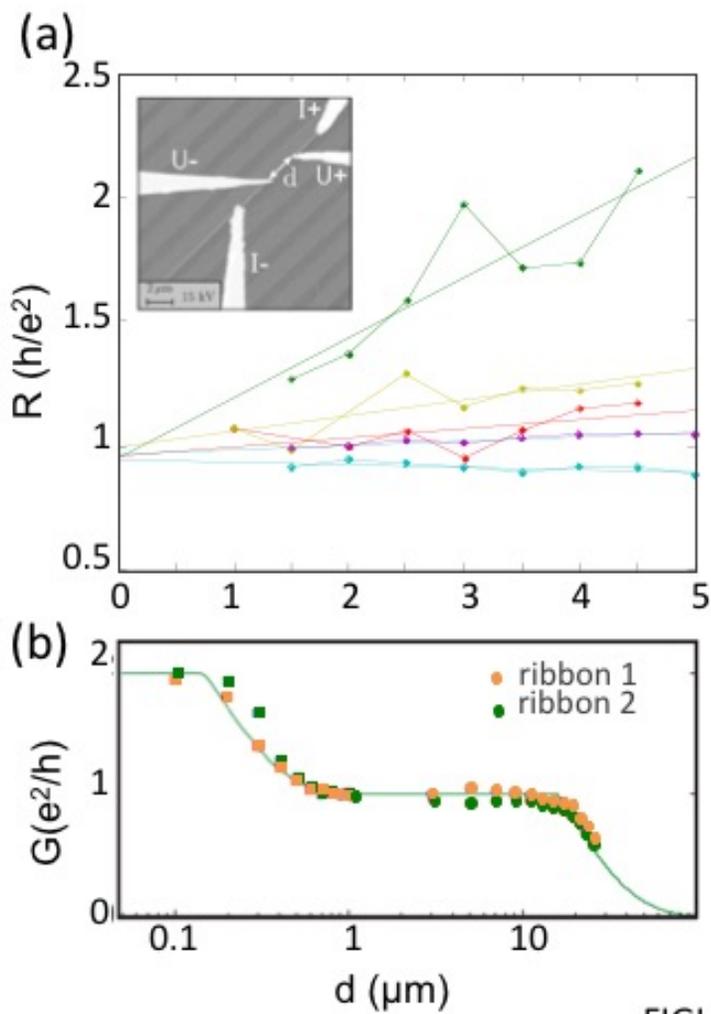
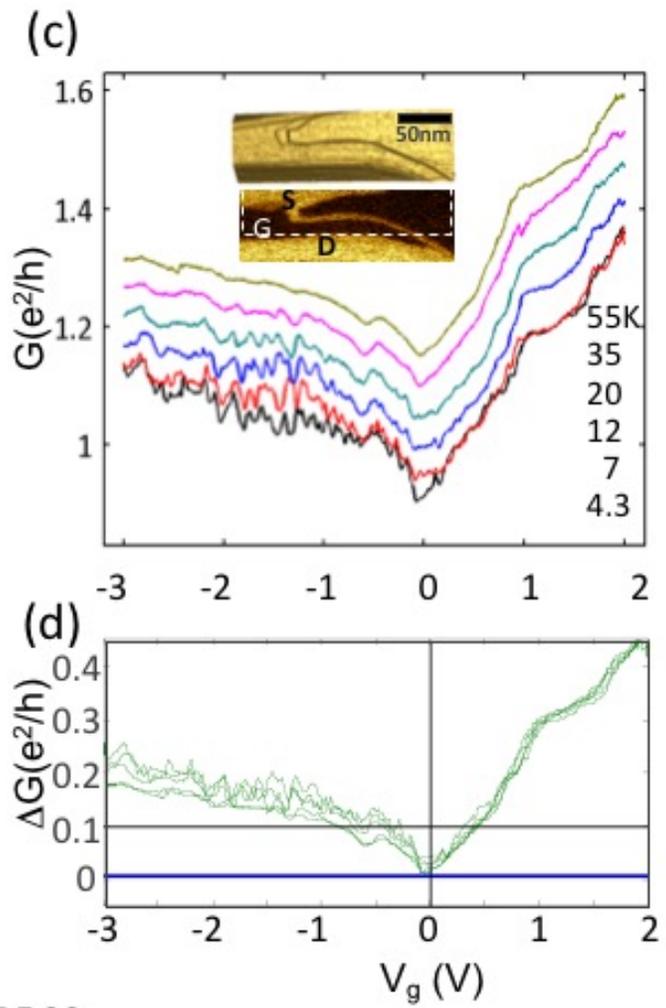

FIGURE 33

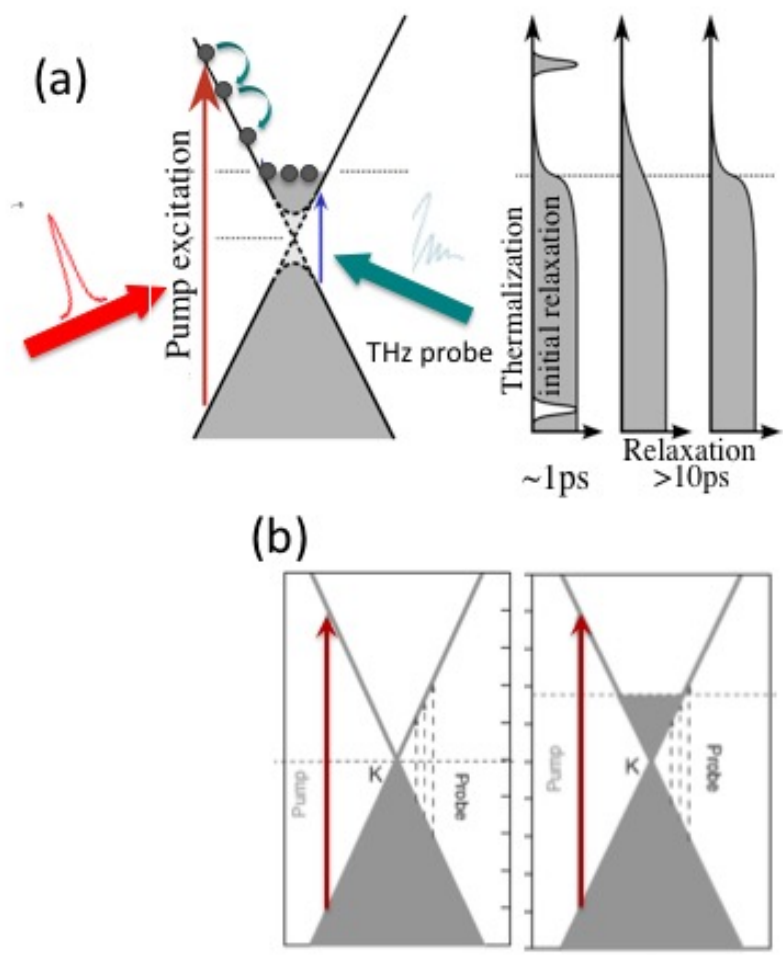

**FIGURE 34**

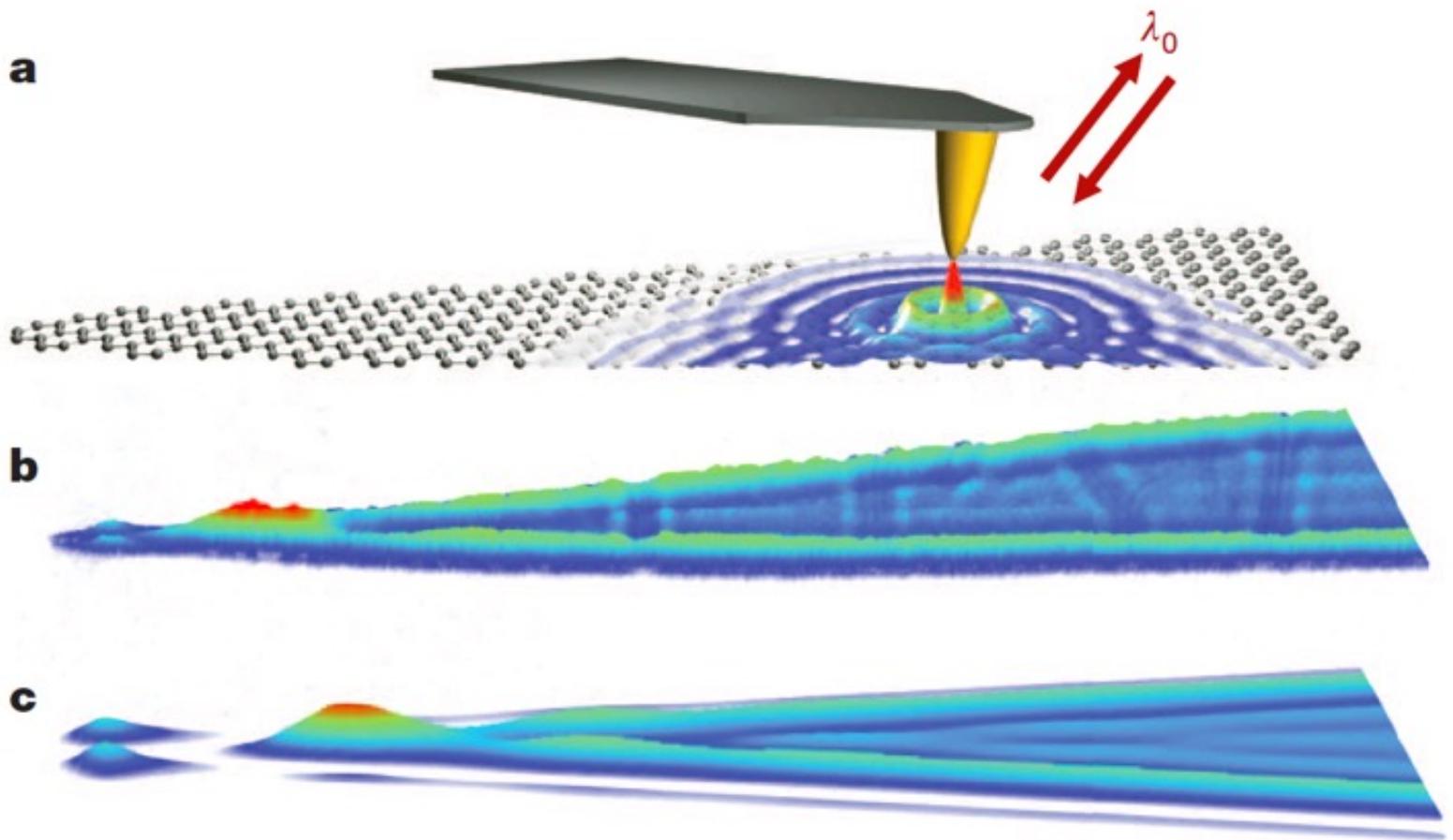

**FIGURE 35**